\documentclass[%
  preprint,
superscriptaddress,
showpacs,preprintnumbers,
nofootinbib,
 amsmath,amssymb, 
 aps,
 longbibliography,
tightenlines
]{revtex4-1}

\usepackage{multirow}
\usepackage{cancel}
\usepackage{accents}
\usepackage{mciteplus,slashed}
\usepackage{amssymb,cancel,amsmath}
\usepackage{dcolumn}% Align table columns on decimal point
\usepackage{bm}% bold math
\usepackage[caption=false]{subfig}
\usepackage{appendix}
\usepackage[T1]{fontenc}	
\usepackage{csvsimple}
\usepackage[colorlinks=true,citecolor=purple,linkcolor=blue, allcolors=violet]{hyperref}
\usepackage[section]{placeins}
\usepackage[capitalise]{cleveref}
\usepackage{booktabs}
\usepackage{graphicx}
\usepackage{mathrsfs}
\usepackage[utf8]{inputenc}

\usepackage[dvipsnames]{xcolor}
\usepackage[normalem]{ulem}

\newcommand{\eg}{\emph{e.g.}}

\newcommand{\lsim}{\mathrel{\hbox{\rlap{\lower.55ex\hbox{$\sim$}} \kern-.3em \raise.4ex \hbox{$<$}}}}
\newcommand{\gsim}{\mathrel{\hbox{\rlap{\lower.55ex\hbox{$\sim$}} \kern-.3em \raise.4ex \hbox{$>$}}}}

\usepackage{tikz}

\newcommand{\ie}{\emph{i.e.~}}

\begin{document}

% PREPRINT NUMBERS
\preprint{\hfill MIT-CTP/5157}
\preprint{\hfill FERMILAB-PUB-19-628-A}

\title{A Systematic Study of Hidden Sector Dark Matter:\\ Application to the Gamma-Ray and Antiproton Excesses}

\author{Dan Hooper}
\thanks{{\scriptsize Email}: \href{mailto:dhooper@fnal.gov}{dhooper@fnal.gov}; {\scriptsize ORCID}: \href{http://orcid.org/0000-0001-8837-4127}{0000-0001-8837-4127}}
\affiliation{\footnotesize Fermilab, Fermi National Accelerator Laboratory, Batavia, IL 60510, USA}
\affiliation{\footnotesize University of Chicago, Kavli Institute for Cosmological Physics, Chicago, IL 60637, USA}
\affiliation{\footnotesize University of Chicago, Department of Astronomy and Astrophysics, Chicago, IL 60637, USA}

\author{Rebecca K. Leane}
\thanks{{\scriptsize Email}: \href{mailto:rleane@mit.edu}{rleane@mit.edu}; {\scriptsize ORCID}: \href{http://orcid.org/0000-0002-1287-8780}{0000-0002-1287-8780}}
\affiliation{\footnotesize Center for Theoretical Physics, Massachusetts Institute of Technology, Cambridge, MA 02139, USA}

\author{\\ Yu-Dai Tsai}
\thanks{{\scriptsize Email}: \href{mailto:ytsai@fnal.gov}{ytsai@fnal.gov}; {\scriptsize ORCID}: \href{https://orcid.org/0000-0002-5763-5758}{0000-0002-5763-5758}}
\affiliation{\footnotesize Fermilab, Fermi National Accelerator Laboratory, Batavia, IL 60510, USA}

\author{Shalma Wegsman}
\thanks{{\scriptsize Email}: \href{mailto:swegsman@uchicago.edu}{swegsman@uchicago.edu}; {\scriptsize ORCID}: \href{http://orcid.org/0000-0002-4378-842X}{0000-0002-4378-842X}}
\affiliation{\footnotesize University of Chicago, Department of Physics, Chicago, IL 60637, USA}

\author{Samuel J. Witte}
\thanks{{\scriptsize Email}: \href{mailto:sam.witte@ific.uv.es}{sam.witte@ific.uv.es}; {\scriptsize ORCID}: \href{http://orcid.org/0000-0003-4649-3085}{0000-0003-4649-3085}}
\affiliation{\footnotesize Instituto de Fisica Corpuscular (IFIC), CSIC-Universitat de Valencia, Spain}

\date{\today}

\begin{abstract}

In hidden sector models, dark matter does not directly couple to the particle content of the Standard Model, strongly suppressing rates at direct detection experiments, while still allowing for large signals from annihilation. In this paper, we conduct an extensive study of hidden sector dark matter, covering a wide range of dark matter spins, mediator spins, interaction diagrams, and annihilation final states, in each case determining whether the annihilations are $s$-wave (thus enabling efficient annihilation in the universe today). We then go on to consider a variety of portal interactions that allow the hidden sector annihilation products to decay into the Standard Model. We broadly classify constraints from relic density requirements and dwarf spheroidal galaxy observations. In the scenario that the hidden sector was in equilibrium with the Standard Model in the early universe, we place a lower bound on the portal coupling, as well as on the dark matter's elastic scattering cross section with nuclei. We apply our hidden sector results to the observed Galactic Center gamma-ray excess and the cosmic-ray antiproton excess. We find that both of these excesses can be simultaneously explained by a variety of hidden sector models, without any tension with constraints from observations of dwarf spheroidal galaxies.

\end{abstract}

\maketitle
\tableofcontents
\newpage

\section{Introduction}

Weakly interacting thermal relics have long constituted the most widely studied class of candidates for dark matter~\cite{Bertone:2016nfn}. In recent years, however, many examples of such candidates have been excluded by the null results of direct detection experiments~\cite{Aprile:2017iyp,Akerib:2016vxi,Cui:2017nnn}, as well as accelerators~\cite{Sirunyan:2018dub,Sirunyan:2018xlo,Sirunyan:2018fpy,Sirunyan:2018wcm,Sirunyan:2018gka,Aaboud:2018xdl,Sirunyan:2017leh,Aaboud:2017phn,Aaboud:2017rzf,Aaboud:2017bja} (see Refs.~\cite{Escudero:2016gzx,Arcadi:2017kky,Balazs:2017ple,Roszkowski:2017nbc,Blanco:2019hah,Leane:2018kjk} for reviews). In light of these developments, it is well motivated to consider models in which the dark matter does not couple directly to the particle content of the Standard Model (SM), but instead annihilates to produce other particles which then decay through small couplings to the SM. Such hidden sector models have been studied extensively in the literature, increasing in interest in recent years~\cite{Pospelov:2007mp,ArkaniHamed:2008qn,Berlin:2016gtr,Dror:2016rxc,Berlin:2016vnh,Blanco:2019eij,Patt:2006fw,McDonald:1993ex,Macias:2015cna,Falkowski:2011xh,Gonzalez-Macias:2016vxy,Macias:2015cna,Cline:2014dwa,Bell:2016fqf,Bell:2016uhg,Batell:2017rol,Campos:2017odj,Bell:2017irk,Escudero:2016ksa,Escudero:2016tzx,Escudero:2017yia,Evans:2017kti,Evans:2019vxr}.

In this paper, we explore a wide range of annihilating dark matter models, covering an extensive combination of dark matter spins, mediator spins, interaction types, and annihilation final states. For each of these combinations, we determine whether the dark matter annihilates through an $s$-wave amplitude (enabling the dark matter annihilation cross section to not vanish at low-velocities) and present this information in our Tables~\ref{tab:lorentzfermion} and~\ref{tab:lorentzboson}. We also consider the decays of hidden sector particles through a variety of portal interactions that connect the hidden sector to the SM (or to an extension of the SM).

Within the context of hidden sector models, we evaluate the ability of annihilating dark matter particles to produce the Galactic Center gamma-ray excess~\cite{Goodenough:2009gk,Hooper:2010mq,Hooper:2011ti,Abazajian:2012pn,Gordon:2013vta,Hooper:2013rwa,Daylan:2014rsa,Calore:2014xka,TheFermi-LAT:2015kwa,TheFermi-LAT:2017vmf}, as well as the more recently identified cosmic-ray antiproton excess~\cite{Cuoco:2016eej,Cui:2016ppb,Cholis:2019ejx,Cuoco:2019kuu}. Although two groups in 2015 claimed that the data favored the existence of a population of astrophysical point sources which was likely responsible for the Galactic Center gamma-ray excess~\cite{Lee:2015fea,Bartels:2015aea}, more recent work has shown these results to be problematic. At this point in time, the available gamma-ray data does not favor the presence of a significant unresolved point source population near the Galactic Center~\cite{Leane:2019xiy,Zhong:2019ycb},  revitalizing interest in dark matter interpretations of this signal. The antiproton excess has been a more recent development in the field, and is inherently subject to a larger range of uncertainties than its gamma ray counterpart (arising \eg~ from the effects of the Solar wind and the production cross section, among others). While various studies have found the excess to be robust to robust to these uncertainties~\cite{Cholis:2019ejx,Cuoco:2019kuu}, at least one group has claimed the excess could be consistent with a secondary origin~\cite{Boudaud:2019efq}. 

In this study, we show that a wide variety of hidden sector dark matter models can simultaneously accommodate both of these excesses (for earlier related work, see Refs.~\cite{Escudero:2017yia,Berlin:2014pya,Abdullah:2014lla,Martin:2014sxa,Tang:2015coo,Hooper:2012cw,Folgado:2018qlv}). We attempt to broadly characterize the ingredients of a hidden sector model that would be required in order to (1) produce s-wave annihilation (and thus produce a sufficient number of gamma-rays and anti-protons), (2) have a final state gamma-ray and anti-proton spectrum peaked in the energy ranges of the respective excesses, (3) evade current constraints, and (4) achieve kinetic equilibrium in the early Universe. As a consequence of living in a hidden sector, these ingredients can to large degree be addressed individually. For example, the first point above concerns (\ie the existence of s-wave annihilation) only the spin of the dark matter, the spin of dark mediator, and their respective interaction vertex, while the second point (\ie producing the correct spectra) constrains the relative masses of the dark matter and dark mediator, and the interaction between the dark mediator and the standard model. We study these requirements individually, which when collectively considered contain all the ingredients to explain both excesses. We emphasize, however, that most of these results are quite general, and can in part be applied to hidden sectors in other contexts as well.

The remainder of this paper is structured as follows. In Sec.~\ref{models}, we consider dark matter annihilation within a wide range of models, determining which scenarios allow for efficient low-velocity annihilation, and thus to potentially observable indirect detection signatures. In Sec.~\ref{portals} we consider several portals through which hidden sector particles could decay to the SM, calculating in each case the corresponding branching fractions. In Sec.~\ref{char} we discuss the Galactic Center gamma-ray excess and the cosmic-ray antiproton excess, describing the observed characteristics of each of these signals. In Sec.~\ref{spectragen} we describe how we calculate the gamma-ray and antiproton spectra in these models, and in Sec.~\ref{dwarfs} we discuss the constraints derived from gamma-ray observations of dwarf spheroidal galaxies. In Sec.~\ref{results}, we present our main results, showing the range of masses and other parameters that can accommodate the observed features of the gamma-ray and antiproton excesses. In Sec.~\ref{cosmo}, we discuss the cosmological considerations in regard to these models and comment on the prospects for future direct detection experiments. We draw conclusions and summarize our results in Sec.~\ref{summary}.

%\newpage

\section{Hidden Sector Dark Matter Annihilation}
\label{models}

In this section, we consider the annihilation of dark matter particles. Although we are primarily interested in this study in dark matter that resides within a hidden sector, the contents of this section can be applied to non-hidden sector scenarios as well. In order for annihilating dark matter to generate signals that are consistent with the observed gamma-ray and antiproton excesses, they must proceed predominantly through $s$-wave processes. Here we will enumerate the varieties of dark matter models that can give rise to $s$-wave dark matter annihilation. Since a truly comprehensive list would be near impossible to enumerate, we restrict our attention to renormalizable interactions between fermions, vectors, and scalars that lead to $2\leftrightarrow 2$ annihilations in the hidden sector. For each operator listed below, we calculate the $\sigma \cdot v$, and expand to the leading orders, to determine if it is a s-wave process or not. 

We begin by considering the case of fermionic dark matter. Generically, spin-$1/2$ particles can interact with spin-1 states through an arbitrary combination of vectorial and axial couplings,\footnote{Majorana fermions interact with spin-1 states only through axial couplings.} and to spin-0 states through any combination scalar and pseudoscalar couplings. To be concrete, the Lagrangian could contain any of the following interactions:
\begin{eqnarray}
\label{fermionints}
    \mathcal{L}^{(V)} & \supset & g_V \overline{\chi}_i \gamma^\mu \chi_j Z^\prime_\mu + \textit{h.c.} \\ \mathcal{L}^{(A)} & \supset & g_A \, \overline{\chi}_i \gamma^\mu \gamma^5 \chi_j Z^\prime_\mu + \textit{h.c.} \nonumber \\ 
    \mathcal{L}^{(S)} & \supset & g_S \overline{\chi}_i \,  \chi_j \, \phi + \textit{h.c.} \nonumber \\
    \mathcal{L}^{(P)} & \supset & g_P \, i \, \overline{\chi}_i \gamma^5 \chi_j \phi + \textit{h.c.} \, , \nonumber
\end{eqnarray}
where $\chi_i$, $Z'_{\mu}$ and $\phi$ denote spin-1/2, spin-1 and spin-0 states, respectively. In these expressions, we remain fully general by considering the possibility that $i \neq j$, unless the fermions are assumed to account for dark matter, in which case we enforce $i = j$, limiting ourselves to the case of a single dark matter species. 

Annihilation diagrams can also include vertices which contain multiple spin-0 particles:
\begin{eqnarray}
    %\mathcal{L}^{(\phi^3)} & = & \lambda_{\phi^3}\, \phi_i^3 \\
    %\mathcal{L}^{(\phi^4)} & = & \lambda_{\phi^4} \, \phi_i^4 \nonumber \\
    %\mathcal{L}^{(\phi_{i^2j})} & = & \lambda_{\phi_{i^2j}} \, \phi_i^2 \, \phi_j \nonumber \\
    %\mathcal{L}^{(\phi_{i^2j^2})} & = & \lambda_{\phi_{i^2j^2}} \, \phi_i^2 \, \phi_j^2 \nonumber \\
    \mathcal{L}^{(\phi_{ijk})} & \supset & \lambda_{\phi_{ijk}} \phi_i \, \phi_j \, \phi_k \\
    \mathcal{L}^{(\phi_{ijkl})} & \supset & \lambda_{\phi_{ijkl}} \, \phi_i \,\phi_j \, \phi_k \, \phi_l \, , \nonumber 
\end{eqnarray}
where $\phi$'s are real scalar fields. Note that the repeated indices do not indicate summation in these expressions. Alternatively, we can consider the following interactions involving gauge bosons:
\begin{eqnarray}\label{eq:scalar-vector_int}
    \mathcal{L}^{(\phi_{ij}z_{km})} & \supset & \lambda_{\phi_{ij}z_{km}}\, \phi_i \, \phi_j \, Z_k^{\prime \, \mu} \, Z_{m, \, \mu}^\prime +\textit{h.c.}\\
    \mathcal{L}^{(\phi z_{km})} & \supset & \lambda_{\phi z_{km}} \, \phi_i \, Z_k^{\prime \, \mu} \, Z_{m, \, \mu}^\prime +\textit{h.c.} \nonumber \\
    \mathcal{L}^{(\phi_{ij} z)} & \supset & \lambda_{\phi_{ij} z} \, \phi_i \, \overset{\text{$\leftrightarrow$}}{\partial^\mu} \, \phi_j  \, Z_{\mu}^\prime +\textit{h.c.}  \nonumber 
\end{eqnarray}

Note that some of these interactions while considered alone, may not be gauge invariant in their current form. In complete UV theories, it may be that several interactions are required in order to not violate unitarity (see e.g. Refs.~\cite{Bell:2015sza, Bell:2015rdw, Bell:2016fqf,Duerr:2016tmh,Bell:2016uhg,Bell:2016ekl,Cui:2017juz}).  More importantly, when constructing portals between the dark sector and the Standard Model, we will define interactions that respect gauge invariance of the Standard Model.

In addition, we can also consider interactions involving complex scalars. Although this allows for the possibility of many different interactions, we include in our study only the following case which can potentially lead to $s$-wave dark matter annihilation:
\begin{eqnarray}
    \mathcal{L}^{(\Phi z)} & \supset & \lambda_{\Phi z} \, i \Phi^*_i \, \overset{\text{$\leftrightarrow$}}{\partial^\mu} \, \Phi_i \, Z_{\mu}^\prime +\textit{h.c.} \, ,
\end{eqnarray}
where $\Phi_i$ is a complex scalar field. 

Lastly, we consider interactions that take place exclusively between spin-$1$ particles. Such interactions can naturally appear in hidden sector models which feature more complex gauge symmetries (see, for example, Refs.~\cite{Hambye:2008bq,Hambye:2009fg,Arina:2009uq,Carone:2013wla,Boehm:2014bia}). Specifically, we allow for the following interactions:
\begin{eqnarray}
\label{vectorints}
    \mathcal{L}^{(z_{ijk})} & \supset & \lambda_{z_{ijk}} \, \partial^\mu \, Z_{i,\mu}^\prime \, Z_{j}^{\prime \, \nu} \, Z_{k,\nu}^\prime +\textit{h.c.} \\ \mathcal{L}^{(z_{ijkm})} & \supset & \lambda_{z_{ijkm}} \, Z_{i}^{\prime \, \mu} \,  Z_{j,\mu}^\prime \, Z_{k}^{\prime \, \nu} \, Z_{m,\nu}^\prime +\textit{h.c.}  \nonumber
\end{eqnarray}
Note that in the first line of this expression, the derivative could act on a vector that is an initial state, final state, or mediator. 

In Tables~\ref{tab:lorentzfermion} and~\ref{tab:lorentzboson}, we list each of the processes through which a fermionic or bosonic dark matter particle could annihilate through an $s$-wave amplitude (see also, Ref.~\cite{Kumar:2013iva,Berlin:2014tja}). We use the labels as indicated in Eqs.~\ref{fermionints}-\ref{vectorints} to denote the types of interactions that in each case leads to an $s$-wave annihilation amplitude. An entry containing $V \otimes A$, for example, should be understood to lead to $s$-wave annihilation if the two vertices of the diagram correspond to vectorial and axial interactions, respectively (as defined in Eq.~\ref{fermionints}). The presence of multiple rows within the same table entry implies that there are multiple interaction combinations that can lead to $s$-wave annihilation. Those cases denoted with a `-' do not correspond to any renormalizable model within our framework. Asterisks indicate cases in which the amplitude is $s$-wave but (if the dark matter is its own antiparticle) the cross section is helicity suppressed ($\sigma v \propto m_f^2/m_{\chi}^2$).

\begin{table}[] \small
\begin{tabular}{|c|c|c|c|c|c|c|}
\hline
\multirow{2}{*}{\textbf{\begin{tabular}[c]{@{}c@{}}Dark Matter\end{tabular}}} & \multicolumn{2}{c|}{\multirow{2}{*}{\textbf{Mediator}}} & \multicolumn{4}{c|}{\textbf{Annihilation Products}} \\ \cline{4-7} 
 & \multicolumn{2}{c|}{} & \textbf{$f_1+\overline{f}_2$} & \textbf{$\phi_1+\phi_2$} & \textbf{$Z'_1+Z'_2$} & \textbf{$\phi+Z'$} \\ \hline \hline
 %row_2
\multirow{4}{*}{\textbf{spin-1/2}} & \multirow{2}{*}{\textbf{$s$-channel}} & \textbf{spin-0} & \begin{tabular}[c]{@{}c@{}}\underline{$\Gamma_{\rm DM}\otimes\Gamma_f$:}\\ $P\otimes P$\\ $P\otimes S$\end{tabular} & 
%update_1
\begin{tabular}[c]{@{}c@{}}
\underline{$\Gamma_{\rm DM}\otimes\Gamma_\phi$:}\\ $P\otimes \phi_{ijk}$\end{tabular} & 
%update_2
\begin{tabular}[c]{@{}c@{}}\underline{$\Gamma_{\rm DM}\otimes\Gamma_{Z'}$:}\\ $P\otimes\phi z_{km}$\end{tabular} &
%update_7
\begin{tabular}[c]{@{}c@{}c@{}}
$\underline{\Gamma_{\rm DM}\otimes\Gamma_{\phi Z'}}$:\\ $P\otimes \phi_{ij}z$
\\ $P\otimes \Phi z$ 
\end{tabular} 
\\ \cline{3-7} 
 &  & \textbf{spin-1} & \begin{tabular}[c]{@{}c@{}}\underline{$\Gamma_{\rm DM}\otimes\Gamma_f$:}\\ $V\otimes V$\\ $V\otimes A$\\ $A\otimes A^*$\end{tabular} & 
 %update_8
 \begin{tabular}[c]{@{}c@{}c@{}}
 $\underline{\Gamma_{\rm DM}\otimes\Gamma_\phi}$:\\ $V\otimes\phi_{ij}z $\\
 $V\otimes\Phi z $ \\ $A\otimes\phi_{ij}z$
 \end{tabular}  & 
%update_3
 \begin{tabular}[c]{@{}c@{}}
 \underline{$\Gamma_{\rm DM} \otimes \Gamma_{Z'}$:}\\ $V\otimes z_{ijk}$ \\ $A\otimes z_{ijk}$\end{tabular} & 
 %update_4
 \begin{tabular}[c]{@{}c@{}}\underline{$\Gamma_{\rm DM} \otimes \Gamma_{\phi Z'}$:}\\ 
 $V\otimes \phi z_{km}$\\ 
 $A\otimes \phi z_{km}$\end{tabular} \\ \cline{2-7} 
 & \multirow{2}{*}{\textbf{$t$-channel}} & \textbf{spin-1/2} & - & 
 \begin{tabular}[c]{@{}c@{}}\underline{$\Gamma_{\phi_1}\otimes\Gamma_{\phi_2}$:}\\ $S\otimes P$\end{tabular} 
 & \begin{tabular}[c]{@{}c@{}}\underline{$\Gamma_{Z'_1}\otimes\Gamma_{Z'_2}$:}\\ $V\otimes V$\\ $V\otimes A$\\ $A\otimes A$\end{tabular} & \begin{tabular}[c]{@{}c@{}}
 $\underline{\Gamma_{\phi}\otimes\Gamma_{Z'}}$:\\ $S\otimes V$ \\ $P\otimes V$
 \end{tabular} \\  \cline{3-7}
&  & \textbf{spin-0} & 
%update_4
\begin{tabular}[c]{@{}c@{}c@{}}\underline{$\Gamma_{f_1}\otimes\Gamma_{\bar{{f_2}}}$}:\\
$S\otimes S$\\
$P\otimes P$
\\
$S\otimes P$
\end{tabular}
& - & - & - \\  \cline{3-7}  & & \textbf{spin-1} 
%update_5
&\begin{tabular}[c]{@{}c@{}c@{}c@{}}
\underline{$\Gamma_{f_1}\otimes\Gamma_{\bar{f}_2}$}: 
\\ 
$V\otimes V$
\\ 
$A\otimes A$
\\ 
$V\otimes A$
\end{tabular}& - & - & -\\
\hline
\end{tabular} 
\caption{For each annihilation diagram, mediator spin, and choice of annihilation products, we identify the combinations of interactions that could enable a fermionic dark matter candidate to annihilate with an $s$-wave amplitude. We use the labels as indicated in Eqs.~\ref{fermionints}-\ref{eq:scalar-vector_int} to denote the types of interactions that lead to an $s$-wave annihilation amplitude. Those cases denoted with a `-' do not correspond to any renormalizable model within our framework. Asterisks indicate cases in which the amplitude is $s$-wave but (if the dark matter is its own antiparticle) the cross section is helicity suppressed ($\sigma v \propto m_f^2/m_{\chi}^2$).}
\label{tab:lorentzfermion}
\end{table}

\begin{table}[] \small
\begin{tabular}{|c|c|c|c|c|c|c|}
\hline
\multirow{2}{*}{\textbf{\begin{tabular}[c]{@{}c@{}}Dark Matter\end{tabular}}} & \multicolumn{2}{c|}{\multirow{2}{*}{\textbf{Mediator}}} & \multicolumn{4}{c|}{\textbf{Annihilation Products}} \\ \cline{4-7} 
 & \multicolumn{2}{c|}{} & \textbf{$f_1+\overline{f}_2$} & \textbf{$\phi_1+\phi_2$} & \textbf{$Z'_1+Z'_2$} & \textbf{$\phi+Z'$} \\ 
 \hline\hline
\multirow{3}{*}{\textbf{spin-0}} & \textbf{$s$-channel} & \textbf{spin-0} & \begin{tabular}[c]{@{}c@{}c@{}}\underline{$\Gamma_{\rm DM}\otimes\Gamma_f$:}\\ $\phi_{ijk}\otimes S$\\
$\phi_{ijk}\otimes P$ 
\end{tabular} & 
%update
\begin{tabular}[c]{@{}c@{}}
\underline{$\Gamma_{\rm DM}\otimes\Gamma_\phi$:}\\ $\phi_{ijk}\otimes \phi_{ijk}$\end{tabular} 
%update
& 
\begin{tabular}[c]{@{}c@{}}
\underline{$\Gamma_{\rm DM}\otimes\Gamma_{Z'}$:}\\ $\phi_{ijk}\otimes \phi z_{km}$\end{tabular} 
%update
& 
\begin{tabular}[c]{@{}c@{}c@{}}
\underline{$\Gamma_{\rm DM}\otimes\Gamma_{\phi Z'}$:}\\ $\phi_{ijk}\otimes \phi_{ij} z$
\end{tabular} 
\\ \cline{2-7} 
& \textbf{$s$-channel} & \textbf{spin-1} & None 
& 
None
& 
None
& 
None \\ \cline{2-7} 
 & \multirow{2}{*}{\textbf{$t$-channel}} & \textbf{spin-0} & - & 
 \begin{tabular}[c]{@{}c@{}}\underline{$\Gamma_{\phi_1}\otimes\Gamma_{\phi_2}$:}\\ $\phi_{ijk}\otimes \phi_{ijk}$\end{tabular}  
 & 
 \begin{tabular}[c]{@{}c@{}c@{}}\underline{$\Gamma_{Z'_1}\otimes\Gamma_{Z'_2}$:}\\ $\phi_{ij}z\otimes \phi_{ij}z$
 \\ $\Phi z\otimes \Phi z$
 \end{tabular}  
 & 
 \begin{tabular}[c]{@{}c@{}c@{}}\underline{$\Gamma_{\phi}\otimes\Gamma_{Z'}$:}\\ $\phi_{ijk} \otimes \phi_{ij}z$
 \end{tabular}  
 \\ \cline{3-7} 
 &  & \textbf{spin-1/2} & 
 \begin{tabular}[c]{@{}c@{}c@{}}\underline{$\Gamma_{f_1}\otimes\Gamma_{\bar{f}_2}$\::}\\
$S\otimes S$\\
$P\otimes P$ 
\\
$S\otimes P$
\end{tabular} & - & - & - \\ \cline{3-7}  & & \textbf{spin-1}
& - & 
\begin{tabular}[c]{@{}c@{}}
\underline{$\Gamma_{\phi_1}\otimes\Gamma_{\phi_2}$:}\\ $\phi_{ij} z \otimes \phi_{ij} z$ 
\end{tabular}
& 
\begin{tabular}[c]{@{}c@{}}
\underline{$\Gamma_{Z'_1}\otimes\Gamma_{Z'_2}$:}\\ $\phi z_{km}\otimes \phi z_{km}$\end{tabular} 
& 
\begin{tabular}[c]{@{}c@{}}
\underline{$\Gamma_{\phi}\otimes\Gamma_{Z'}$:}\\ $\phi_{ijk} \otimes \phi_{ij} z$
\end{tabular} 
\\ \hline
\hline
\multirow{4}{*}{\textbf{spin-1}} & \multirow{2}{*}{\textbf{$s$-channel}} & \textbf{spin-0} & 
%update
\begin{tabular}[c]{@{}c@{}}\underline{$\Gamma_{\rm DM}\otimes\Gamma_f$:}\\ $\phi z_{km}\otimes S$
\\ $\phi z_{km}\otimes P$ 
\end{tabular} 
& 
\begin{tabular}[c]{@{}c@{}}\underline{$\Gamma_{\rm DM}\otimes\Gamma_\phi$:}\\ $\phi z_{km}\otimes \phi_{ijk}$ 
\end{tabular} 
& 
\begin{tabular}[c]{@{}c@{}}\underline{$\Gamma_{\rm DM}\otimes\Gamma_{Z'}$:}\\ $\phi z_{km}\otimes \phi z_{km}$ 
\end{tabular}  
& 
\begin{tabular}[c]{@{}c@{}}\underline{$\Gamma_{\rm DM}\otimes\Gamma_{\phi Z'}$:}\\ $\phi z_{km}\otimes \phi_{ij} z$  
\end{tabular}  
\\ \cline{3-7} 
 &  & \textbf{spin-1} & 
 \begin{tabular}[c]{@{}c@{}}\underline{$\Gamma_{\rm DM}\otimes\Gamma_f$:}\\ $ z_{ijk}\otimes V^*$ 
 \\ $ z_{ijk}\otimes A^*$
 \end{tabular}  
 & None 
 & 
 \begin{tabular}[c]{@{}c@{}}\underline{$\Gamma_{\rm DM}\otimes\Gamma_{Z'}$:}\\ $z_{ijk}\otimes z_{ijk}$ 
\end{tabular}  
 &
\begin{tabular}[c]{@{}c@{}}\underline{$\Gamma_{\rm DM}\otimes\Gamma_{\phi Z'}$:}\\ $z_{ijk}\otimes \phi z_{km}$ 
\end{tabular}  
\\ \cline{2-7} 
 & \multirow{2}{*}{\textbf{$t$-channel}} & \textbf{spin-0} & - 
 & 
 \begin{tabular}[c]{@{}c@{}@{}}
 \underline{$\Gamma_{\phi_1}\otimes\Gamma_{\phi_2}$:}\\ $\phi_{ij} z\otimes \phi_{ij} z$
 \\ $\Phi z \otimes \Phi z$
\end{tabular}
 & 
 \begin{tabular}[c]{@{}c@{}}\underline{$\Gamma_{Z'_1}\otimes\Gamma_{Z'_2}$:}\\ $\phi z_{km}\otimes \phi z_{km}$ 
\end{tabular}  
 & 
 \begin{tabular}[c]{@{}c@{}}\underline{$\Gamma_{\phi}\otimes\Gamma_{Z'}$:}\\ $\phi_{ij} z\otimes \phi z_{km}$  
\end{tabular} 
 \\ \cline{3-7} 
&  & \textbf{spin-1} & - & 
\begin{tabular}[c]{@{}c@{}}\underline{$\Gamma_{\phi_1}\otimes\Gamma_{\phi_2}$:}\\ $\phi z_{km}\otimes \phi z_{km}$ 
\end{tabular}  
& 
\begin{tabular}[c]{@{}c@{}}\underline{$\Gamma_{Z'_1}\otimes\Gamma_{Z'_2}$:}\\ $z_{ijk}\otimes z_{ijk}$ 
\end{tabular}  
& 
\begin{tabular}[c]{@{}c@{}}\underline{$\Gamma_{\phi}\otimes\Gamma_{Z'}$:}\\ $\phi z_{km}\otimes z_{ijk}$ 
\end{tabular}  
\\ \cline{3-7}  & & \textbf{spin-1/2} & \begin{tabular}[c]{@{}c@{}c@{}c@{}c@{}}
\underline{$\Gamma_f\otimes\Gamma_{\bar{f}}$\::}\:
\\ 
$V\otimes V$
\\
$A\otimes A$
\\ 
$V\otimes A$
\end{tabular} & - & - & - \\ \hline 
\end{tabular}
\caption{As in Table~\ref{tab:lorentzfermion}, but for spin-0 or spin-1 dark matter candidates. We use the labels as indicated in Eqs.~\ref{fermionints}-\ref{vectorints} to denote the types of interactions that lead to an $s$-wave annihilation amplitude. In those entries labeled ``none'', we found that none of the interactions described in this section lead to an $s$-wave amplitude for dark matter annihilation.}
%Daggers ($^{\dagger}$) indicate cases in which the amplitude is $s$-wave but the cross section is still suppressed by two powers of velocity ($\sigma v \propto v^2$).
\label{tab:lorentzboson}
\end{table}

%Since we consider the dark matter initial states to be one species, the inelastic coupling term in Eq. \ref{eq:scalar-vector_int} would vanishes (see \cite{Berlin:2014tja} for detailed discussions). The initial dark matter state is thus a complex scalar in this case in Table \ref{tab:lorentz}. In other situations, the real scalar interactions capture the processes of interest, but we still include the s-wave processes involving this specific vertex.

Although the range of interactions described in Tables~\ref{tab:lorentzfermion} and~\ref{tab:lorentzboson} does not strictly cover all of the possibilities that could give rise to $s$-wave annihilation (in particular, our treatment may leave unaddressed some models featuring composite particles~\cite{Kribs:2016cew} or $2\rightarrow3$ processes~\cite{Bell:2017irk}), it is quite general and captures a broad range of phenomenological possibilities.

\section{Portals Between the Hidden Sector and the Standard Model}
\label{portals}

In the previous section, we identify a wide range of hidden sector models in which the dark matter annihilates through an $s$-wave amplitude. While this is a necessary condition to generate the gamma-ray and antiproton excesses, it is not necessarily sufficient. In addition, these annihilations must produce spectra of gamma rays and antiprotons that are consistent with the signals measured by \textit{Fermi} and \textit{AMS}. Within the context of hidden sector models, this occurs through the decays of the dark matter annihilation products into SM states through a small ``portal'' interaction. In this section, we discuss a range of such hidden sector portals and their prospects for generating the Galactic Center gamma-ray and cosmic-ray antiproton excesses.

In the subsections below, we will discuss a number of simple and well-motivated portals capable of connecting a hidden sector to the particle content of the SM. In Fig.~\ref{figpie}, we summarize the branching fractions that result from several of these portal couplings. It is important to note that in order to produce gamma-rays and antiprotons with the spectra observed by Fermi and AMS-02, it is typically necessary to have mediator masses in the range $20 \lesssim m_{\rm med} \lesssim 100$ GeV. This range of masses is sufficiently above the b-quark threshold and below the threshold for electroweak gauge boson contributions that the branching ratios shown in Fig.~\ref{figpie} are, to a large degree, independent of mass. Nevertheless, we include the mass dependent branching ratios in all calculations. While this list should not be considered extensive, it should be considered as representative of the final states that may appear in viable hidden sector models.

\subsection{Hypercharge Portal}

One of the renormalizable portals to the SM is the so-called hypercharge portal, arising from the kinetic mixing between SM hypercharge, $U(1)_Y$, and a new $U(1)_D$ gauge symmetry. This portal corresponds to a term in the Lagrangian of the form $\epsilon F_{\mu\nu}F^{\prime \, \mu\nu}$, where $F_{\mu\nu}$ and $F^{\prime \, \mu\nu}$ are the hypercharge and dark field strength tensors, respectively~\cite{Pospelov:2007mp,Krolikowski:2008qa,Holdom:1985ag,Okun:1982xi}. Upon restoring the kinetic normalization, the dark gauge boson develops an $\epsilon$-suppressed coupling to the hypercharge current. Consequently, the decay width of the $Z'$ to SM fermions is given by:
  \begin{equation}
   \Gamma_{Z'}= \sum_f \frac{m_{Z'}N_c}{12\pi}\sqrt{1-\frac{4m_f^2}{m_{Z'}^2}}\,\left[g_{fV}^2\left(1+\frac{2m_f^2}{m_{Z'}^2}\right)+g_{fA}^2\left(1-\frac{4m_f^2}{m_{Z'}^2}\right)\right],
   \label{eq:zppartial}
 \end{equation}
where $N_c$ is the color factor corresponding to fermion, $f$. Following Ref.~\cite{Hoenig:2014dsa}, the vector and axial couplings are given by $g_{fV,fA}\equiv (g_{f_R} \pm g_{f_L})/2$, where 
\begin{equation}
g_{f_R,f_L} = \epsilon \bigg(\frac{m^2_{Z'} g_Y Y_{f_{R,L}}-m^2_Z g \sin \theta_W Q_f}{m^2_Z-m^2_{Z'}}\bigg).    
\end{equation}
Here, $g_Y$ and $g$ are the SM gauge couplings, $Y_{f_{R,L}}$ and $Q_f$ are the SM hypercharge and electric charge assignments, and $\theta_W$ is the weak mixing angle. 

Models in which the dark matter is part of a hidden sector that is connected to the SM through the hypercharge portal have been previously shown to provide a good fit to the Galactic Center gamma-ray excess~\cite{Escudero:2017yia}.

\subsection{$B-L$ Portal}

Although kinetic mixing with hypercharge is the only renormalizable portal capable of connecting a hidden sector $Z'$ to the particle content of the SM, one could also consider scenarios in which the SM is extended by one or more additional gauge symmetries, leading to additional vectors which could mix with a hidden sector $Z'$. In this and the following two subsections, we will consider several examples of such scenarios.

A well-motivated extension of the SM arises from gauging the combination of baryon number minus lepton number, $B-L$, which is anomaly free after the inclusion of three right-handed neutrinos. This model has been explored extensively over the years as it provides a simple framework for the implementation of the seesaw mechanism, and is thus capable of explaining the smallness of the neutrino masses (see, e.g., \cite{Mohapatra:1980qe,WETTERICH1981343,Buchmuller:1991ce}). 

Here, we consider a scenario in which the gauge boson associated with the group $U(1)_{B-L}$ undergoes kinetic mixing with a gauge boson within a hidden sector, $Z'$. Similar to the case of the hypercharge portal, upon restoring the canonical normalization of the kinetic terms, the dark gauge boson develops the following couplings to the $B-L$ current:
\begin{eqnarray}
\label{BLeq}
g_{fV} &\cong& \epsilon \, g_{B-L} \, (B-L)_f \, \bigg|\frac{m^2_{Z_{B-L}} + m^2_{Z'}} {m^2_{Z_{B-L}}-m^2_{Z'}}\bigg| \, ,  \\
g_{fA} &=& 0 \, , \nonumber
\end{eqnarray}
where $g_{B-L}$ is the gauge coupling associated with $U(1)_{B-L}$ and $(B-L)_f$ is the baryon number minus lepton number of fermion, $f$. The decay width of the $Z'$ can then be calculated using Eq.~\ref{eq:zppartial} with couplings as given above in Eq.~\ref{BLeq}~\cite{Escudero:2018fwn}.\footnote{In scenarios in which a hidden sector $Z'$ mixes with $Z_{B-L}$ (or with $Z_B$ or $Z_{L_i-L_j}$, as discussed in the followings subsections), hypercharge will also participate in the mixing. In generality, we expect the branching fractions of such a $Z'$ to be an admixture of these two cases.}

%
%\begin{equation}
%    \mathcal{L} \supset \, \epsilon \, Z^D_{\mu} J^\mu_{B-L}
%\end{equation}
%with
%\begin{equation}
%    J^\mu_{B-L} = g_{B-L}\left( \sum_i \frac{1}{3} \overline{q}_i \gamma^\mu q_i - \sum_j \overline{\ell}_j \gamma^\mu \ell_j\right) \, .
%\end{equation}

\subsection{Baryon Portal}

Next, we consider the case in which we gauge the baryon number (see, e.g., \cite{carone1995possible,perez2010baryon,duerr2013gauge}). The kinetic mixing between the gauge boson associated with this symmetry, $Z_B$, and a hidden sector $Z'$ leads to couplings similar to those shown in Eq.~\ref{BLeq}:
\begin{eqnarray}
\label{Beq}
g_{fV} &\cong& \epsilon \, g_{B} \, B_f \, \bigg|\frac{m^2_{Z_{B}} + m^2_{Z'}} {m^2_{Z_{B}}-m^2_{Z'}}\bigg|\, , \\
g_{fA} &=& 0, \nonumber
\end{eqnarray}
where $g_B$ is the gauge coupling associated with $U(1)_B$ and $B_f$ is the baryon number of fermion, $f$. Again, the decay width of the $Z'$ can then be calculated using Eq.~\ref{eq:zppartial} with couplings from Eq.~\ref{Beq}.

\subsection{$L_i-L_j$ Portal}

Lastly, we consider the case in which we gauge the difference of two lepton families, $L_e-L_{\mu}$, $L_{\mu}-L_{\tau}$ or $L_e-L_{\tau}$. This class of extensions of the SM is particularly attractive given that they do not require any additional particle content to cancel anomalies~\cite{He:1990pn,He:1991qd}. The kinetic mixing between $L_i-L_j$ and a hidden sector $Z'$ leads to couplings that are similar to those shown in Eqs.~\ref{BLeq} and~\ref{Beq}~\cite{Escudero:2019gzq}:
\begin{eqnarray}
g_{fV} &\cong& \epsilon \, g_{L_i-L_j} \, (L_i-L_j)_f \, \bigg|\frac{m^2_{Z_{L_i-L_j}} + m^2_{Z'}} {m^2_{Z_{L_i-L_j}}-m^2_{Z'}}\bigg|\, , \\
g_{fA} &=& 0 \, , \nonumber
\label{BL}
\end{eqnarray}
where $g_{L_i-L_j}$ is the gauge coupling associated with $U(1)_{L_i-L_j}$. Given that these couplings lead the $Z'$ to decay entirely to leptons (at tree level~\cite{Fox:2008kb,Bell:2014tta,DEramo:2017zqw}), this portal cannot generate the measured flux of the cosmic-ray antiproton excess. We include it here only for completeness.

\subsection{Higgs Portal}

A hidden sector scalar can decay to the SM through mass mixing with the SM Higgs boson (see, e.g., \cite{McDonald:1993ex,Burgess:2000yq}). 
As the Higgs couples to fermions proportionally to their mass, the decay of the $Z'$ will in this case
 be predominantly to $b$ quarks in the mass range that we are considering, here. In our calculations, however, we include all SM decay products. As loop decays and higher-order corrections can be
 relevant in this case, we use the {\sc Fortran} package {\sc
   HDecay} \cite{Djouadi:1997yw} to calculate these branching fractions, which takes these effects into account.

\subsection{Two-Higgs Doublet Portal}

While the Higgs portal (described above) provides perhaps the simplest way for a hidden sector scalar to decay to the SM, it is also straightforward to construct models in which a hidden sector scalar decays to one or more particle species within an extended Higgs sector. A model that is comparable in complexity to the Higgs portal, but that allows for a richer phenomenology, is that in which a hidden sector scalar mixes with the scalars found within a two-Higgs doublet model (2HDM). In this case, the inclusion of a second Higgs doublet allows for a spin-$0$ particle to couple, for example, asymmetrically to up-like and down-like fermions, preferentially to charged leptons, or even in a gauge-phobic manner. Rather than exploring the full range of possibilities here, we will focus on some of the limiting cases in which the two Higgs doublets couple in a manner that is noticeably different from that found in the case of the Higgs portal, with the understanding that a 2HDM portal could also easily mimic the behavior found in the conventional Higgs portal. A more extensive review of this model can be found in Ref.~\cite{Branco:2011iw}.

{\bf Type I:} In this case, the couplings of fermions to the lighter Higgs boson are modified by a factor of $\cos\alpha / \sin\beta$, while the couplings to $W^+ W^-$ and $ZZ$ are modified by $\sin(\alpha - \beta)$, where $\alpha$ and $\beta$ are free parameters of the theory. For the heavy Higgs, the corresponding factors are $\sin\alpha / \sin\beta$ and $\cos(\alpha-\beta)$, respectively. Particularly interesting limits of such models exist in which one of the Higgs states couples preferentially to gauge bosons (fermio-phobic) or to fermions (gauge-phobic).

{\bf Type II (MSSM-like):} In this model, the couplings of the light Higgs to up-type quarks are similar to those found in the Type-I model. The couplings of the light Higgs to down-type quarks and to leptons, in contrast, are modified relative to that of the SM Higgs by a factor of $-\sin\alpha / \cos\beta$. Again, there are a number of interesting limits of this model to consider, giving rise to different branching fractions. First, one can quite easily enhance the coupling to $b$ quarks, leading to phenomenology similar to that found in the standard Higgs portal scenario. Alternatively, one can suppress the coupling to $b$ quarks while enhancing the branching fraction to $\tau^+ \tau^-$, or to $c\bar{c}$, $\tau^+ \tau^-$ and $gg$~\cite{Arhrib:2009hc}.

\subsection{Neutrino Portal}

In the case of the neutrino portal, a spin-1/2 hidden sector particle, $N$, can decay through mixing with an SM neutrino, resulting in the production of a lepton along with an on- or off-shell gauge or Higgs boson. If $m_N$ is large enough to enable on-shell decays, the partial widths are given by~\cite{Pilaftsis:1991ug}:
\begin{eqnarray}
    \Gamma(N \rightarrow W^\pm \ell^\mp_\alpha) & = & \frac{g^2}{64\pi} |U_{\alpha N}|^2 \, \frac{M_N^3}{M_W^2}\left(1 - \frac{M_W^2}{M_N^2} \right)^2\,\left(1 + \frac{2 M_W^2}{M_N^2} \right) \, ,\\
    \Gamma(N \rightarrow Z \nu_\alpha) & = & \frac{g^2}{64\pi c_w^2} |C_{\alpha N}|^2 \, \frac{M_N^3}{M_Z^2}\left(1 - \frac{M_Z^2}{M_N^2} \right)^2\,\left(1 + \frac{2 M_Z^2}{M_N^2} \right) \, , \nonumber \\
    \Gamma(N \rightarrow h \nu_\alpha) & = & \frac{g^2}{64\pi} |C_{\alpha N}|^2 \, \frac{M_N^3}{M_W^2}\left(1 - \frac{M_h^2}{M_N^2} \right)^2\, . \nonumber
\end{eqnarray}

Alternatively, if $m_N \lsim m_W$, the decays will be dominated by three-body final states:
\begin{eqnarray}
    \Gamma(N \rightarrow \nu \overline{q}q) &=& 3\, A \, C_{NN} \left[2 (a_u^2 + b_u^2) + 3(a_d^2 +b_d^2) \right]f(z)\, , \\
    \Gamma(N \rightarrow 3\nu) &=&  A \, C_{NN} \left[\frac{3}{4}f(z) + \frac{1}{4}g(z)\right] \, , \nonumber \\
    \Gamma(N \rightarrow \ell \overline{q}q) &=& 6\, A \, C_{NN} f(\omega, 0) \, , \nonumber \\
    \Gamma(N \rightarrow \nu \overline{\ell}\ell) &=&  A \, C_{NN} \left[3 (a_e^2 + b_e^2)f(z) + 3f(\omega) - 2 a_e g(z,\omega) \right] \, ,  \nonumber
\end{eqnarray}
where 
\begin{eqnarray}
    A & \equiv & \frac{G_F^2 M_N^5}{192\pi^3}\, , \,\,
    C_{ij}  \equiv  \sum_{\alpha = 1}^3 \, U_{\alpha i}U^*_{\alpha j} \, , \,\,  
    z  \equiv  \left( \frac{M_N}{M_Z} \right)^2 \, , \,\,
    \omega  \equiv \left( \frac{M_N}{M_W} \right)^2 \, ,
\end{eqnarray}
and $a_f,b_f$ are the left and right neutral current couplings to fermion, $f$, and the functions $f(z), f(\omega, 0)$, and $g(z,\omega)$ are given in Ref.~\cite{Dittmar:1989yg}.

As this portal typically involves 3-body boosted decays, we do not include an analysis of this case in our study. We do note, however, that this model has previously been shown to be capable of fitting the Galactic Center gamma-ray excess~\cite{Tang:2015coo,Campos:2017odj,Batell:2017rol,Folgado:2018qlv} as well as the cosmic-ray antiproton excess~\cite{Folgado:2018qlv}.

\begin{figure*}[t!]
\centering
\subfloat{\includegraphics[width=0.49\columnwidth]{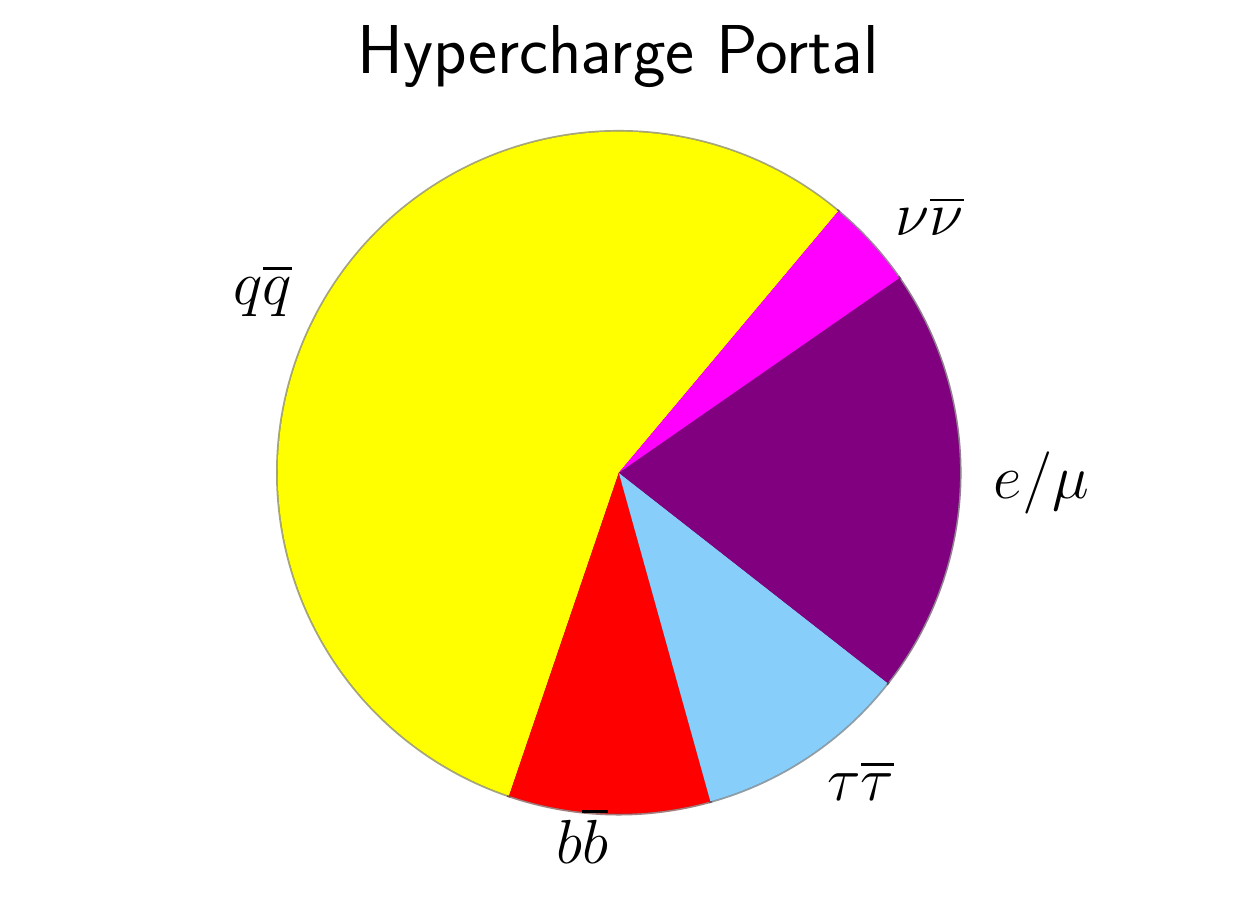}}
\subfloat{\includegraphics[width=0.49\columnwidth]{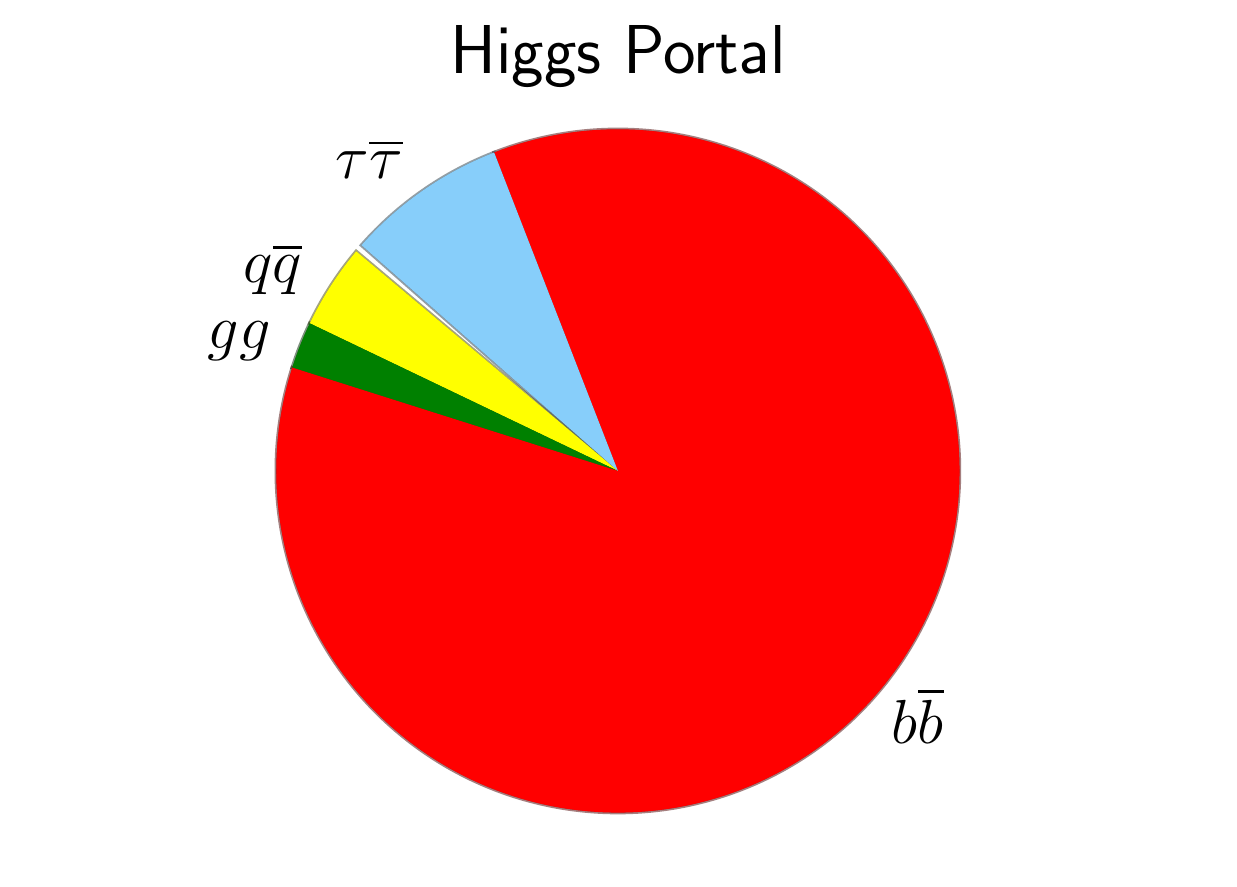}}\\
\vspace{8mm}
\subfloat{\includegraphics[width=0.49\columnwidth]{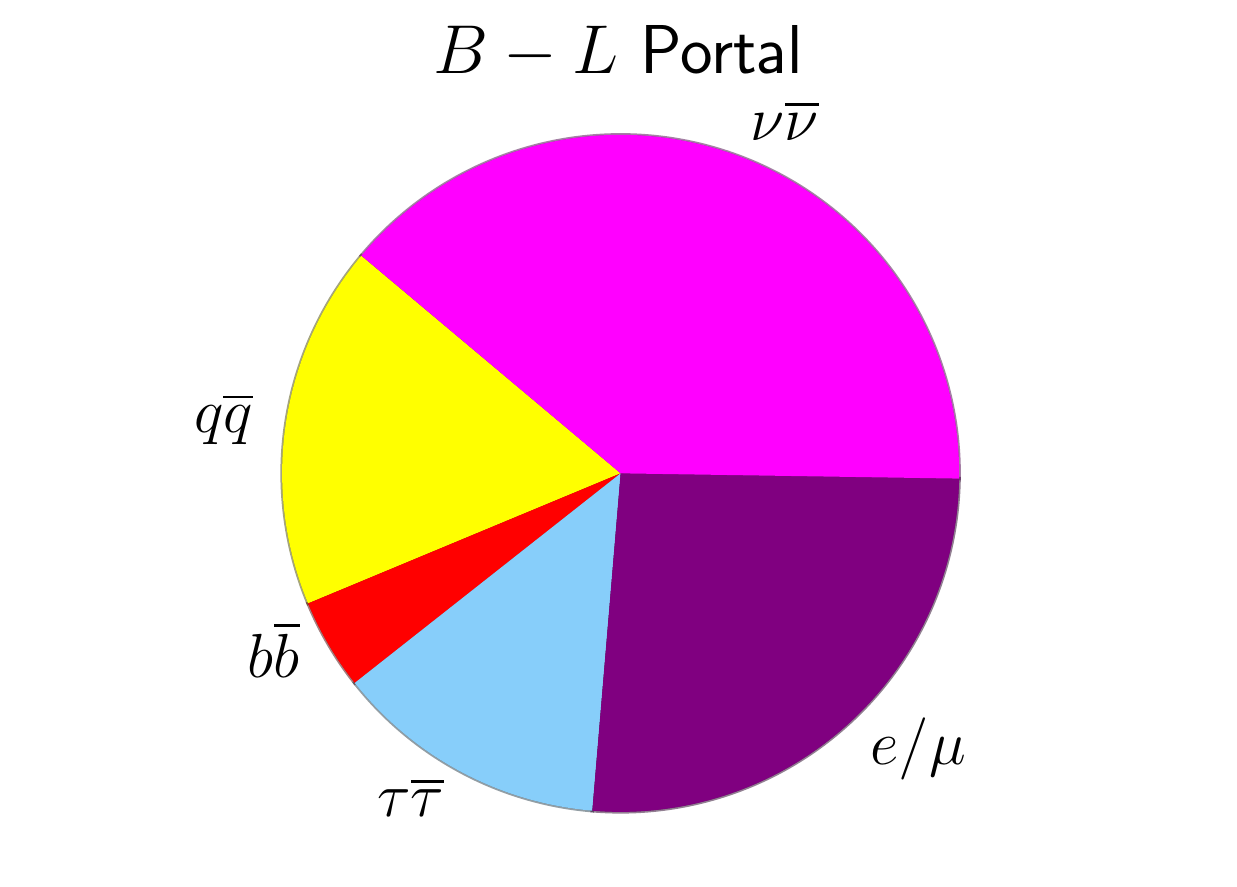}}
\subfloat{\includegraphics[width=0.49\columnwidth]{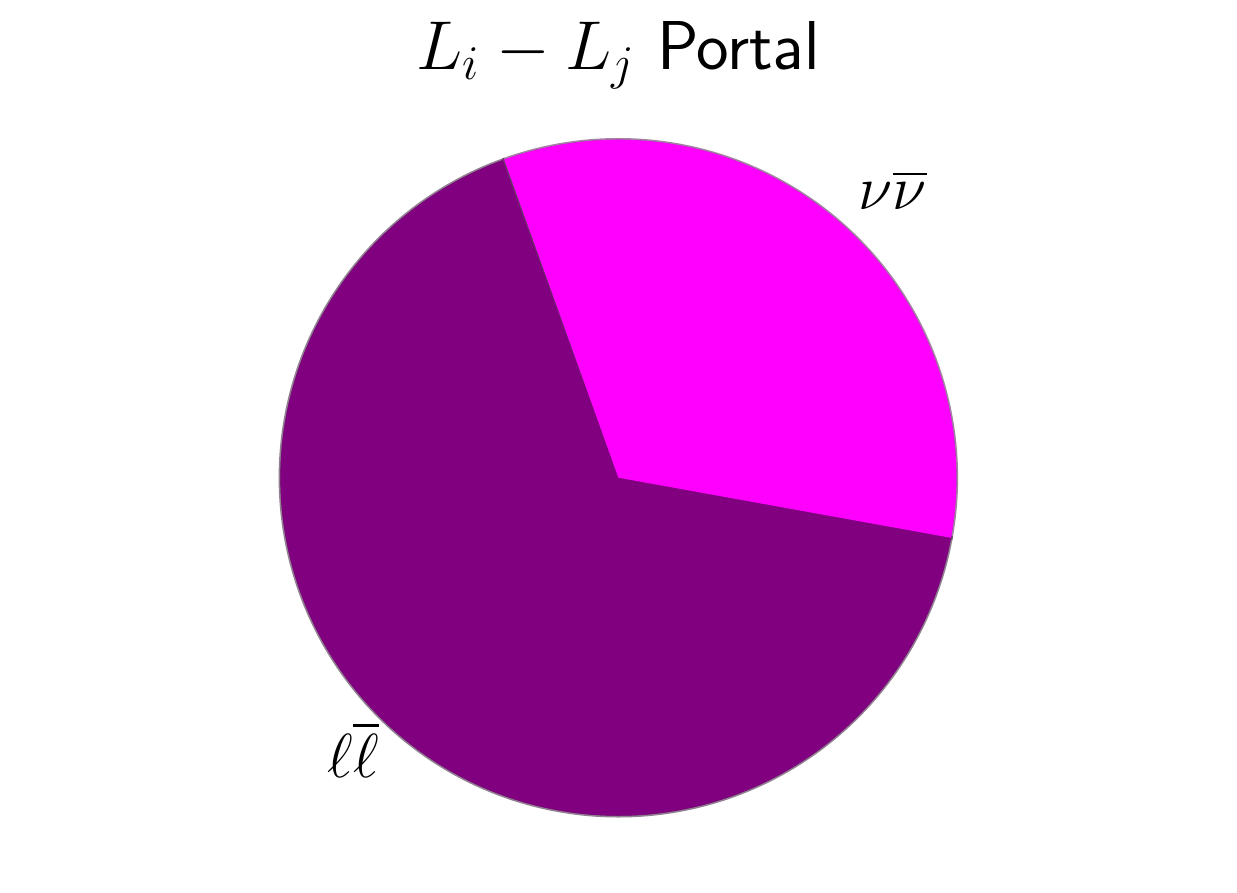}}\\
\vspace{8mm}
\subfloat{\includegraphics[width=0.49\columnwidth]{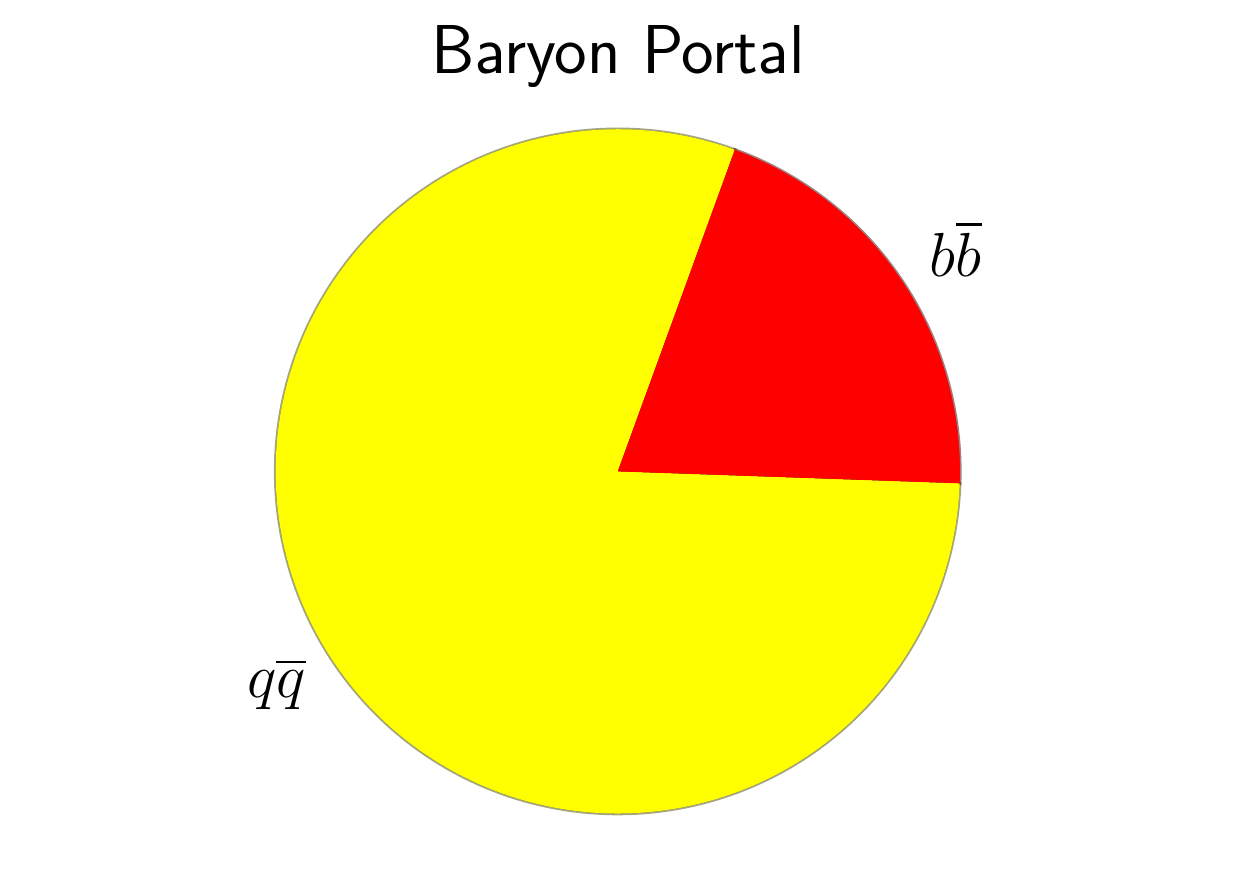}}
\subfloat{\includegraphics[width=0.49\columnwidth]{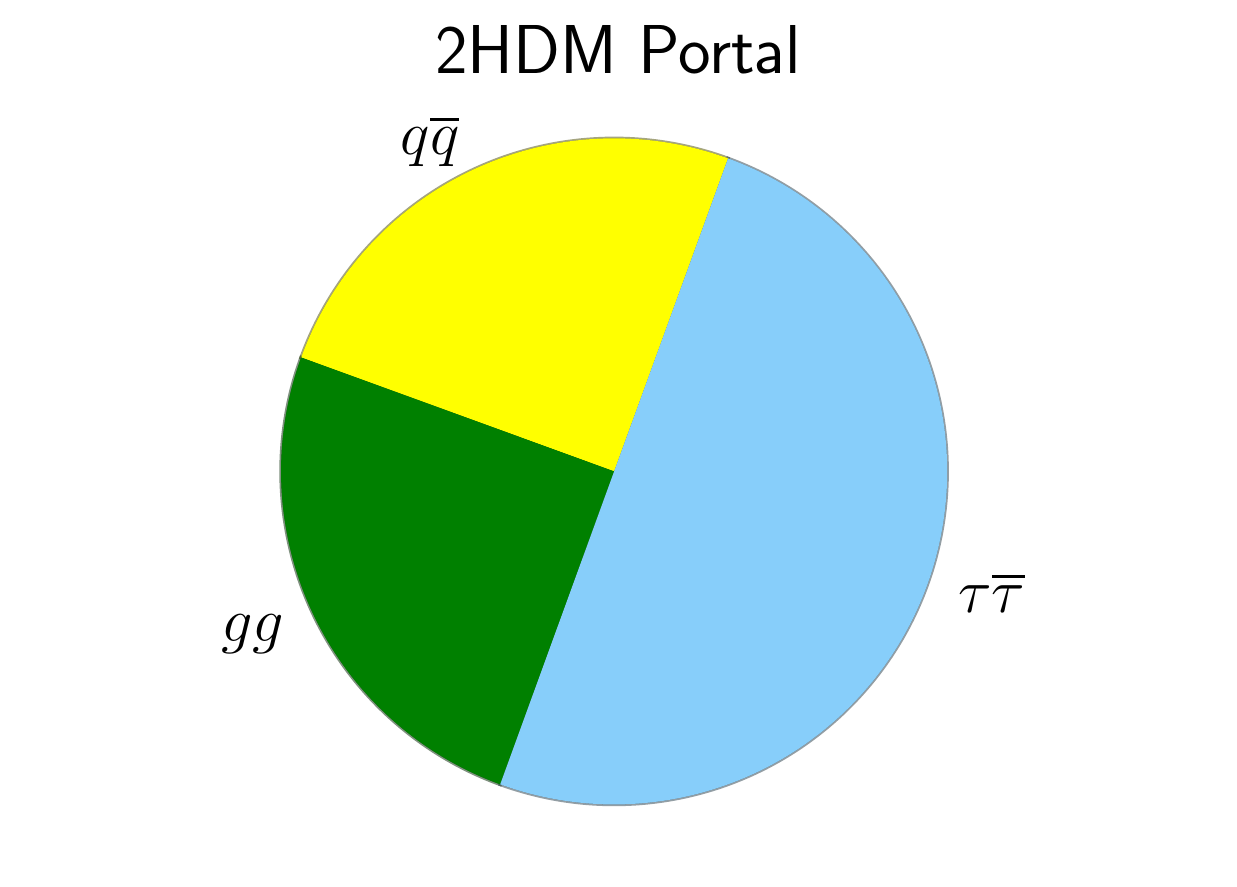}}
\caption{Branching fractions of hidden sector particles decaying through various portal interactions. In each case, we have adopted a mass of 50 GeV for the decaying particle. In the case of the two-Higgs doublet model, we show results for a Type-II model with $\tan \beta =1$, $\sin \alpha =0$. Note that $q\bar{q}$ in this figure denotes any of the four lightest quark species ($u$, $d$, $s$, $c$). \label{figpie}}
\end{figure*}

\section{Characteristics of the Gamma-Ray and Antiproton Excesses}
A number of ingredients are necessary in order for a hidden sector to be capable of explaining the astrophysical excesses. As discussed in previous sections, the dark matter must annihilate to dark mediators through a s-wave process (see Section \ref{models}), and the mediator must decay to SM particles through a weak portal coupling (see Section \ref{portals}). The final requirement is that the final state gamma-ray and antiproton spectra resemble the observed features of each excess. In general, the final state spectra is independent of the interaction vertex that lead to the s-wave annihilation, but it does strongly depend on the mediator mass and the branching fractions to final state particles. In the following section we will address the extent to which the portals introduced in Section \ref{portals} can produce the robust features of these anomalies, however before doing so we must define the features of each excess which have been identified as robust -- that is the purpose of this section.  

\label{char}

\subsection{The Galactic Center Gamma-Ray Excess}

A bright excess of GeV-scale gamma-rays has been observed from the region surrounding the Galactic Center, with a spectral shape, morphology and intensity that are each consistent with the signal predicted from annihilating dark matter~\cite{Goodenough:2009gk,Hooper:2010mq,Hooper:2011ti,Abazajian:2012pn,Gordon:2013vta,Hooper:2013rwa,Daylan:2014rsa,Calore:2014xka,TheFermi-LAT:2015kwa,TheFermi-LAT:2017vmf}. Although this signal has received a great deal of interest within the context of dark matter~\cite{Ipek:2014gua,Izaguirre:2014vva,Agrawal:2014una,Berlin:2014tja,Alves:2014yha,Boehm:2014hva,Huang:2014cla,Cerdeno:2014cda,Okada:2013bna,Freese:2015ysa,Fonseca:2015rwa,Bertone:2015tza,Cline:2015qha,Berlin:2015wwa,Caron:2015wda,Cerdeno:2015ega,Liu:2014cma,Hooper:2014fda,Arcadi:2014lta,Cahill-Rowley:2014ora,Ko:2014loa,McDermott:2014rqa,Abdullah:2014lla,Martin:2014sxa,Berlin:2014pya,Hooper:2012cw,TheFermi-LAT:2017vmf,Carena:2019pwq,Karwin:2016tsw,Escudero:2017yia,Tang:2015coo,Escudero:2016kpw}, astrophysical explanations of this emission have also been  extensively considered. In particular, scenarios have been proposed in which the gamma-ray excess is generated by a large population of unresolved millisecond pulsars~\cite{Cholis:2014lta,Petrovic:2014xra,Brandt:2015ula,Lee:2015fea,Hooper:2015jlu,Bartels:2015aea,Hooper:2016rap,Hooper:2010mq,Abazajian:2010zy,Abazajian:2012pn,Hooper:2013nhl,Gordon:2013vta,Abazajian:2014fta}, or by a series of recent cosmic-ray outbursts~\cite{Cholis:2015dea,Petrovic:2014uda,Carlson:2014cwa}. Outburst scenarios, however, have been shown to require a large degree of tuning in their parameters in order to produce the observed features of this signal~\cite{Cholis:2015dea}, leaving pulsars as the primary astrophysical explanation for this excess.

The pulsar interpretation of the GeV excess was elevated substantially in 2015, when two independent groups presented evidence that the gamma-ray emission from the direction of the Inner Galaxy contains a significant degree of small-scale power, suggestive of a point source origin for the excess emission~\cite{Lee:2015fea,Bartels:2015aea}. It has recently been shown, however, that these analyses are each subject to significant limitations, which bring their conclusions into considerable doubt. In particular, in Ref.~\cite{Leane:2019xiy}, it was shown that the non-Poissonian template fit technique utilized in Ref.~\cite{Lee:2015fea} can misattribute smooth gamma-ray signals (such as that predicted from annihilating dark matter) to point source populations. While it was recently shown that this in part was likely due to mismodeling of the diffuse model~\cite{Buschmann:2020adf}, more serious issues with this technique have been identified~\cite{Leane:2020nmi,Leane:2020pfc}. It has been shown that systematics arising from mismodeling give rise to spurious point source evidence for the GCE, and once correcting for this systematic, the evidence for point sources disappears~\cite{Leane:2020nmi,Leane:2020pfc}. Given such systematics, the Fermi data cannot be said to favor (or disfavor) a pulsar interpretation, as previously claimed. Also recently, the authors of Ref.~\cite{Zhong:2019ycb} showed that when updated point source catalogs are taken into account, the wavelet technique employed in Ref.~\cite{Bartels:2015aea} does not favor the presence of an additional unresolved point source population. Instead, strong constraints can be placed on the luminosity function of any such source population that might exist. It is now clear that if a population of millisecond pulsars does generate this excess, it must feature a very different luminosity function (containing far fewer bright members) than those observed in globular clusters or in the field of the Milky Way~\cite{Hooper:2016rap,Hooper:2015jlu,Cholis:2014lta,Bartels:2017xba}. These considerations, along with the low number of low-mass X-ray binaries observed in the Inner Galaxy~\cite{Haggard:2017lyq}, increasingly disfavor pulsar interpretations of the GeV excess.

The spectrum, morphology and overall intensity of the Galactic Center gamma-ray excess have each been found to be in good agreement with the expectations from annihilating dark matter~\cite{Daylan:2014rsa,Calore:2014xka,TheFermi-LAT:2017vmf}. In particular, the angular distribution of the excess is approximately azimuthally symmetric with respect to the Galactic Center, and is consistent with arising from annihilating dark matter with a halo profile with an inner slope of $\gamma \sim 1.1-1.3$~\cite{Daylan:2014rsa,Abazajian:2014fta,Calore:2014xka,TheFermi-LAT:2015kwa,Linden:2016rcf,TheFermi-LAT:2017vmf}, only slightly steeper than the canonical Navarro-Frenk-White~profile~\cite{Navarro:1995iw,Navarro:1996gj} and consistent with recent dynamical determinations based on Gaia data~\cite{2019arXiv191104557C}. We also note that while it has been argued that the morphology of the gamma-ray excess prefers the shape of the stellar bulge over that of a dark matter-like template~\cite{Macias:2016nev,Bartels:2017vsx,Macias:2019omb}, this preference is sensitive to the details of the background model adopted and on spatial tails of the excess. Additionally, the spectral shape of this excess is uniform throughout the Inner Galaxy, without any detectable variations~\cite{Calore:2014xka}, peaking at an energy of $\sim$\,1-5 GeV and falling off at both higher and lower energies (in $E^2 dN/dE$ units). If interpreted as the products of dark matter annihilation, the spectral shape of this signal implies a dark matter candidate with a mass in the range of $\sim$\,40-70 GeV (for the case of annihilations to $b\bar{b}$). Additionally, the overall intensity of this excess is consistent with that expected from a dark matter candidate with an annihilation cross section on the order of $\langle \sigma v \rangle \sim 10^{-26}$ cm$^3/$s.

\begin{figure}
 \resizebox{5.0in}{!}
 {
      \includegraphics{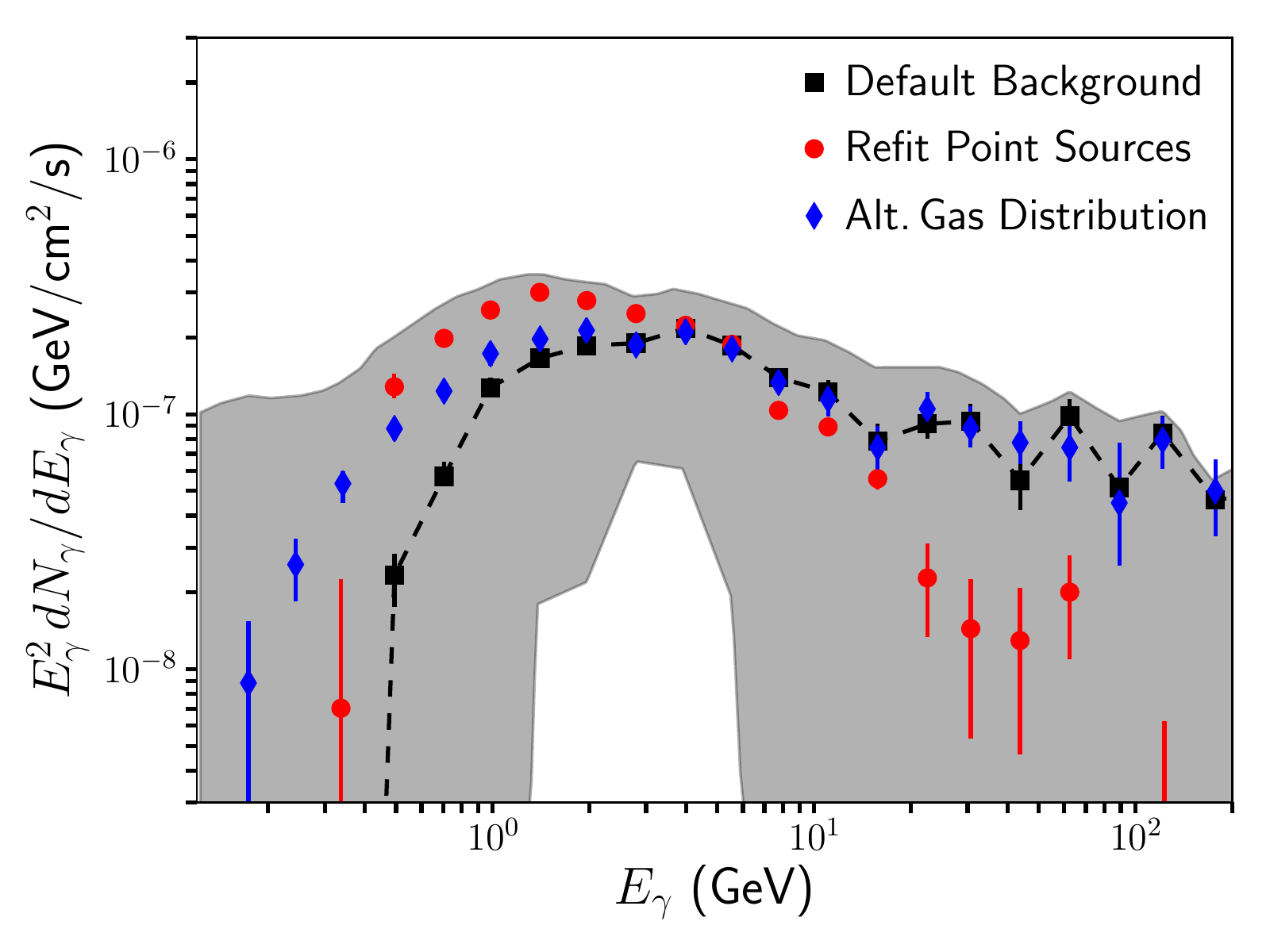}
 }
\caption{The spectrum of the Galactic Center gamma-ray excess as reported in Ref.~\cite{TheFermi-LAT:2017vmf}. The black squares (and dashed line) is the result found using their default background model, while the red circles and blue diamonds represent that found for representative variations in this model (see text for more details). The grey band represents the envelope of the results found across all of the background models considered in Ref.~\cite{TheFermi-LAT:2017vmf}.}
\label{Fig:GCE}
\end{figure}

Although the gamma-ray excess described above has by now been quite robustly detected, the precise spectrum of this signal is subject to significant systematic uncertainties associated with the astrophysical backgrounds. In Fig.~\ref{Fig:GCE}, we show the spectrum of this excess as reported by the \textit{Fermi} Collaboration in Ref.~\cite{TheFermi-LAT:2017vmf}. The black squares and dashed line denote the spectrum of the excess, as found using their default model for the astrophysical backgrounds (referred to in Ref.~\cite{TheFermi-LAT:2017vmf} as the ``Sample Model''). Reasonable variations in this model, however, can substantially alter the shape of the spectrum that is extracted from the data. As examples, we show as red circles in Fig.~\ref{Fig:GCE} the spectrum that is found when the spectra of the known point sources are allowed to float in the fit. This has the effect of shifting the peak of the spectrum downward and substantially reducing the intensity of the excess at energies above $\sim$10 GeV. We also show, as blue diamonds, an example in which an alternative gas distribution was used to derive the diffuse emission model, leading to a greater flux of excess emission below $\sim$1 GeV. The grey band in this figure represents the envelope of the results found across of all background models considered in Ref.~\cite{TheFermi-LAT:2017vmf}.

With these substantial systematic uncertainties in mind, it seems prudent to keep a fairly open mind regarding the precise spectral shape of the excess emission. Throughout this study, we simply require that in order to provide a viable explanation for the gamma-ray excess, a given dark matter model must produce a spectrum that peaks in the range of 1.4 to 4.0 GeV (in $E^2 \, dN/dE$ units), a feature that is found nearly universally across the range of background models considered in Ref.~\cite{TheFermi-LAT:2017vmf} and elsewhere in the literature~\cite{Daylan:2014rsa,Calore:2014xka}.

\subsection{The Cosmic-Ray Antiproton Excess}

Over the past several years, a number of groups~\cite{Cuoco:2016eej,Cui:2016ppb,Cholis:2019ejx,Cuoco:2019kuu} have reported the presence of an excess of $\sim$10-20 GeV antiprotons in the cosmic-ray spectrum as measured by \textit{AMS-02}~\cite{Aguilar:2016kjl} relative to that predicted from secondary production. Remarkably, it has been shown that this excess could be generated by annihilating dark matter particles with masses and cross sections in the same range as to those required to produce the Galactic Center gamma-ray excess (see, for example, Fig.~8 of Ref.~\cite{Cholis:2019ejx}).
Although this excess appears to be statistically significant, estimates of the systematic uncertainties have varied. The authors of Refs.~\cite{Cuoco:2019kuu,Cholis:2019ejx} have each argued that the presence of the antiproton excess is robust to astrophysical uncertainties, while acknowledging that the uncertainties associated with the antiproton production cross section~\cite{Cholis:2019ejx,Cuoco:2019kuu,Reinert:2017aga} and the impact of the Solar Wind on the cosmic-ray spectra observed at Earth~\cite{Cholis:2019ejx,Cholis:2015gna} are each difficult to rigorously quantify. In Refs.~\cite{Boudaud:2019efq,Heisig:2020nse}, the authors argued that the systematic uncertainties related to this signal are significantly larger than had previously been estimated, reducing the overall significance of the excess. In Ref.~\cite{Cholis:2019ejx}, it was found that dark matter annihilating directly to $b\bar{b}$ could potentially provide a good fit to this signal for masses in the range of 46-94 GeV, a range over which the injected antiproton spectrum peaks at energies between 6.2 and 12.6 GeV (in $E^2 dN/dE$ units).

With these results in mind, we consider a dark matter candidate to be potentially capable of producing both the gamma-ray and antiproton excesses if its injected spectra of gamma rays and antiprotons fall within the range of 1.4 to 4.0 GeV and 6.2 and 12.6 GeV, respectively. We also require that the low-velocity annihilation cross section falls within the range presented in Refs.~\cite{Benito:2016kyp} and Ref.~\cite{Cholis:2019ejx}, to accommodate the gamma-ray and antiproton excesses, in each case corrected for the spectrum produced per annihilation in a given model (see Sec.~\ref{spectragen}). For the simple case of direct annihilation to $b\bar{b}$ this corresponds to $\langle \sigma v \rangle = (0.6-7) \times 10^{-26}$ cm$^3/s$ for the gamma-ray excess and $\langle \sigma v \rangle = (0.3-20) \times 10^{-26}$ cm$^3/s$ for the antiproton excess. We further require that the ratio of the peak gamma-ray to antiproton flux (at injection, per annihilation) falls within the range implied by Refs.~\cite{Benito:2016kyp,Cholis:2019ejx}, properly accounting for the correlated uncertainties in the overall normalization of the Milky Way's dark matter halo.

\section{Energy Spectra Generation}
\label{spectragen}

In the previous section, we clarified that analyses of the gamma-ray and antiproton data have roughly agreed upon the energy range in which each signal peaks, as well as the excess flux contributed above the expected background. We have yet to clarify, however, whether there exists a dark matter mass, a dark mediator mass, and a portal interaction which is capable of simultaneously producing gamma-ray and antiproton fluxes consistent with both measurements. In this section we outline the procedure by which we identify models consistent with both anomalies, and we defer the results of this analysis to Section \ref{results}.

In order to generate the spectrum of gamma rays and antiprotons produced through dark matter annihilations in a given model, we utilize \textsc{Pythia8.2}. In each case, we create an effective resonance with $E_{\rm CM}=2m_{\rm DM}$ by colliding two back-to-back neutral beams. The energy resonance is then allowed to decay as specified, and a total of 400,000 events per diagram are accepted. For diagrams that can produce two different final states of spin-$X$ and spin-$Y$, we first produce one diagram with two spin-$X$ particles, and then two spin-$Y$ particles, each with effective resonances in their center of mass frames. We then average these results to produce the effective spectra for a given choice of masses. This is a necessary step to ensure the correct fraction of energy is put into each final state particle's decay products. Specifically, this means that the effective resonances for two different final state particles with masses $m_1$ and $m_2$ are given by the following~\cite{Agashe:2014kda}:
\begin{equation}
E^{m_1}_{\rm CM}=\frac{s+m_{1}^2-m_2^2}{2\sqrt{s}}, \hspace{3mm} E^{m_2}_{\rm CM}=\frac{s+m_{2}^2-m_{1}^2}{2\sqrt{s}}.
 \end{equation}
In the case of $m_1=m_2$, the same amount of energy is given to each final state particle's decay products. These particles are then decayed with branching fractions specified by the portal under consideration. 

\begin{figure}
 \resizebox{3.0in}{!}
 {
      \includegraphics{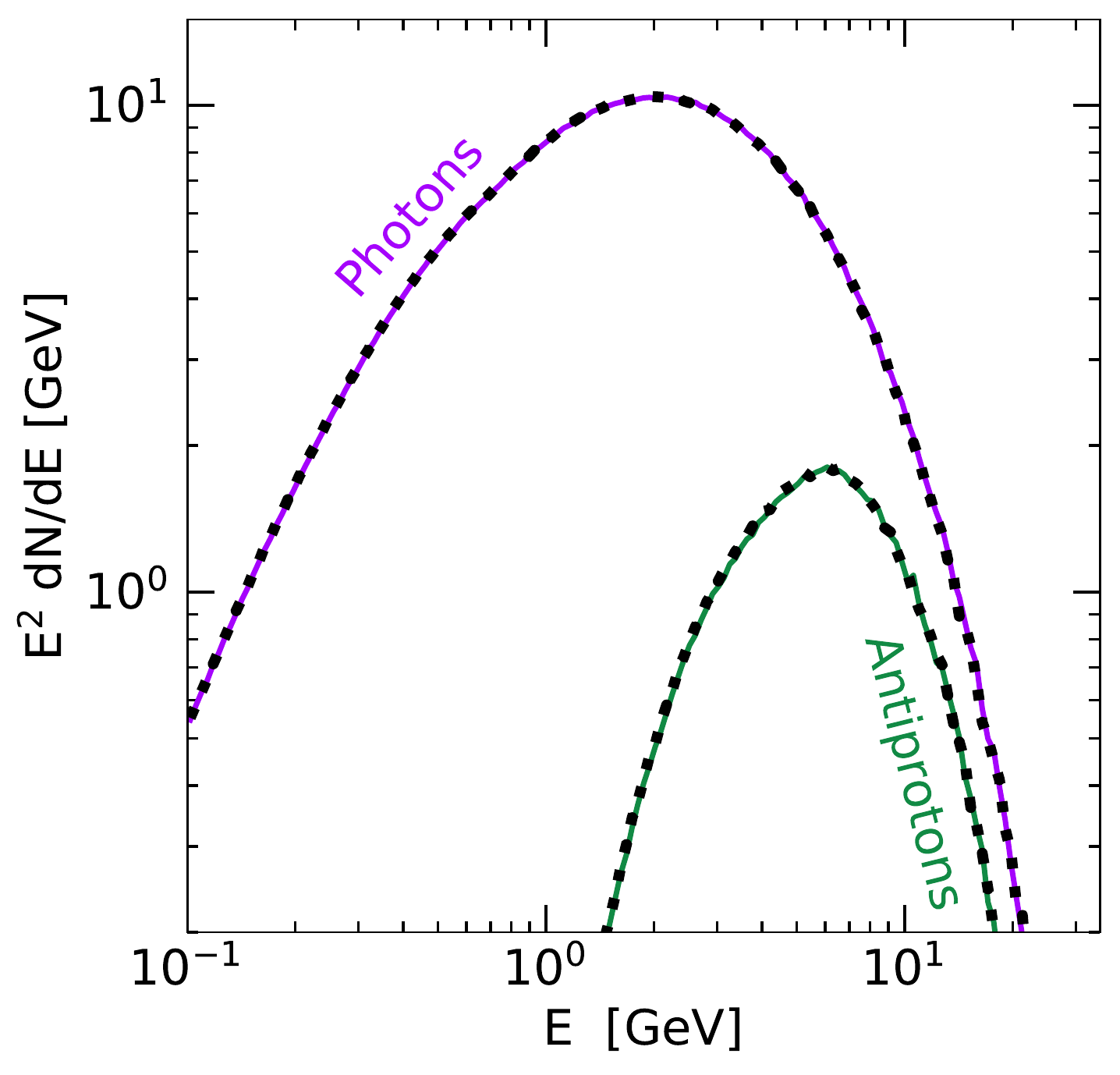}
 }
\caption{An illustration of the minimal impact of the mediator's spin on the gamma-ray and antiproton spectra from dark matter annihilations. The solid and dotted curves represent the spectra predicted for a spin-1 or spin-0 mediator, respectively, for the representative case of 50 GeV dark matter particles annihilating to a pair of 20 GeV hidden sector particles which decay to $b$ quark pairs. The solid and dotted curves are virtually indistinguishable from each other in this case, and across the entirety of the parameter space relevant to this study.}
\label{Fig:spinspec}
\end{figure}

The spectral shape of the gamma rays and antiprotons produced through dark matter annihilation depends significantly on 1) the mass of the dark matter particle, 2) the mass of mediator and 3) the portal connecting the hidden sector to the SM. In principle, these spectra could also depend on the spin of the mediating particle. In practice, however, this dependence is very small, and can safely be neglected~\cite{Elor:2015tva} (see Fig.~\ref{Fig:spinspec}).

\begin{figure*}[t!]
\centering
\subfloat{\includegraphics[width=0.43\columnwidth]{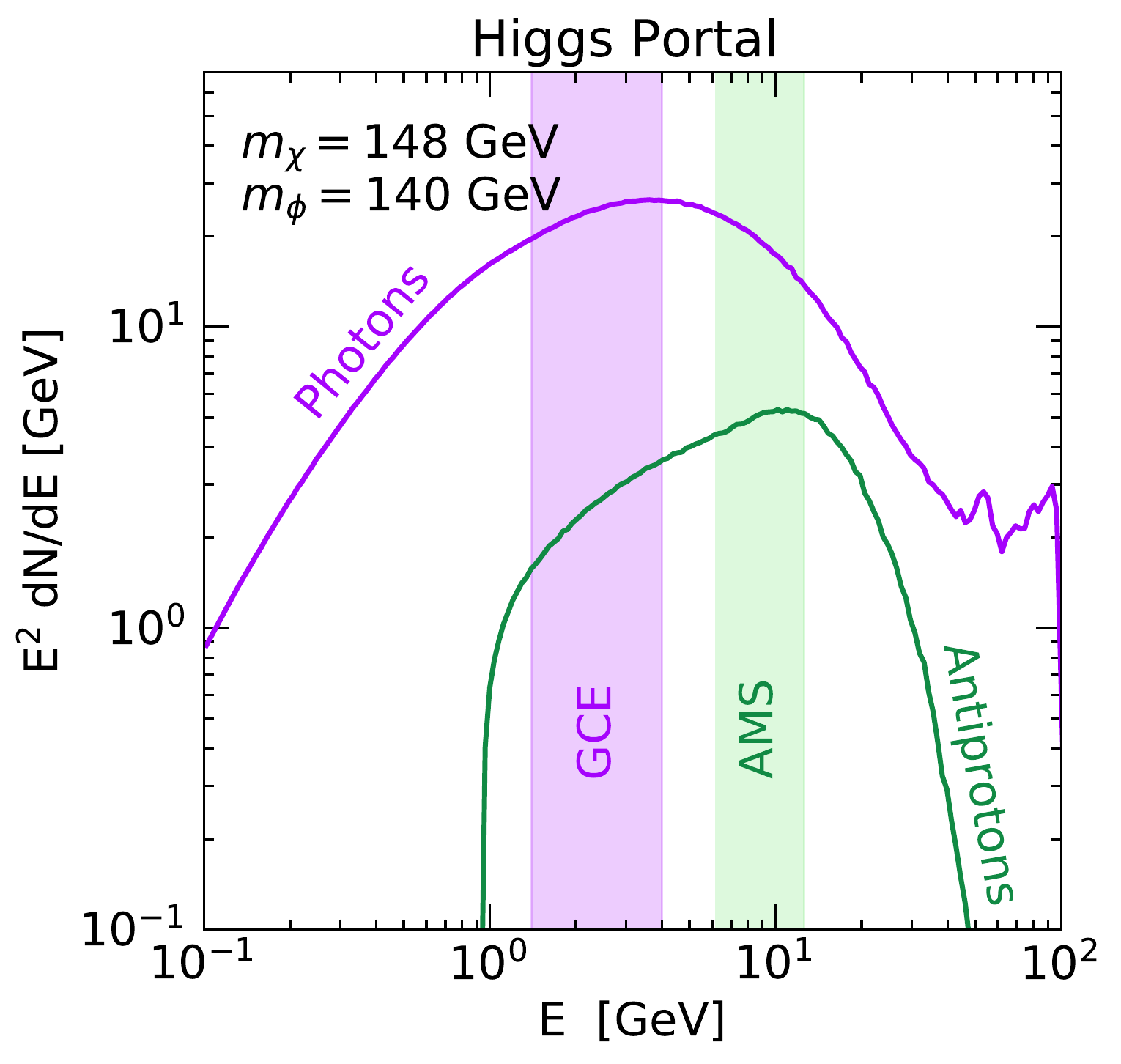}}
\subfloat{\includegraphics[width=0.43\columnwidth]{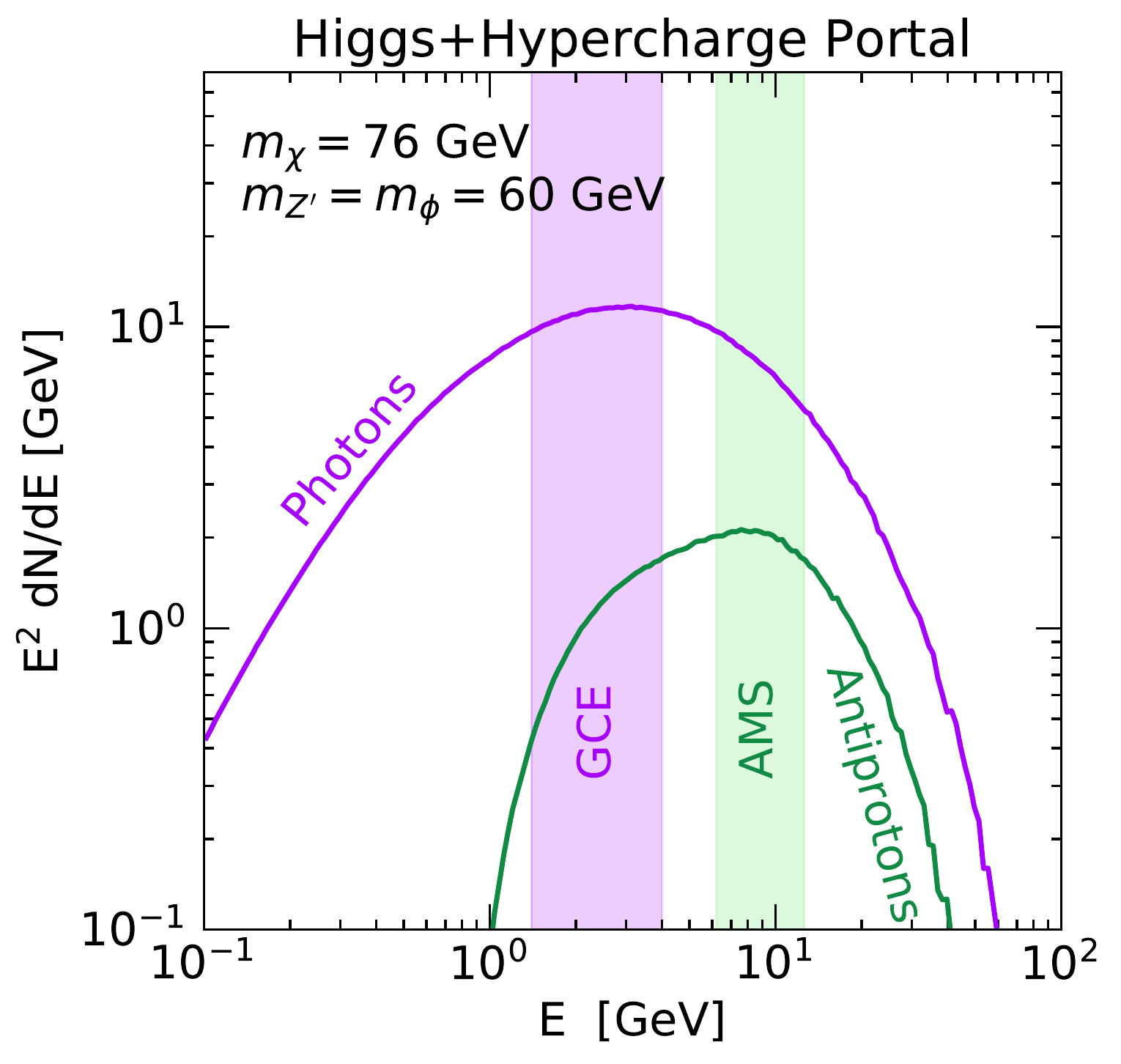}}\\
\subfloat{\includegraphics[width=0.43\columnwidth]{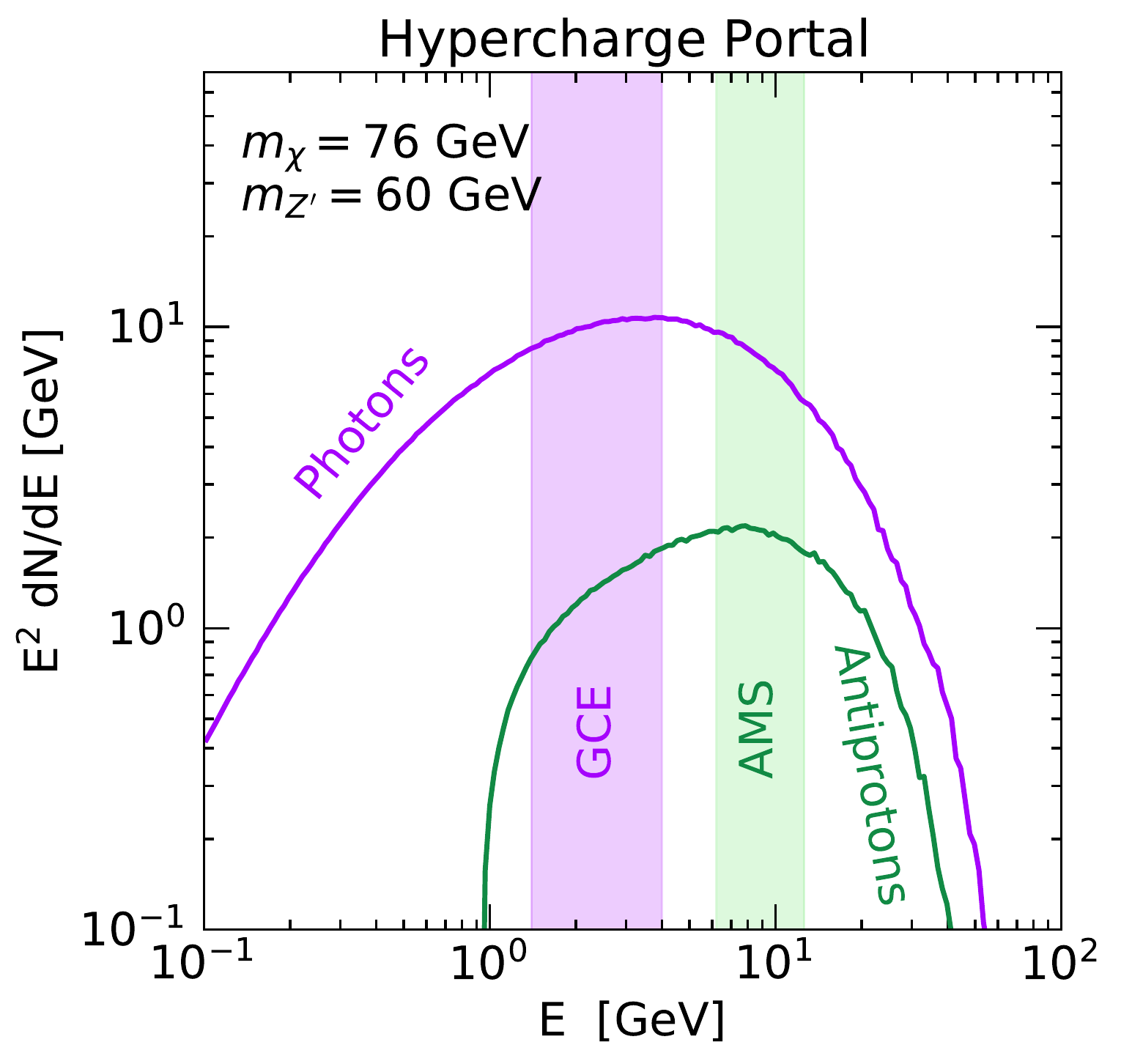}}
\subfloat{\includegraphics[width=0.43\columnwidth]{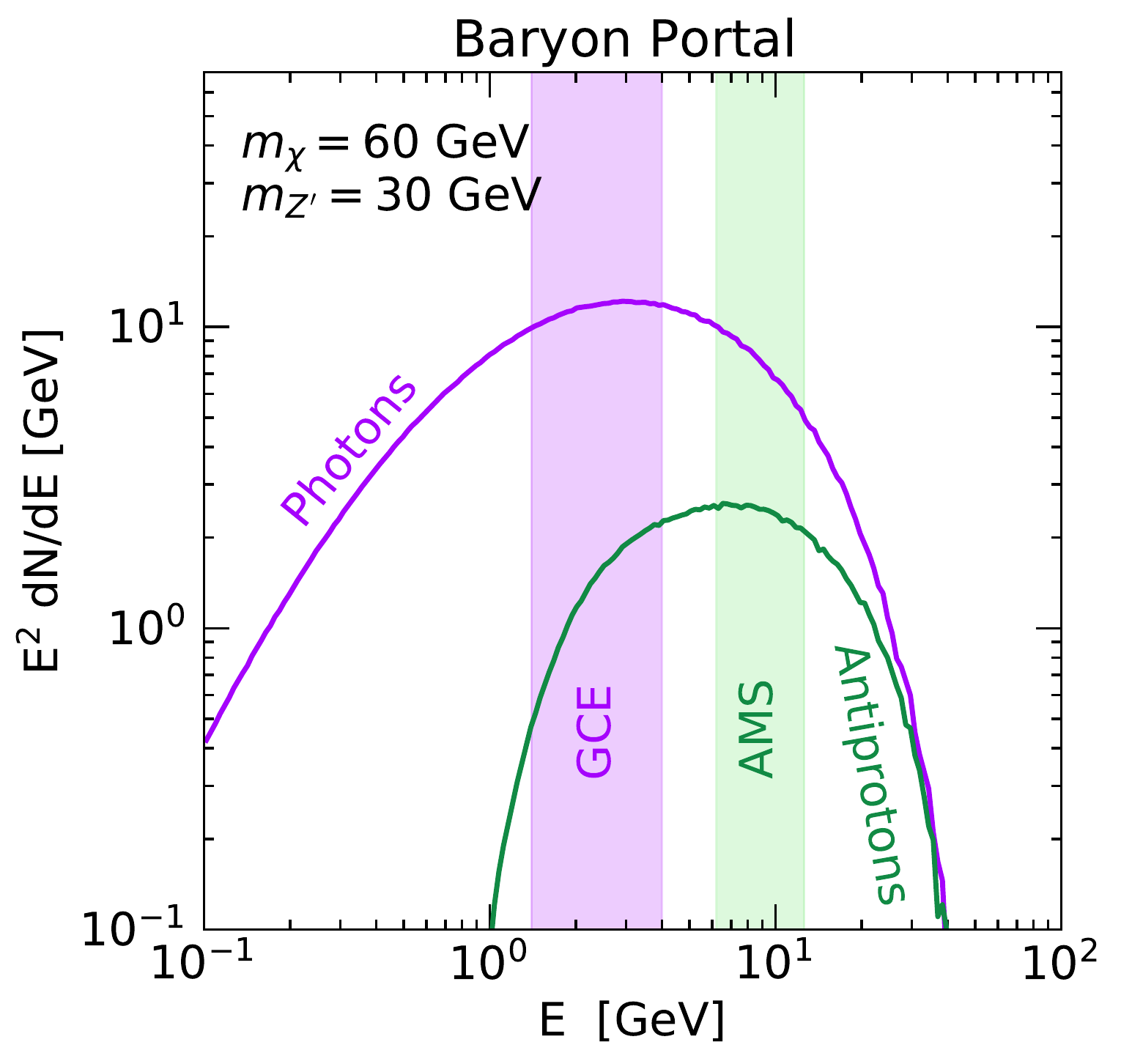}}\\
\subfloat{\includegraphics[width=0.43\columnwidth]{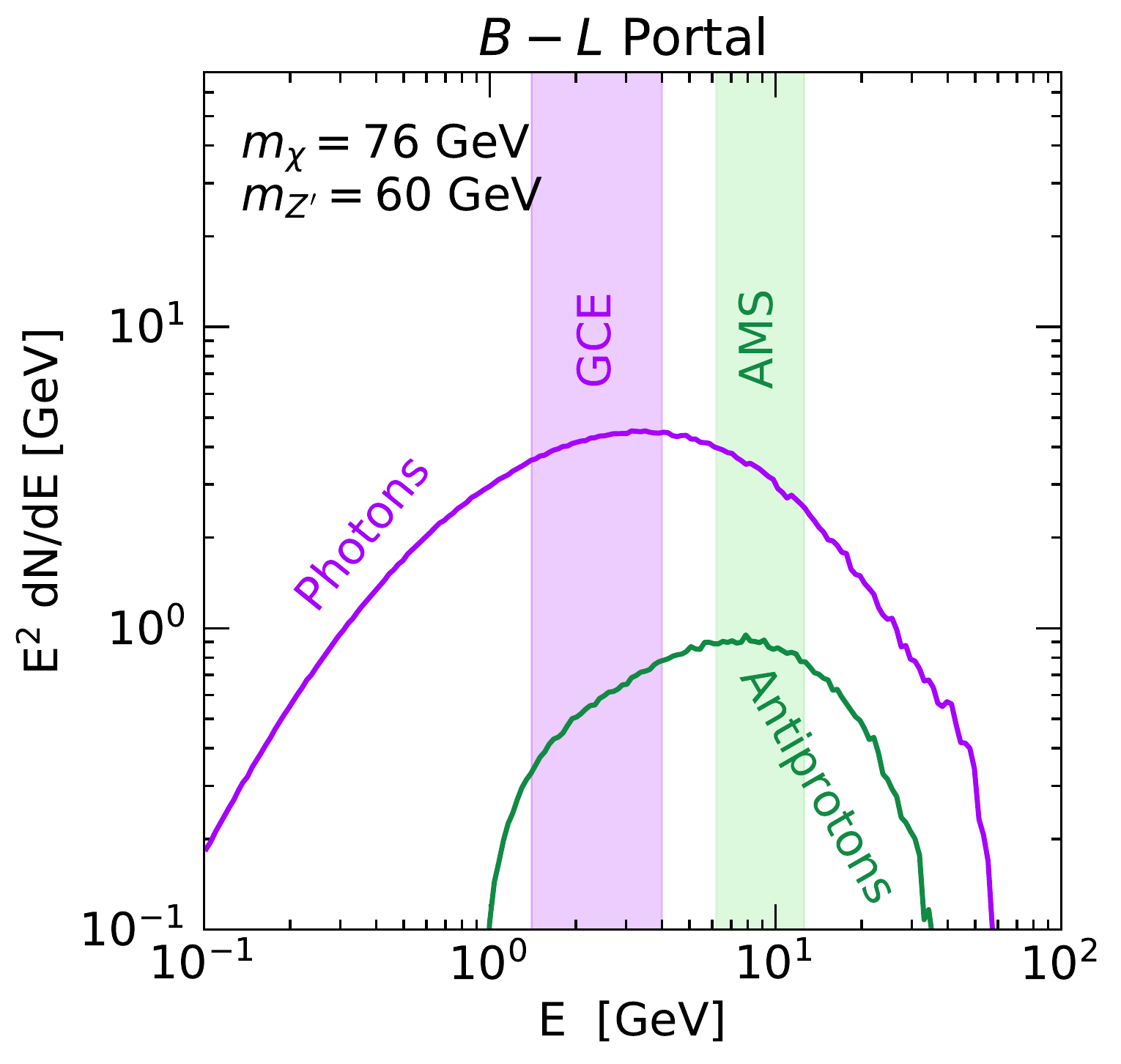}}
\caption{Gamma-ray and antiproton spectra for a selection of representative hidden sector dark matter models, each of which can simultaneously accommodate both the Galactic Center gamma-ray and cosmic-ray antiproton excesses. The target regions that the spectra are required to peak in are shown as the purple and green shaded regions, as described in the text. \label{fig:spec}}
\end{figure*}

In Fig.~\ref{fig:spec}, we plot the spectra of gamma rays and antiprotons predicted in five representative hidden sector dark matter models. As described in Sec.~\ref{char}, we consider a given model to provide an adequate fit to the spectral shapes of the gamma rays and antiprotons peak within the purple ($E_{\gamma}=1.4-4.0$ GeV) and green ($E_{\bar{p}}=6.2-12.6$ GeV) bands, respectively. From this figure, it is clear that the gamma-ray and antiproton excesses can be simultaneously accommodated within a wide range of hidden sector dark matter models. 
Note that as various portals have several decay channels, their total energy spectra will contain a sum of varied final state spectra, which explains i.e. the multi peaks in Fig. \ref{fig:spec}.

\section{Constraints From Dwarf Spheroidal Galaxies}
\label{dwarfs}
Hidden sector models are naturally hidden from collider and direct dark matter searches due to their weak portal interaction with SM particles. Such model cannot, however, hide from indirect searches, since the portal coupling allows does not suppress the annihilation rate. Consequently, it is necessary to determine the extent to which any model identified as capable of producing the observed excesses has been probed by alternative indirect searches.  

The Milky Way's population of dwarf spheroidal galaxies provides one of the most powerful tests of annihilating dark matter, as they are relatively nearby, abundant in dark matter, and emit little gamma-ray background. The \textit{Fermi} Collaboration has published limits on gamma-ray fluxes from a collection of known dwarf spheroidals~\cite{Ackermann:2015zua, Fermi-LAT:2016uux}, identifying no statistically significant signals as of yet.

To derive limits on hidden sector dark matter models, we follow the official \textit{Fermi} analysis on \texttt{Pass 8} LAT data~\cite{Fermi-LAT:2016uux} and consider a total of 41 dwarf galaxies (including both kinematically confirmed dwarfs, as well as unconfirmed but likely galaxies)\footnote{The bin-by-bin likelihoods for each dwarf galaxy can be downloaded from  \href{http://www-glast.stanford.edu/pub_data/1203/}{http://www-glast.stanford.edu/pub$\_$data/1203/}}.
When provided, we use the measured $J$-factor and corresponding uncertainty for each galaxy. 
%This is for 19 dwarf galaxies: Bootes I, Canes Venatici I, 
%Canes Venatici II, Carina, Coma Berenices, Draco, Fornax, Hercules, Leo I, Leo %II, Leo IV, Leo V, Reticulum II, Sculptor, Segue 1, 
%Sextans, Ursa Major I, Ursa Major II, and Ursa Minor.
For those dwarfs without spectroscopic information, we use $J$-factors estimated from their distances, adopting a nominal uncertainty of 0.6 dex (following Ref.~\cite{Fermi-LAT:2016uux}). 

%Dwarfs we consider in this category are Bootes II, Bootes III, Draco II, %Horologium I, Hydra II, Pisces II, 
%Triangulum II, Tucana II, Willman 1, Columba I, Eridanus II, Grus I, Grus II, %Horologium II, Indus II, Pegasus III, 
%Phoenix II, Pictor I, Reticulum III, Sagittarius II, Tucana III, and Tucana IV.

Note that modest ($\sim2\sigma$) gamma-ray excesses have been observed from four of these galaxies (Reticulum II, Tucana III, Tucana IV, Indus II)~\cite{Fermi-LAT:2016uux}. While this could potentially be attributed to dark matter, this potential signal is not globally significant at this time, and we simply use this data to derive upper limits on the dark matter annihilation cross section.

For each of these dwarf galaxies, the \textit{Fermi} Collaboration~\cite{Fermi-LAT:2016uux} provides the likelihood as a function of the integrated energy flux:
\begin{equation}
\Phi_E = \frac{\langle \sigma v \rangle}{8\pi m_{\chi}^2} \left[ \int_{E_{\rm min}}^{E_{\rm max}} E \frac{dN}{dE} dE \right] J_i\,,
\label{eq:Eflux}
\end{equation}
where $J_i$ is the $J$-factor for dwarf, $i$. Following Ref.~\cite{Ackermann:2015zua}, we treat the energy bins as independent, and obtain the full likelihood, $\mathcal{L}_i\left({\mu} \vert \mathcal{D}_i\right)$, which is a function of the model parameters, $\mu$, and data, $D_i$, by multiplying together the likelihoods for each of the 41 dwarfs. The uncertainty in the $J$-factor is included as a nuisance parameter on the global likelihood, modifying the likelihood as follows~\cite{Rolke:2004mj}:
\begin{equation}
\tilde{\mathcal{L}}_i\left({\mu}, J_i \vert \mathcal{D}_i\right) = \mathcal{L}_i\left({\mu} \vert \mathcal{D}_i\right) \times \frac{1}{\ln(10) J_i \sqrt{2\pi} \sigma_i} \, \exp\bigg[ \frac{-(\, \log_{10}(J_i) - \overline{\log_{10}(J_i)}\, )^2} {2 \sigma_i^2} \bigg].
\label{eq:like}
\end{equation}
We use the values of $\overline{\log_{10}(J_i)}$ and $\sigma_i$ provided in Ref.~\cite{Fermi-LAT:2016uux} for the case of a Navarro-Frenk-White profile~\cite{Navarro:1995iw,Navarro:1996gj}. The likelihood is maximized to produce an upper limit on the annihilation cross section at the $95\%$ confidence level.

\section{Results}
\label{results}

In this section, we present the main results of our analysis, identifying which of the hidden sector dark matter models discussed in Secs.~\ref{models} and~\ref{portals} are capable of producing signals that are consistent with the observed features of the Galactic Center gamma-ray excess and the cosmic-ray antiproton excess (see Sec.~\ref{char}). While this section does not directly rely on the results presented in Section \ref{models}, it is important to bear in mind that all viable dark sectors must allow for a vertex list in Section \ref{models} in order to have  s-wave annihilation. 
For the following discussions, the hidden sector dark matter particles we consider are Dirac fermions, unless stated otherwise.

%\subsection{Hypercharge Portal}

\begin{figure*}[t!]
\leavevmode
\centering
\subfloat{\includegraphics[width=0.48\columnwidth]{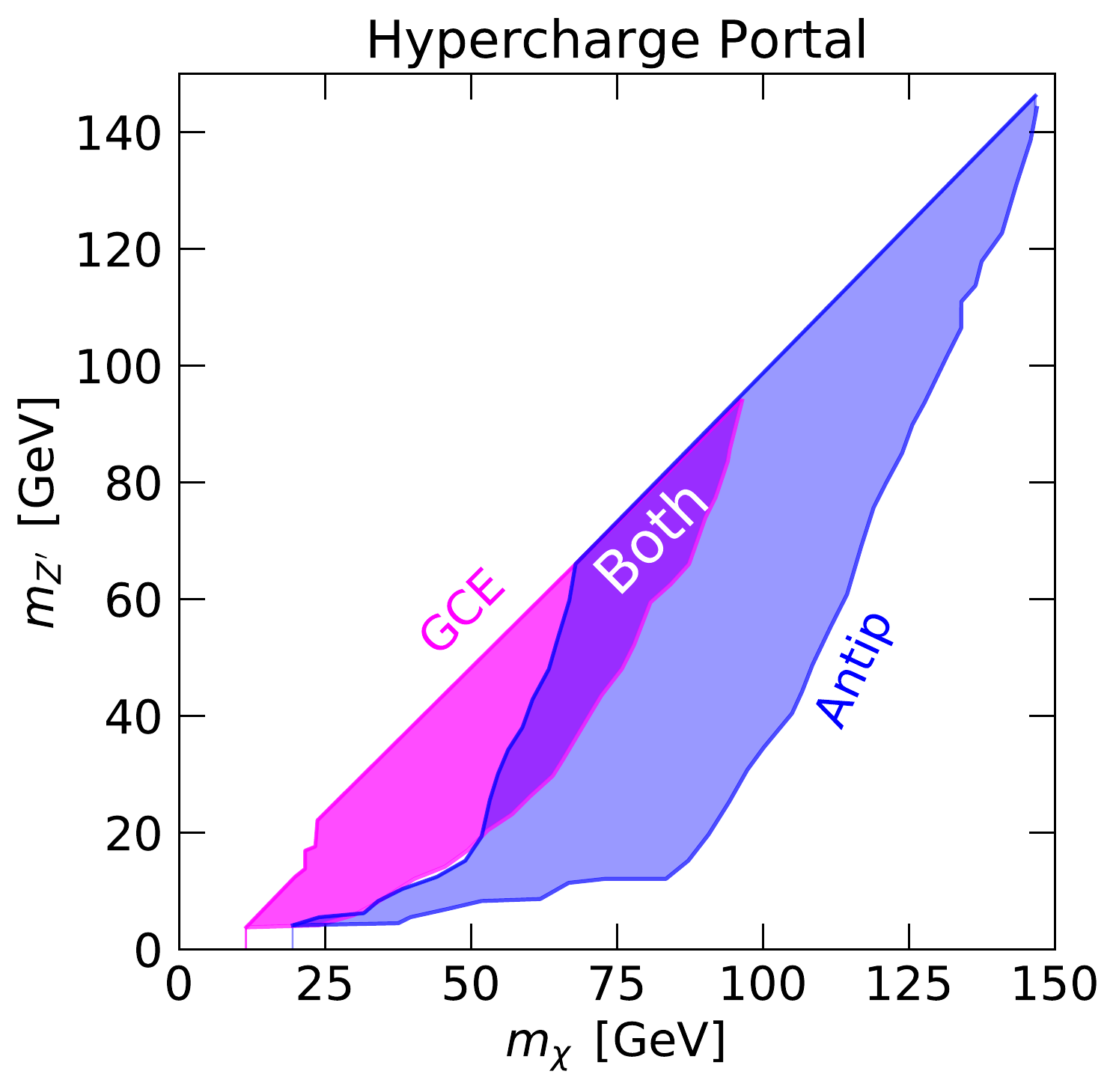}}
\hspace{2mm}
\subfloat{\includegraphics[width=0.49\columnwidth]{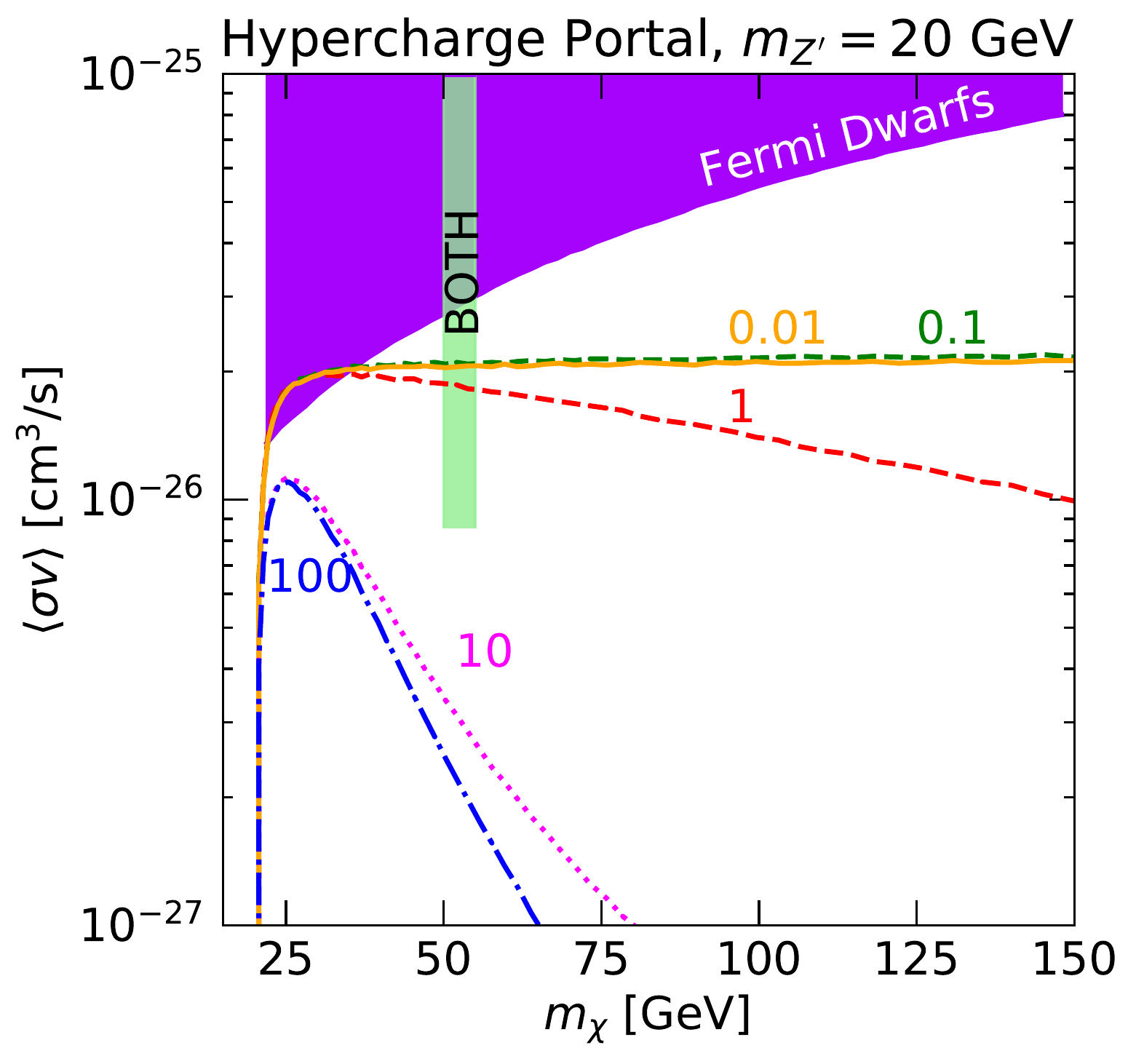}}\\
\subfloat{\includegraphics[width=0.49\columnwidth]{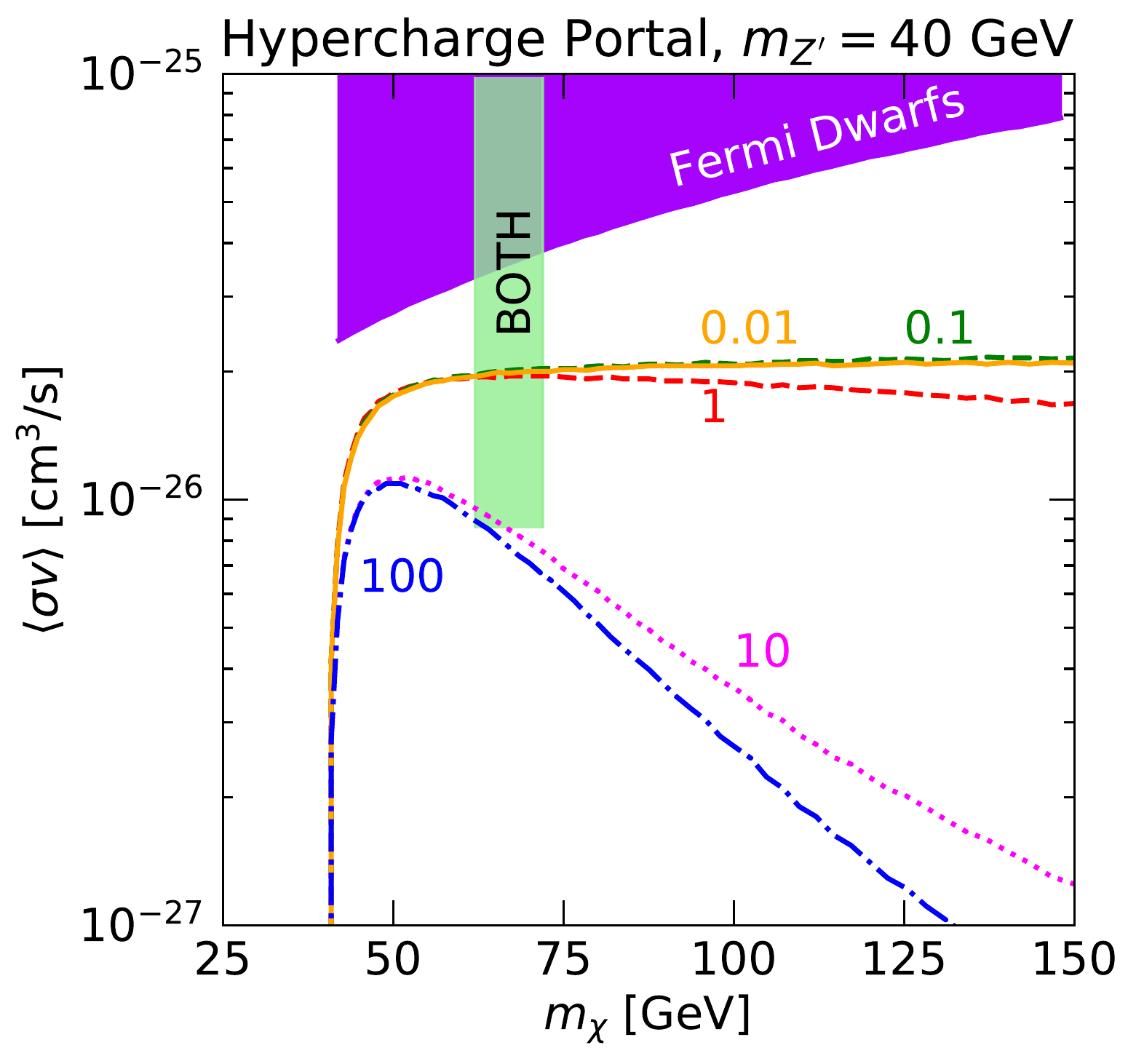}}
\hspace{2mm}
\subfloat{\includegraphics[width=0.49\columnwidth]{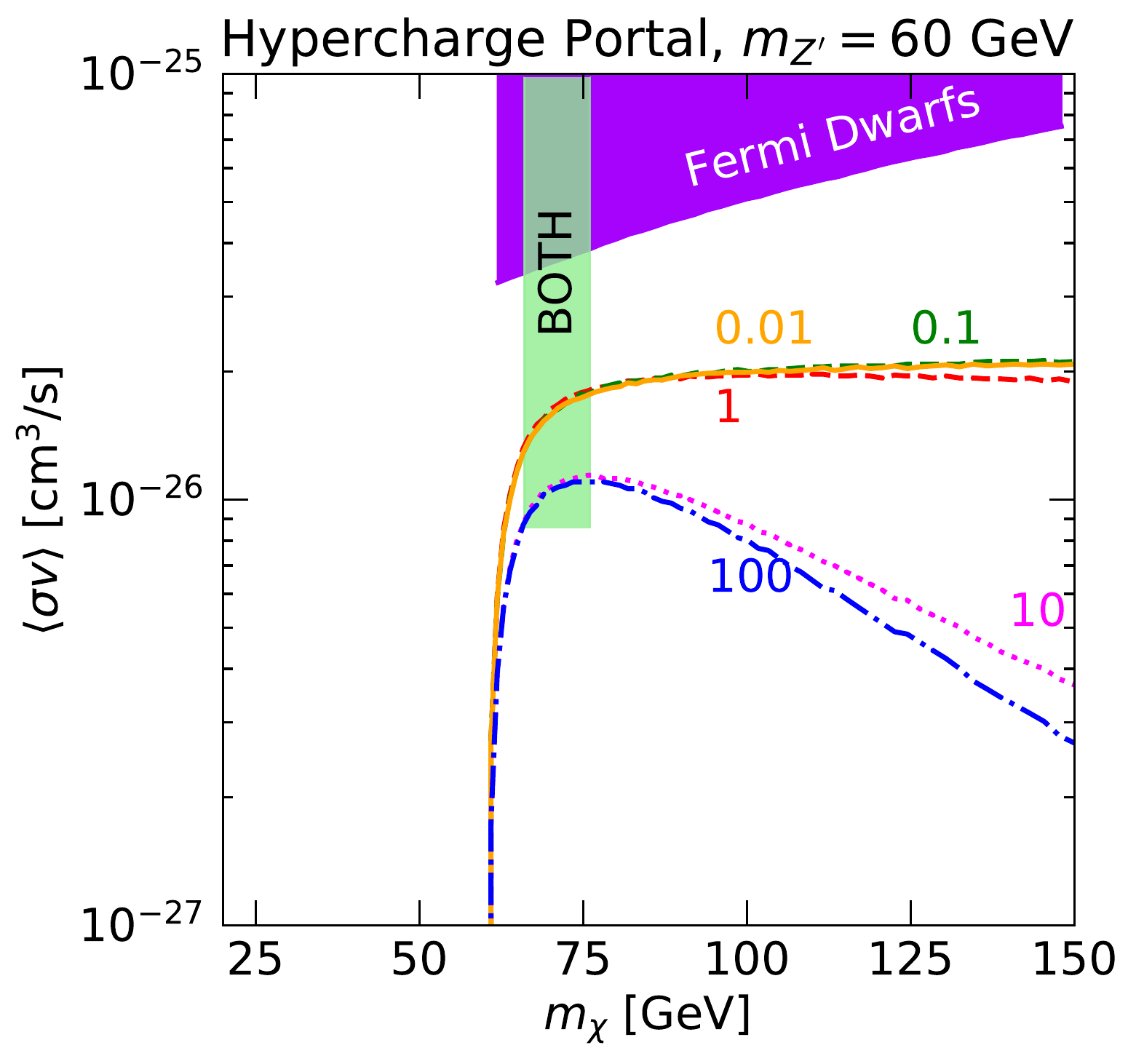}}
\caption{A summary of our results for the case of hidden sector dark matter that annihilates to particles that decay through the hypercharge portal. In the upper left frame, we show the regions of the $m_{\chi}$-$m_{Z'}$ plane that can produce the observed spectrum and intensity of the Galactic Center gamma-ray excess (GCE), the cosmic-ray antiproton excess (Antip), or both. In the remaining three frames, we plot the low-velocity, thermally-averaged annihilation cross section predicted in this model, for three representative values of $m_{Z'}$. Each line corresponds to a different ratio of the vector and axial couplings, $\lambda_v / \lambda_a$. The green shaded regions labeled ``BOTH'' denote the parameter space in which both excesses can be simultaneously accommodated. We also show the regions ruled out by gamma-ray observations of dwarf galaxies.\label{fig:hypercharge}}
\end{figure*}

In Fig~\ref{fig:hypercharge}, we summarize our results for the case of hidden sector dark matter that annihilates into a pair of spin-$1$ particles that decay to the SM through the hypercharge portal. In the upper left frame, we show the regions of the $m_{\chi}$-$m_{Z'}$ plane that can produce the observed spectrum and intensity of the Galactic Center gamma-ray excess (GCE), the cosmic-ray antiproton excess (Antip), or both. In this model, the gamma-ray spectrum is dominated by tau decays, requiring a relatively small value of $m_{\chi}$ in order to accommodate the observed spectral shape of the gamma-ray excess. For these reasons, there is relatively little parameter space in this model in which both excesses can be accommodated, although models with $m_{\chi} \approx 50-100$ GeV and $m_{Z'} \approx 20-100$ GeV can be capable of generating both of these signals.

In the remaining three frames of Fig.~\ref{fig:hypercharge}, we plot the low-velocity, thermally-averaged annihilation cross section predicted in this model (after requiring that the thermal relic abundance equals the measured density of dark matter, which is determined for each model using {\tt micromegas~\cite{Belanger:2010pz}}), for three representative values of $m_{Z'}$. The green shaded regions represent the parameter space in which the spectrum and intensity of the gamma-ray and antiproton excesses can both be accommodated (as described in Sec.~\ref{char}). In each case, we allow for the dark matter to have both vector and axial-vector couplings to the $Z'$, and plot results for values of $\lambda_v / \lambda_{a}$ between 0.01 and 100. We also show the constraints from dwarf galaxies (see Sec.~\ref{dwarfs}), finding no tension with the favored parameter space.

\begin{figure*}[t!]
\centering
\subfloat{\includegraphics[width=0.48\columnwidth]{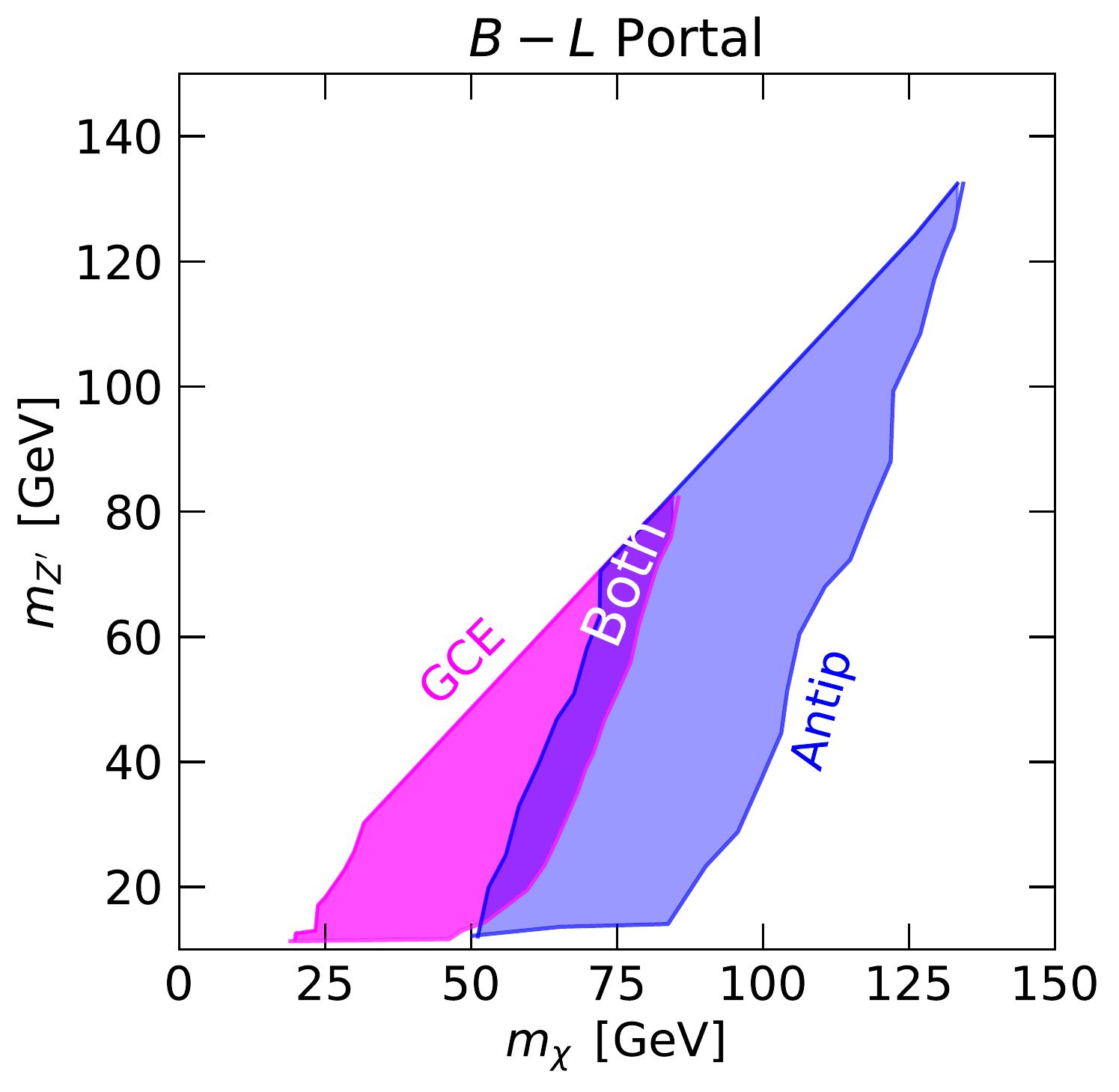}}
\hspace{1mm}
\subfloat{\includegraphics[width=0.49\columnwidth]{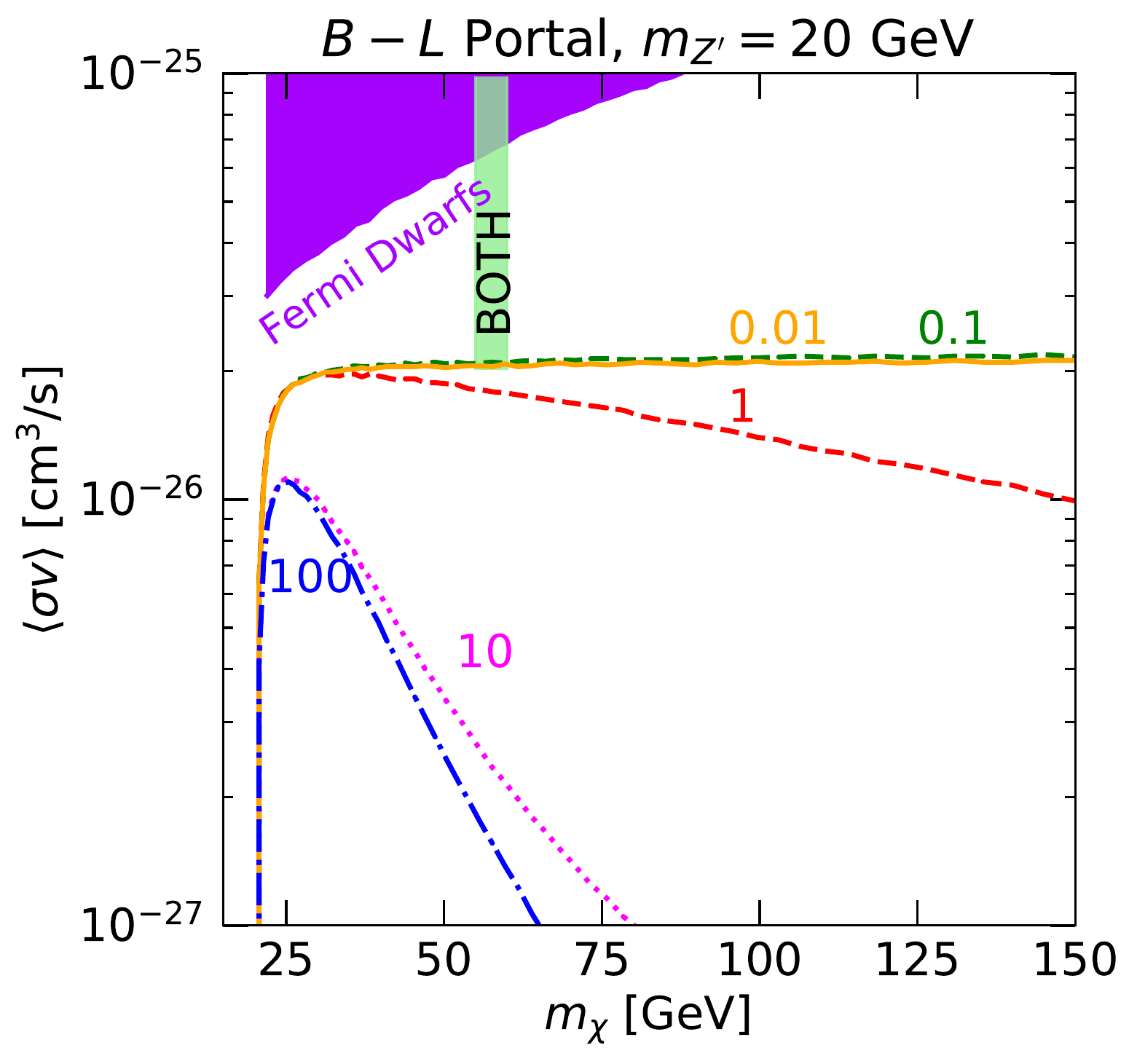}}\\
\subfloat{\includegraphics[width=0.49\columnwidth]{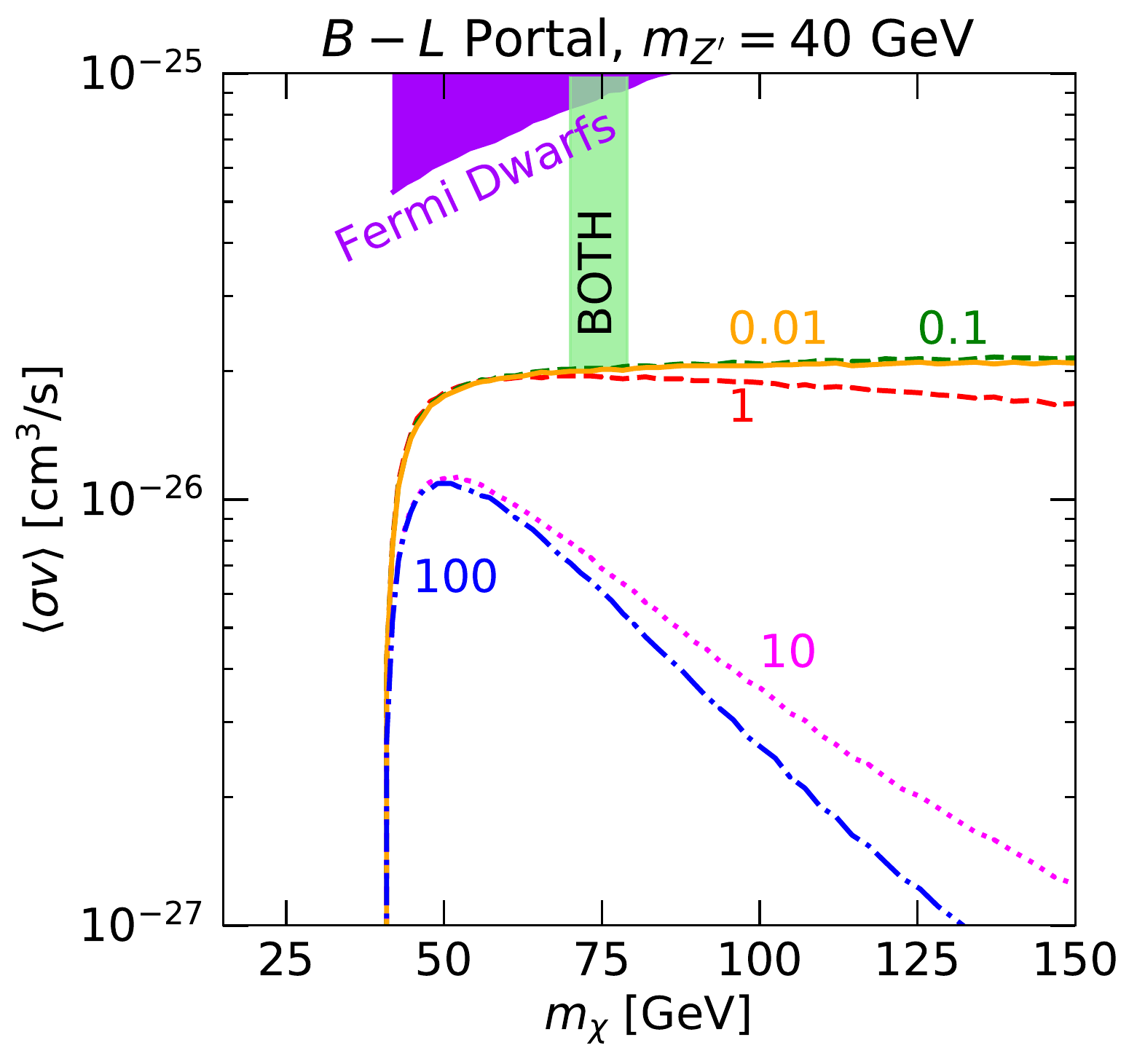}}
\hspace{1mm}
\subfloat{\includegraphics[width=0.49\columnwidth]{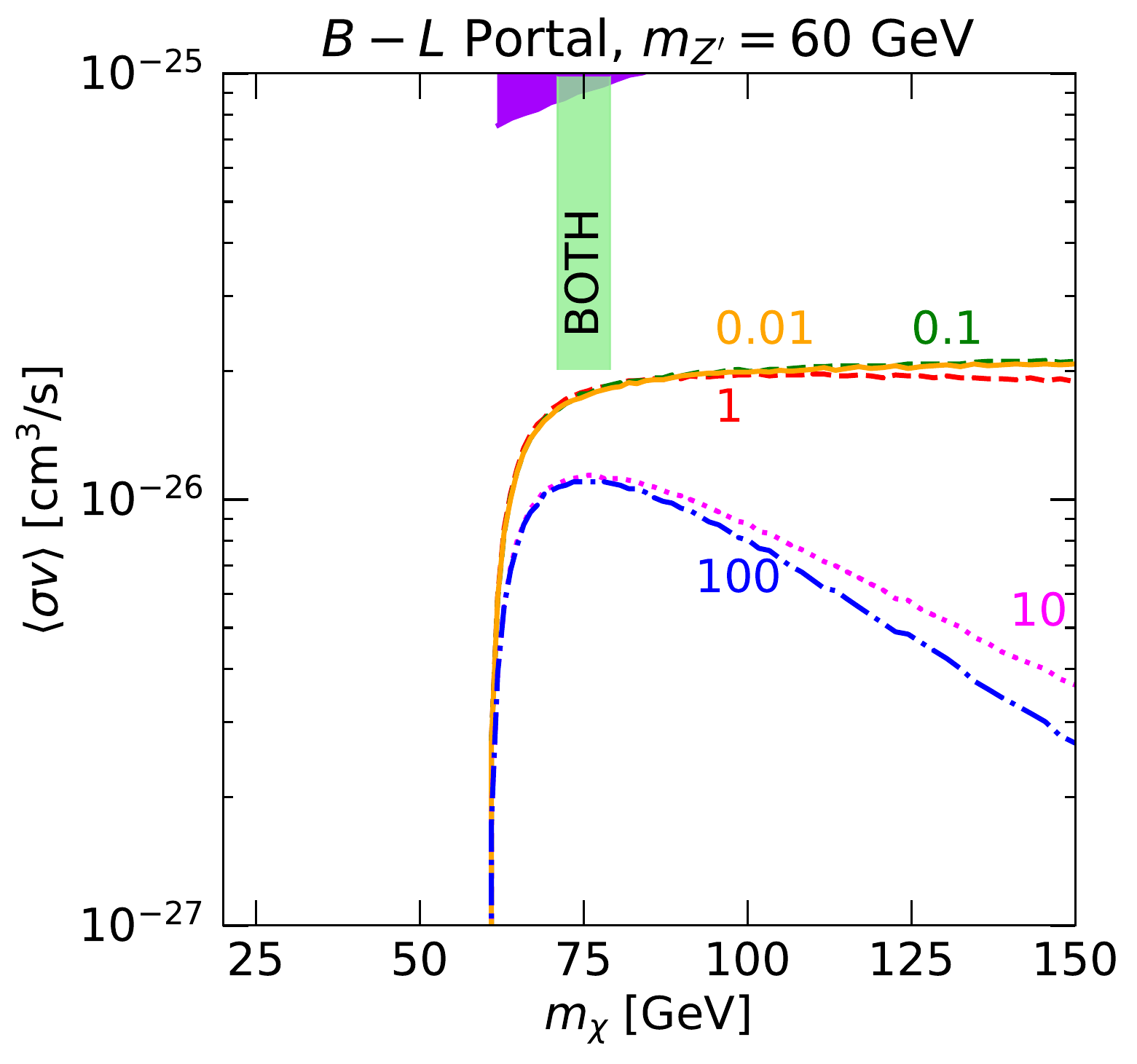}}
\caption{As in Fig.~\ref{fig:hypercharge}, but for the case of hidden sector dark matter that annihilates to particles that decay through the $B-L$ portal.\label{fig:zpbl}}
\end{figure*}

\begin{figure*}[t!]
\centering
\subfloat{\includegraphics[width=0.48\columnwidth]{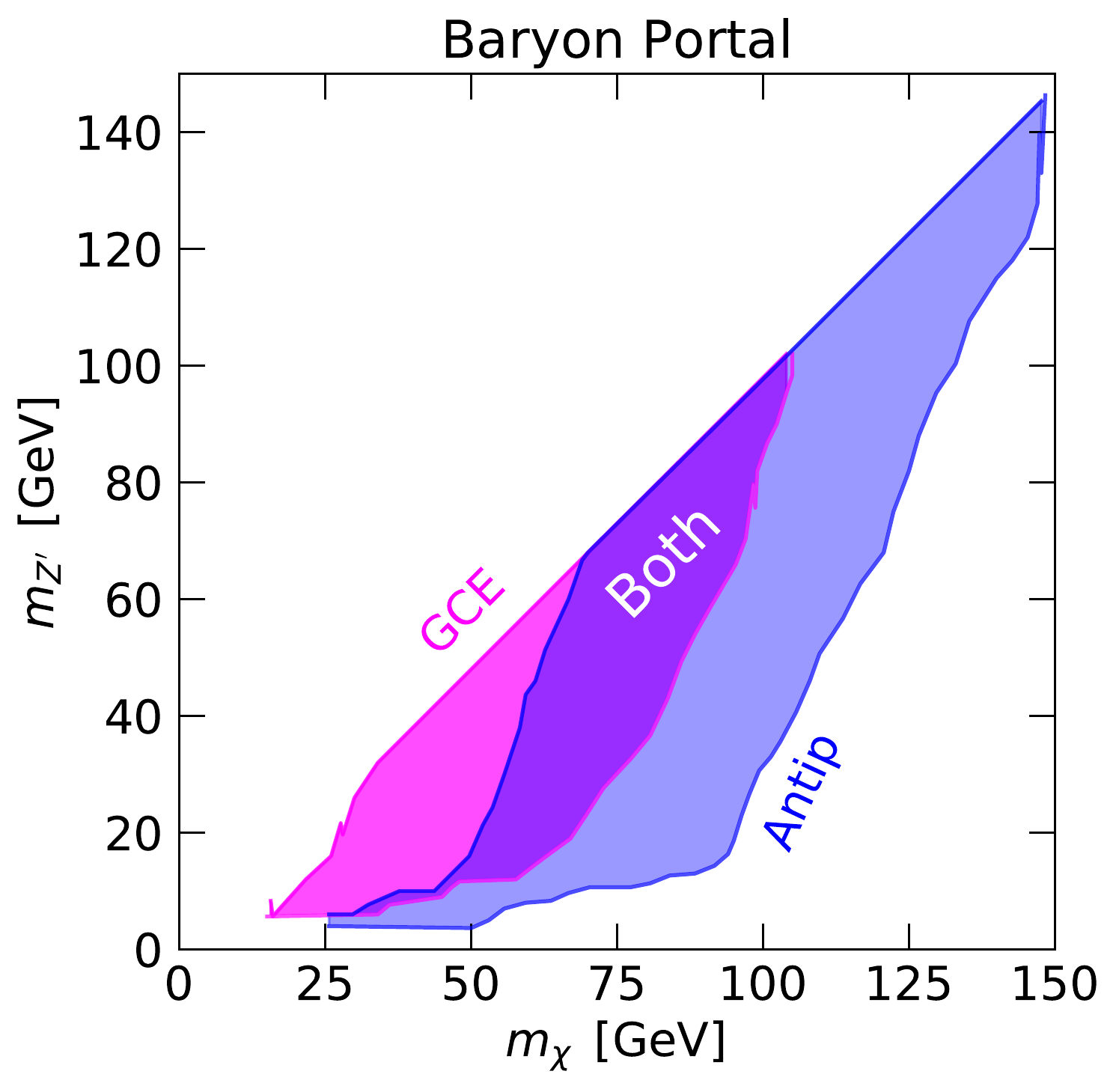}}
\hspace{1mm}
\subfloat{\includegraphics[width=0.49\columnwidth]{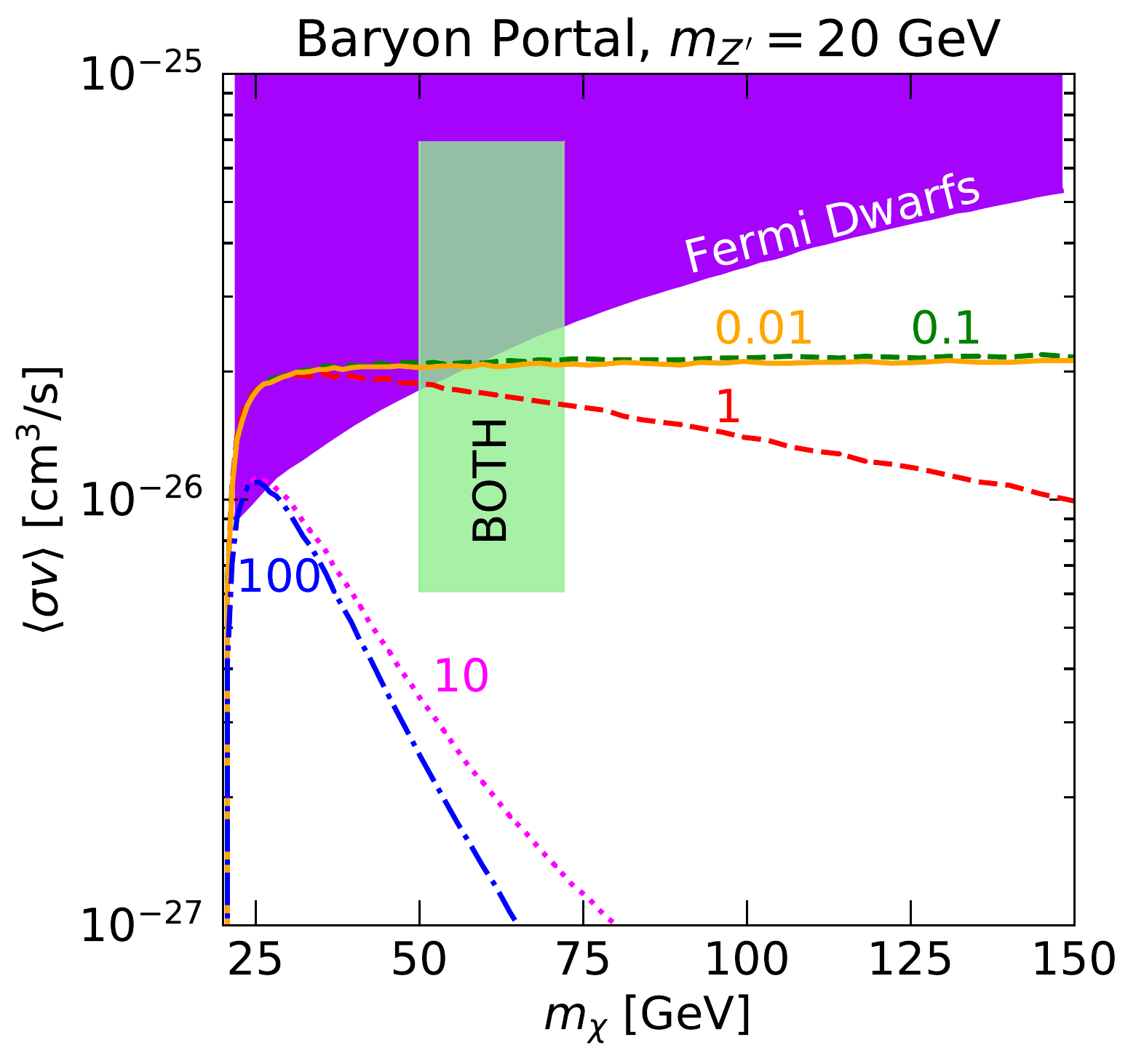}}\\
\subfloat{\includegraphics[width=0.49\columnwidth]{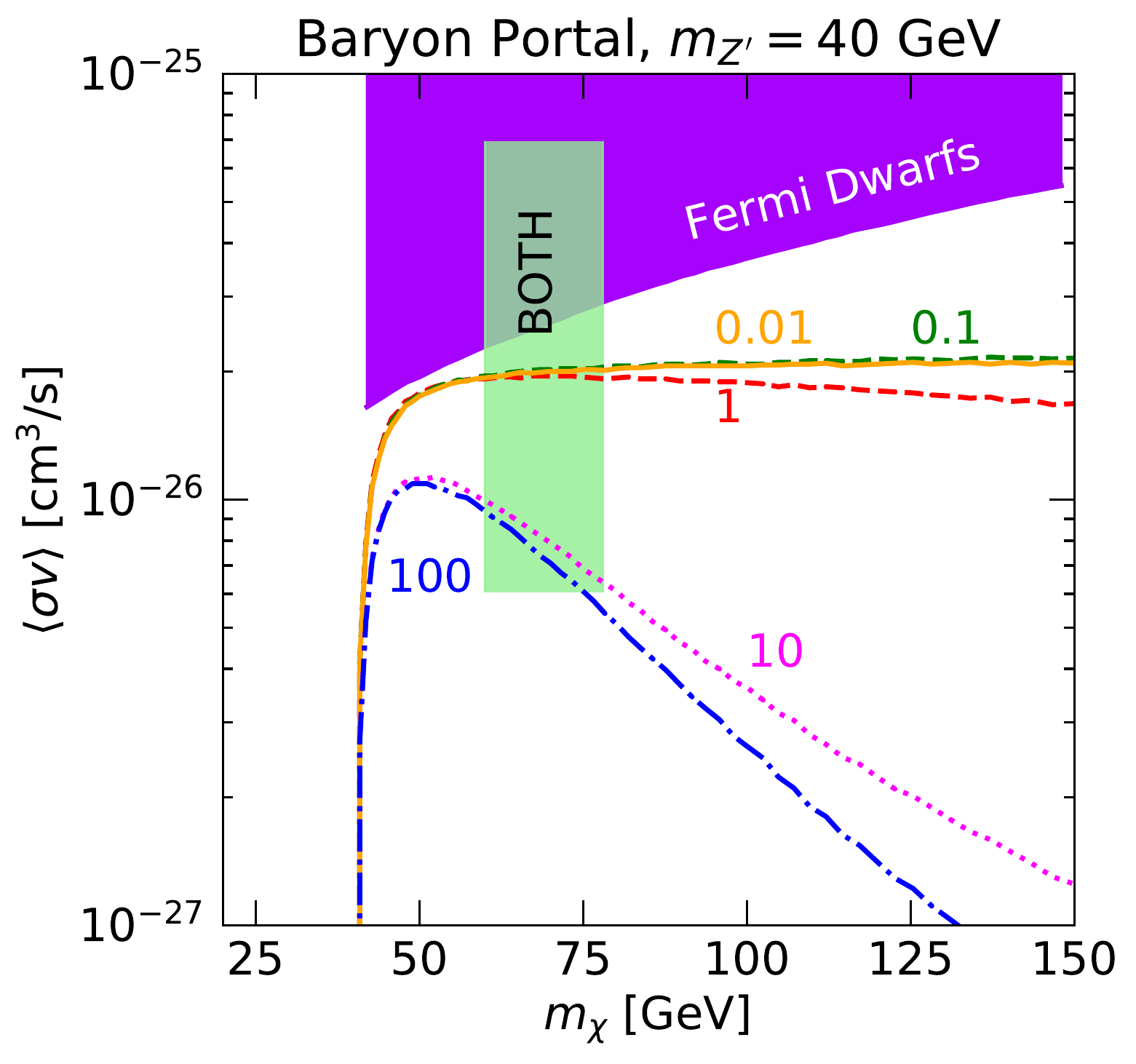}}
\hspace{1mm}
\subfloat{\includegraphics[width=0.49\columnwidth]{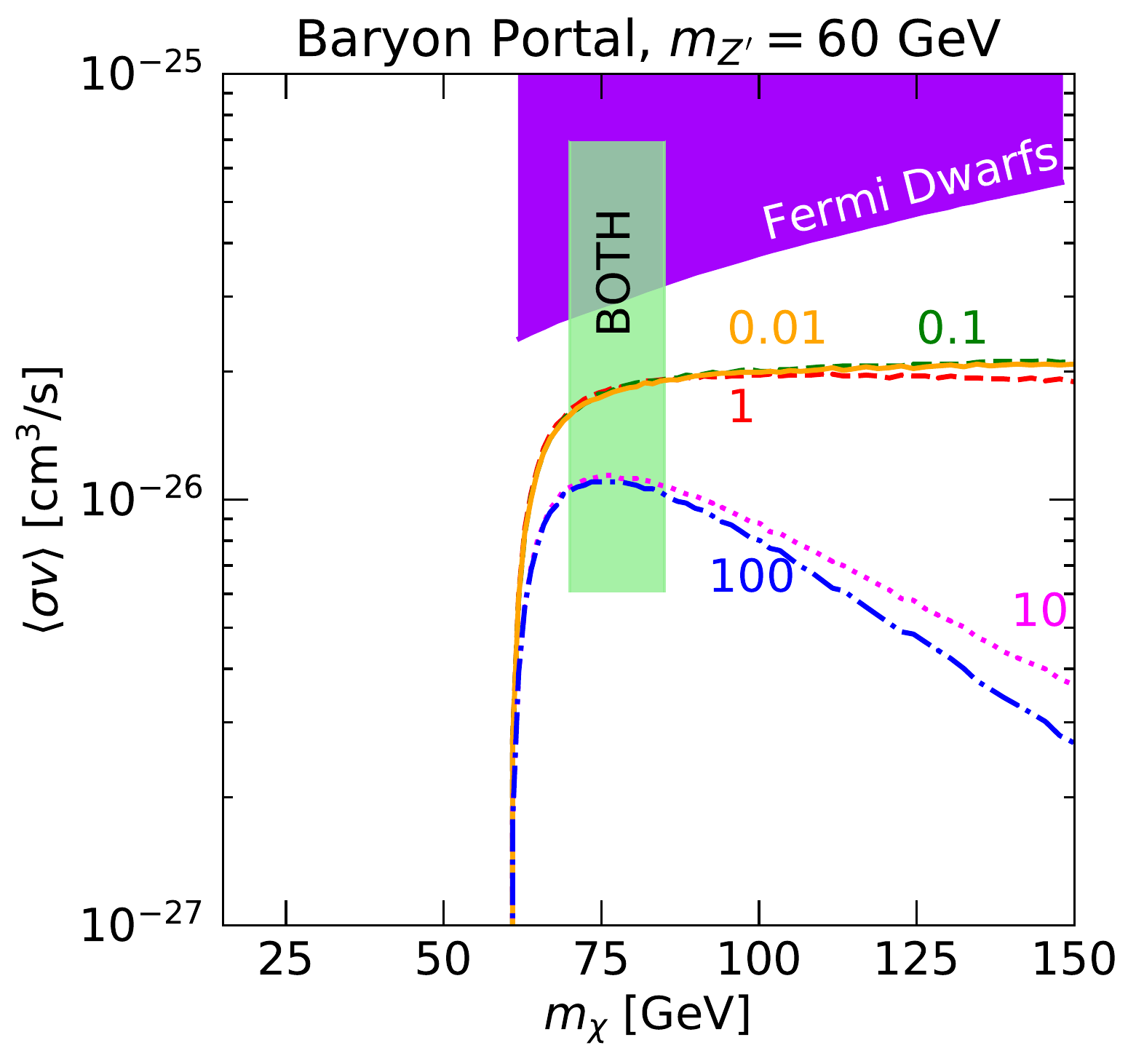}}
\caption{As in Fig.~\ref{fig:hypercharge}, but for the case of hidden sector dark matter that annihilates to particles that decay through the baryon portal.\label{fig:baryons}}
\end{figure*}

We next turn our attention to dark matter candidates that annihilate into particles that decay to the SM through either the $B-L$ portal or through the baryon portal. In Figs.~\ref{fig:zpbl} and~\ref{fig:baryons}, we show our results for this class of scenarios. In the case of the $B-L$ portal, we find only small regions of parameter space that can produce the observed characteristics of the gamma-ray and antiproton excesses. For the baryon portal, larger regions of parameter space (with $m_{\chi} \approx 50-100$ GeV and $m_{Z'} \approx 10-100$ GeV) can produce signals that can accommodate both excesses.

\begin{figure*}[t!]
\leavevmode
\centering
\subfloat{\includegraphics[width=0.46\columnwidth]{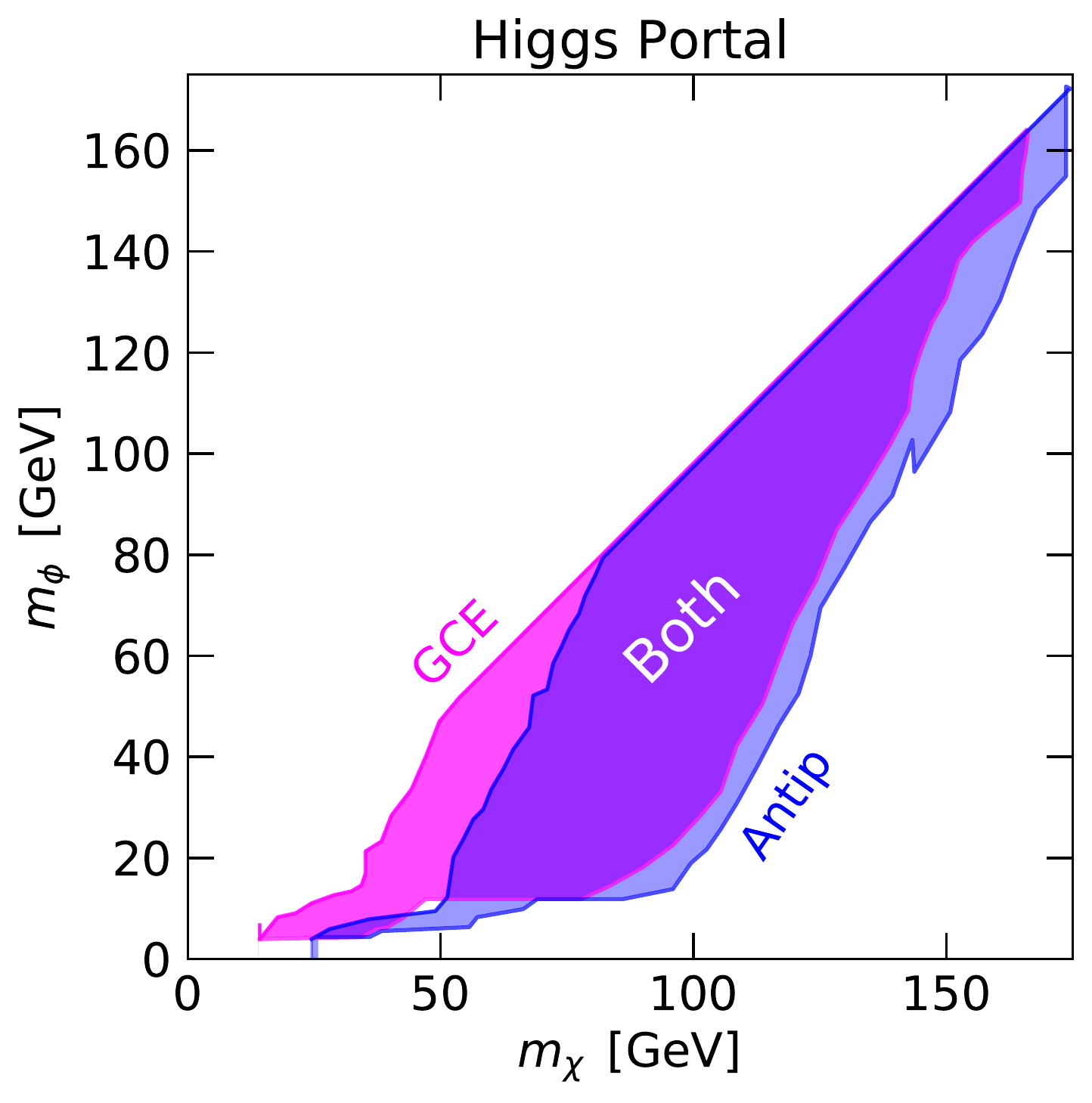}}
\hspace{1mm}
\subfloat{\includegraphics[width=0.49\columnwidth]{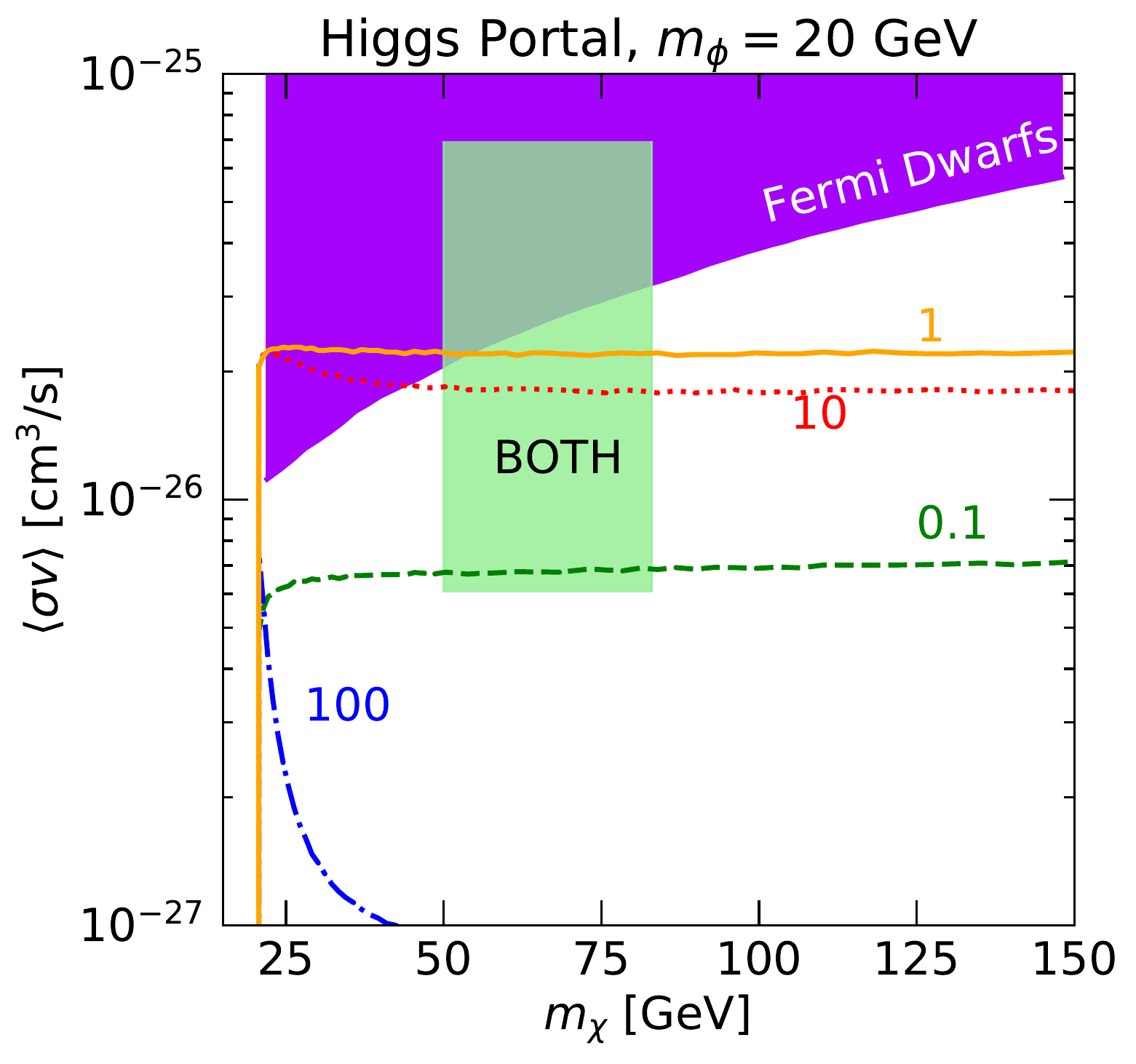}}\\
\subfloat{\includegraphics[width=0.49\columnwidth]{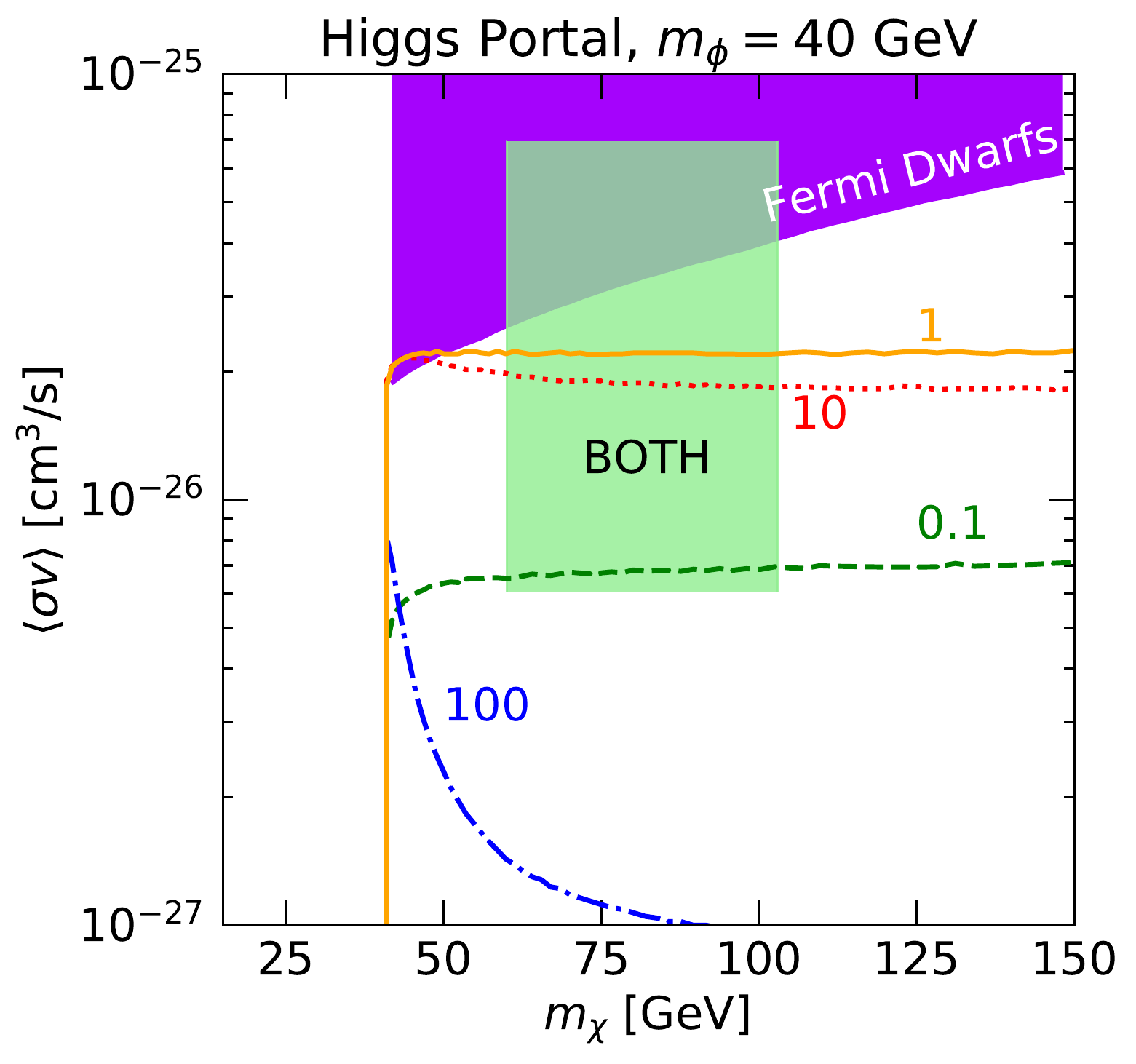}}
\hspace{1mm}
\subfloat{\includegraphics[width=0.49\columnwidth]{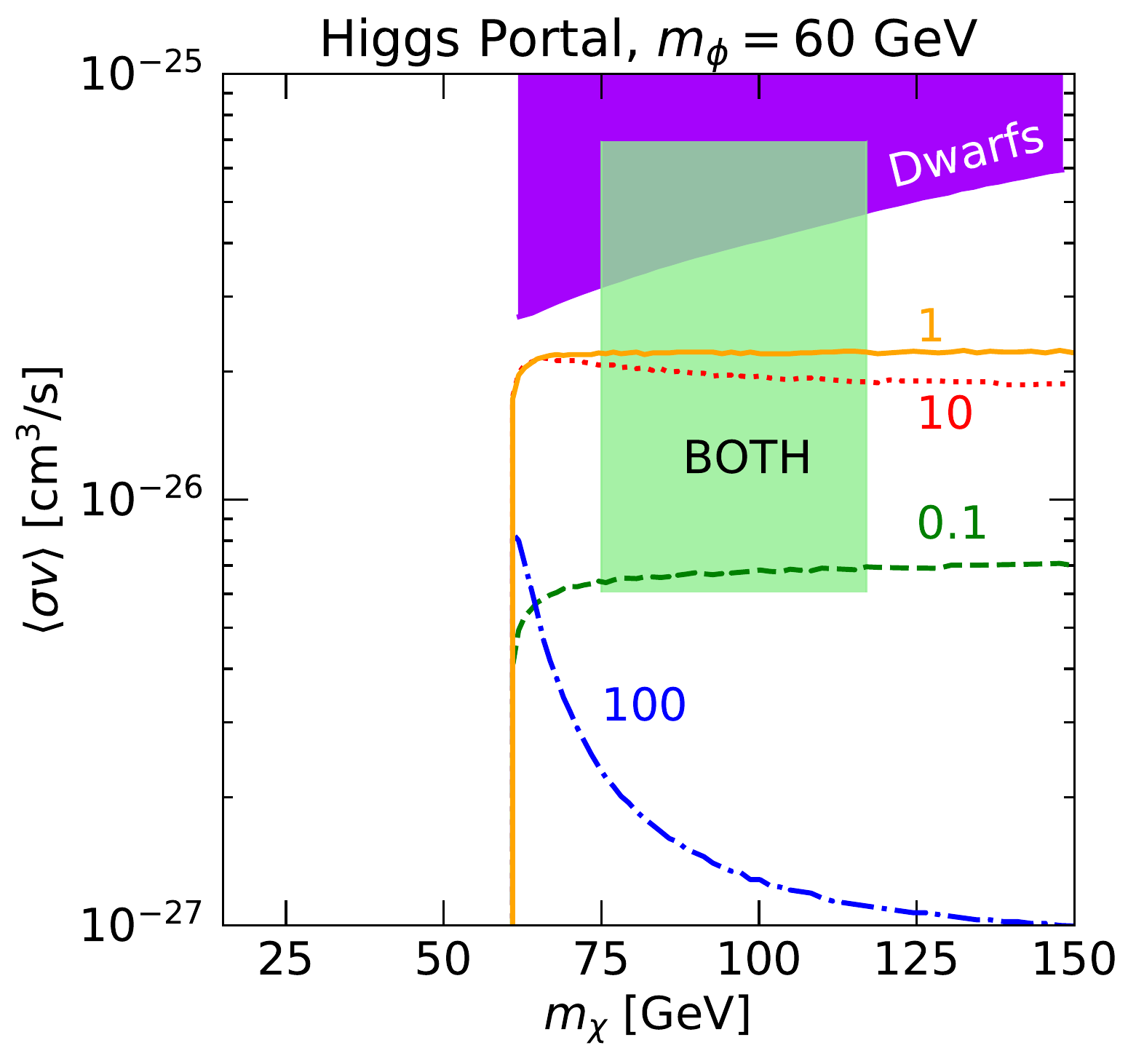}}
\caption{As in the previous three figures, but for the case of hidden sector dark matter that annihilates to particles that decay through the Higgs portal. Each line corresponds to a different ratio of the scalar and pseudoscalar couplings, $\lambda_s / \lambda_{p}$. This class of models provides the largest regions of parameter space that can accommodate both excesses.\label{fig:higgs}}
\end{figure*}

\begin{figure*}[t!]
\centering
{\includegraphics[width=0.48\columnwidth]{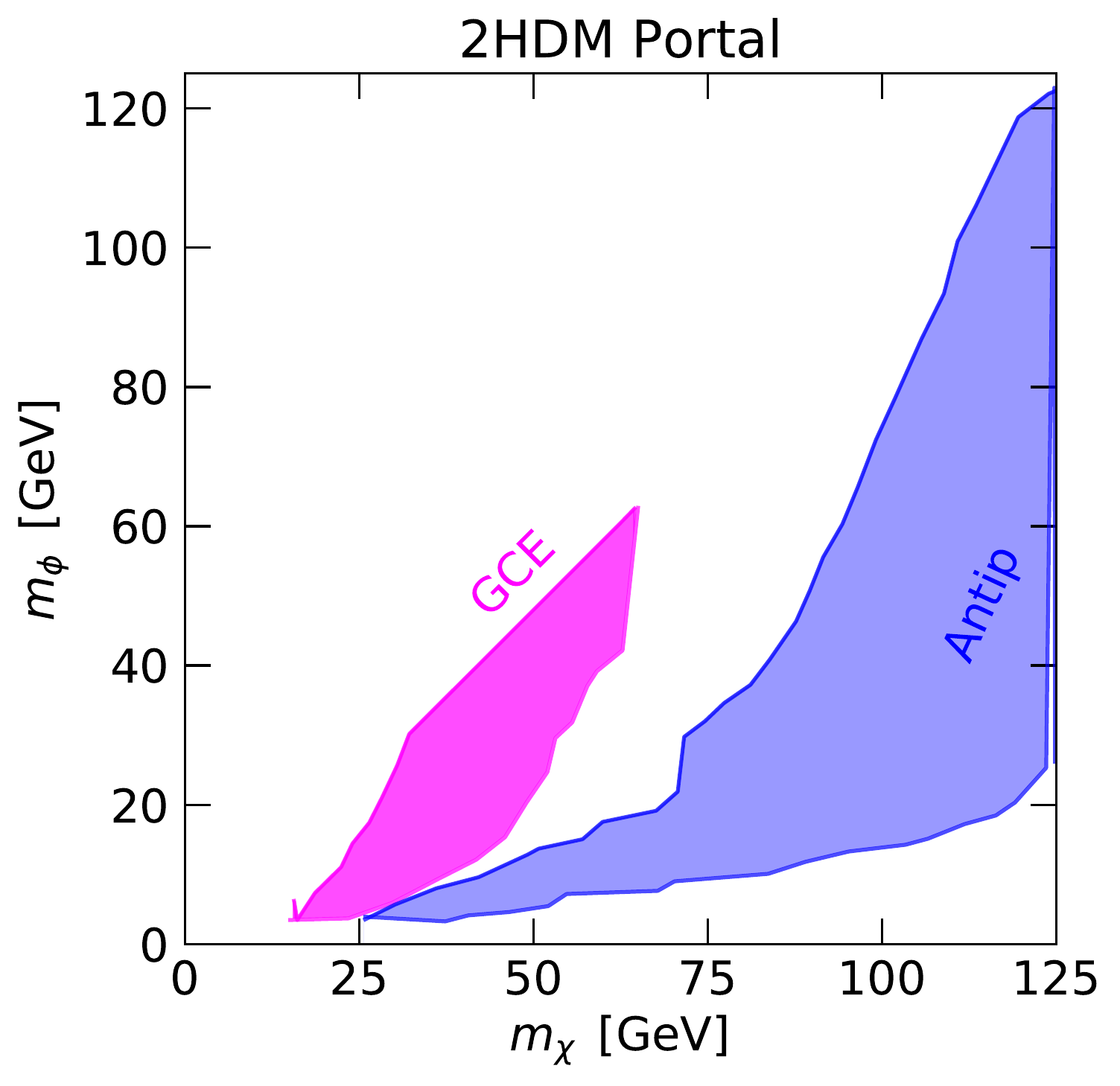}}
\caption{As in Fig.~\ref{fig:hypercharge}, but for the case of hidden sector dark matter that annihilates to particles that decay through the 2HDM portal, considering a Type-II model with $\tan \beta =1$ and $\sin \alpha =0$. In this scenario, there is no parameter space that can simultaneously accommodate the observed spectra of the gamma-ray and antiproton excesses. \label{fig:2hdm}}
\end{figure*}

As shown in Fig.~\ref{fig:higgs}, models in which the dark matter's annihilation products decay through the Higgs portal are particularly well suited to produce the observed features of the gamma-ray and antiproton excesses, favoring parameter space in which $m_{\chi} \approx 50-150$ GeV and $m_{Z'} \approx 10-150$ GeV. In the case of the 2HDM portal, our results depend on the parameters chosen. Across much of the parameter of this scenario, the branching fraction to $b$ quarks is large, leading to results similar to those shown for the case of the Higgs portal in Fig.~\ref{fig:higgs}. Alternatively, we have also considered the limit of the Type-II model in which $\tan \beta=1$ and $\sin \alpha=0$, for which the lightest Higgs decays largely to a combination of $c$ quarks, tau leptons, and gluons (see Fig.~\ref{figpie}). As shown in Fig.~\ref{fig:2hdm}, however, we find no combination of masses in this scenario that can simultaneously accommodate the observed spectra of the gamma-ray and antiproton excesses.

%%%%%%%%%%%%%%%%%%%%%%%%%%%%%%%%%%%%%%%%%%%%%%%%

\begin{figure*}[t!]
\leavevmode
\centering
\subfloat{\includegraphics[width=0.48\columnwidth]{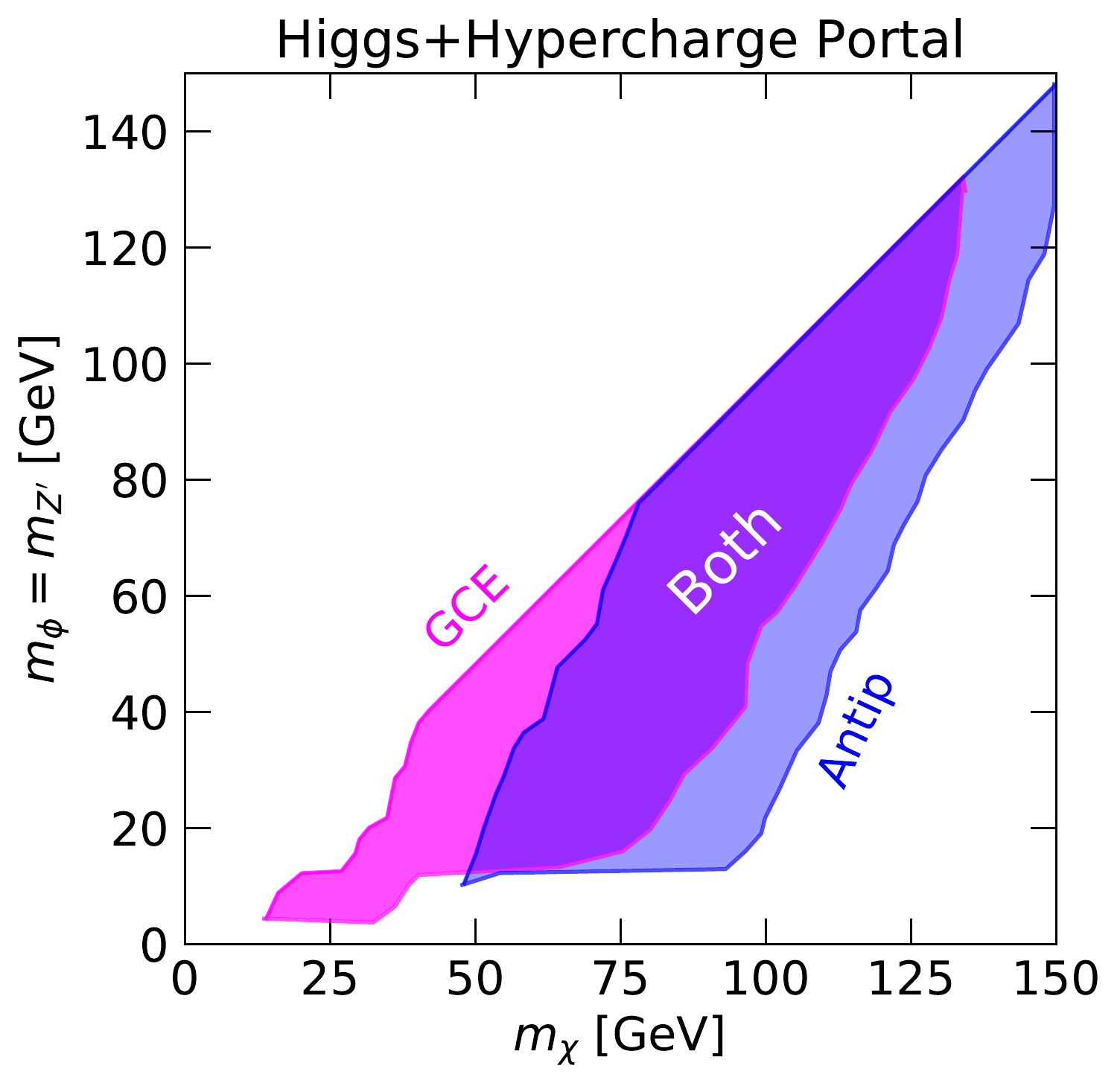}}
\hspace{1mm}
\subfloat{\includegraphics[width=0.49\columnwidth]{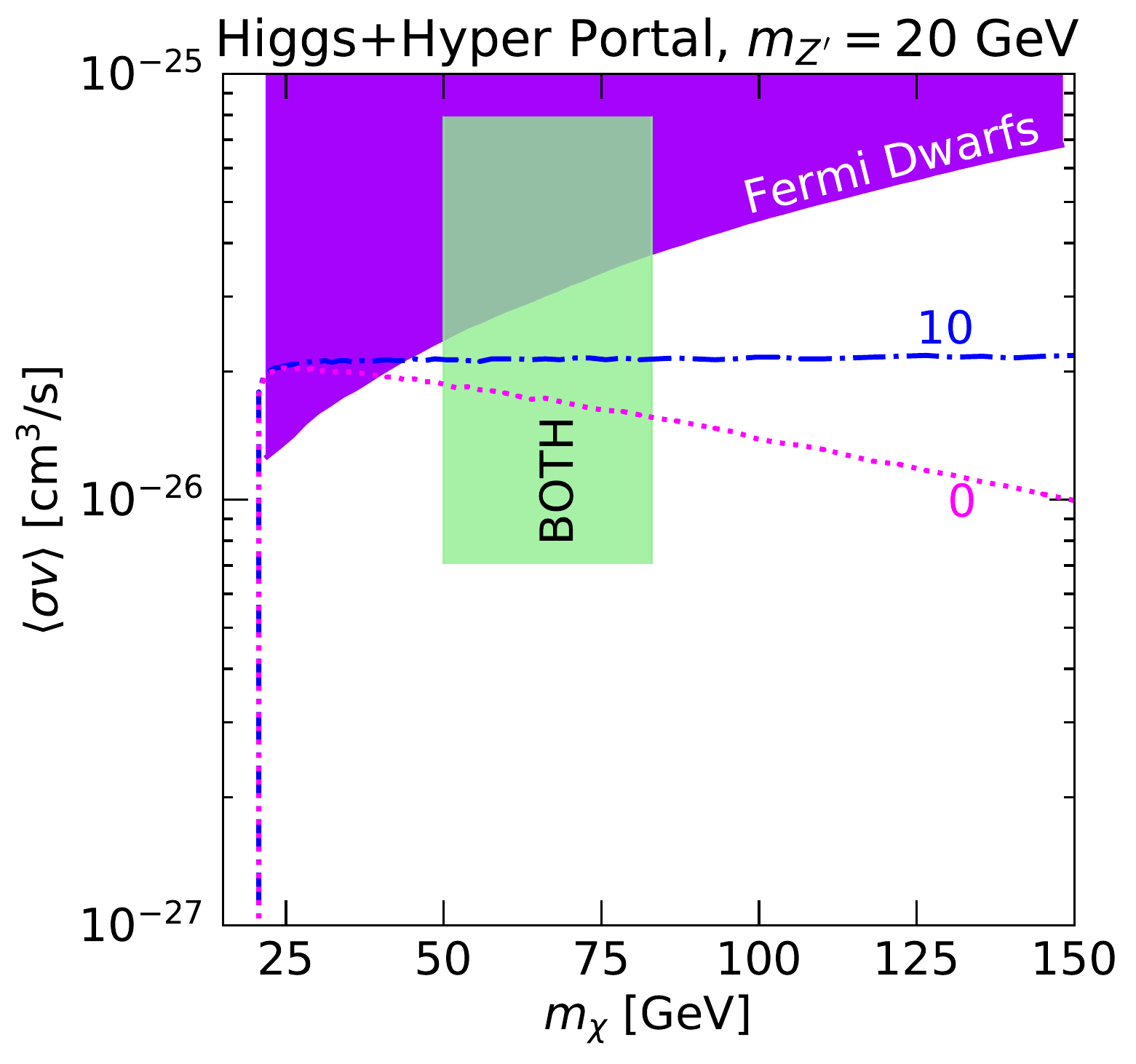}}\\
\subfloat{\includegraphics[width=0.49\columnwidth]{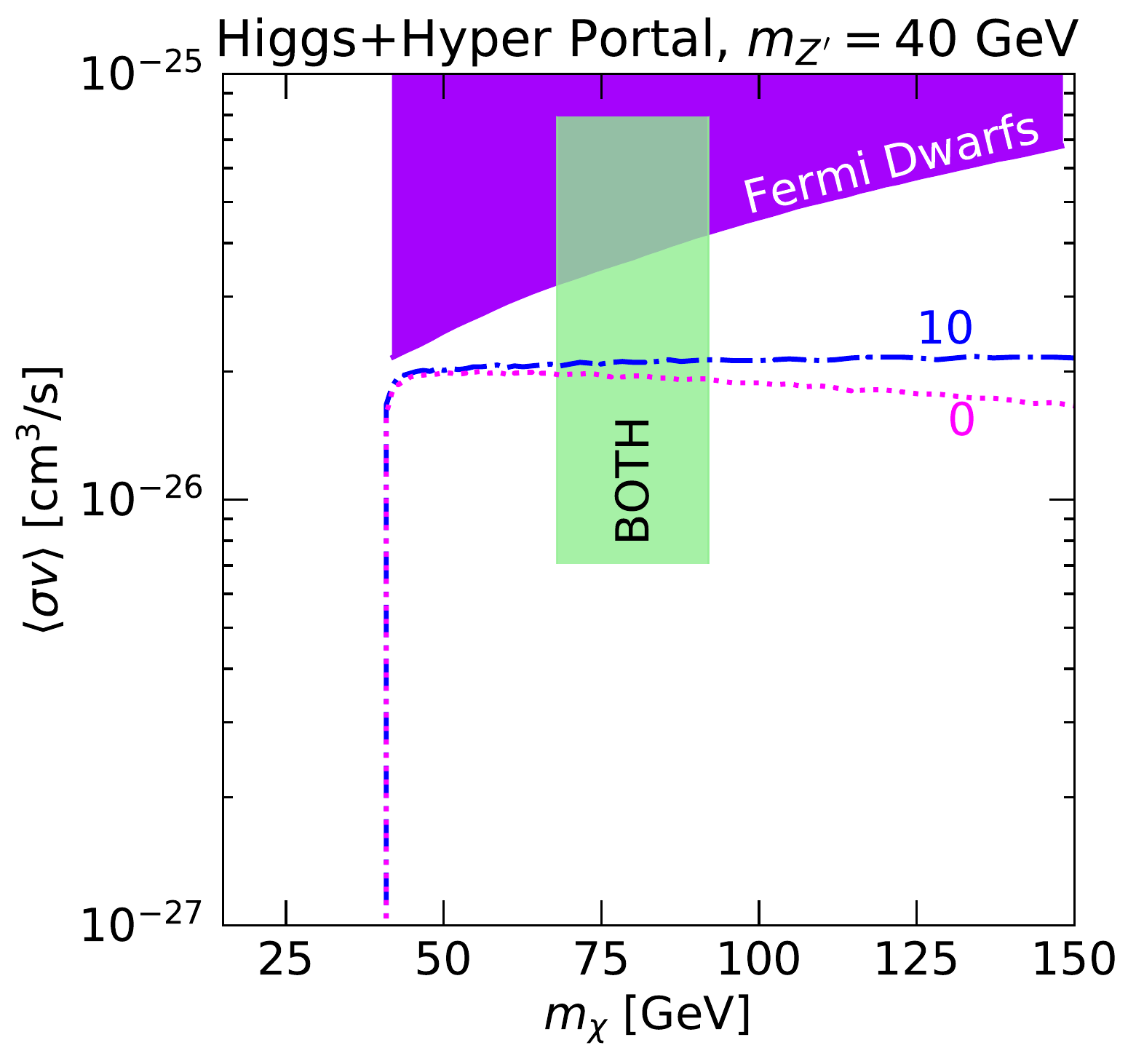}}
\hspace{1mm}
\subfloat{\includegraphics[width=0.49\columnwidth]{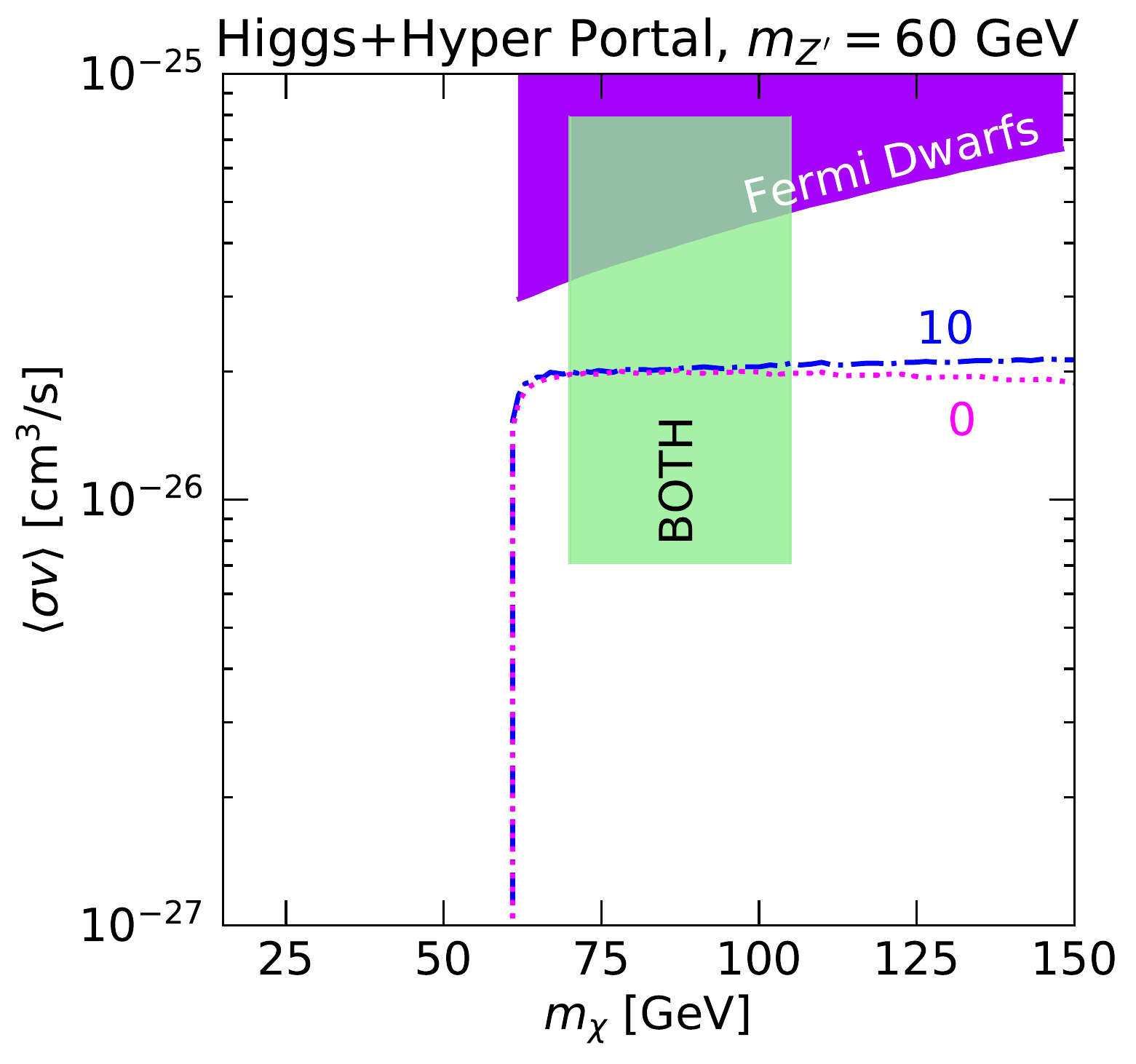}}
\caption{As in the previous figures, but for the case of hidden sector dark matter that annihilates to particles that decay through both the hypercharge and Higgs portals. We have chosen to show results here for the case of $m_{\phi}=m_{Z'}$. For consideration of other values of $m_{\phi}/m_{Z'}$, see Refs.~\cite{Bell:2016uhg,Escudero:2017yia}. In this figure, we have taken the dark matter candidate to be a Dirac fermion with equal vector, axial, scalar, and pseudoscalar couplings. We have also introduced an interaction of the form $\mu \phi Z_\mu Z^\mu$, and show results for $\mu=0$ and 10 GeV. \label{fig:higgshyper}}
\end{figure*}

Lastly, in Fig.~\ref{fig:higgshyper}, we consider models in which the dark matter annihilates into a combination of spin-$1$ and spin-$0$ states, $\chi \chi \rightarrow Z' + \phi$, which then decay to the SM through the hypercharge and Higgs portals, respectively. Such a scenario is well-motivated within the context of generating masses in the dark sector, as well as by the requirement of dark gauge invariance~\cite{Kahlhoefer:2015bea,Bell:2016fqf,Duerr:2016tmh,Bell:2016uhg}. For simplicity, we focus on the case in which $m_{\phi}=m_{Z'}$, and in which the dark matter is a Dirac fermion with equal vector, axial, scalar, and pseudoscalar couplings. We also introduce a $\mu \phi Z_\mu Z^\mu$ interaction between the vector and scalar in the hidden sector. The quantity $\mu$, being dimensionful, is typically expected to be proportional to the hidden sector vev, and we show results for $\mu=$0 or 10 GeV.

Note that as we are considering annihilation of hidden sectors, the results presented in this section rely solely on the dark coupling, rather than the coupling to SM. As such, there are a range of coupling values that can be taken to suppress direct detection or collider constraints without tension. We provide a detailed discussion of potential reach of direct detection in the following section.

\section{Cosmological Considerations and Prospects for Direct Detection}
\label{cosmo}

Throughout this study, we have calculated the dark matter's relic abundance assuming that the hidden sector was in equilibrium with the particle content of the SM during the process of freeze-out. This will be the case only if the scattering rate between the two sectors exceeded the rate of Hubble expansion during or prior to the era in which freeze-out occurred. 

It is possible that the dark matter is part of a hidden sector that was never in equilibrium with the SM. In that case, however, we would have no reason to expect the dark matter's annihilation cross section to be near the range of values that are required to produce the Galactic Center gamma-ray excess or cosmic-ray antiproton excess, $\langle \sigma v \rangle \sim 10^{-26}$ cm$^3/$s. With this in mind, we chose to focus on scenarios in which the portal interaction is strong enough to maintain equilibrium between the two sectors prior to dark matter freeze-out -- that is to say, we impose a minimum on the portal coupling based on the requirement that the hidden sector was in equilibrium at high energies.

\begin{figure*}[t!]
\centering
\subfloat{\includegraphics[width=0.48\columnwidth]{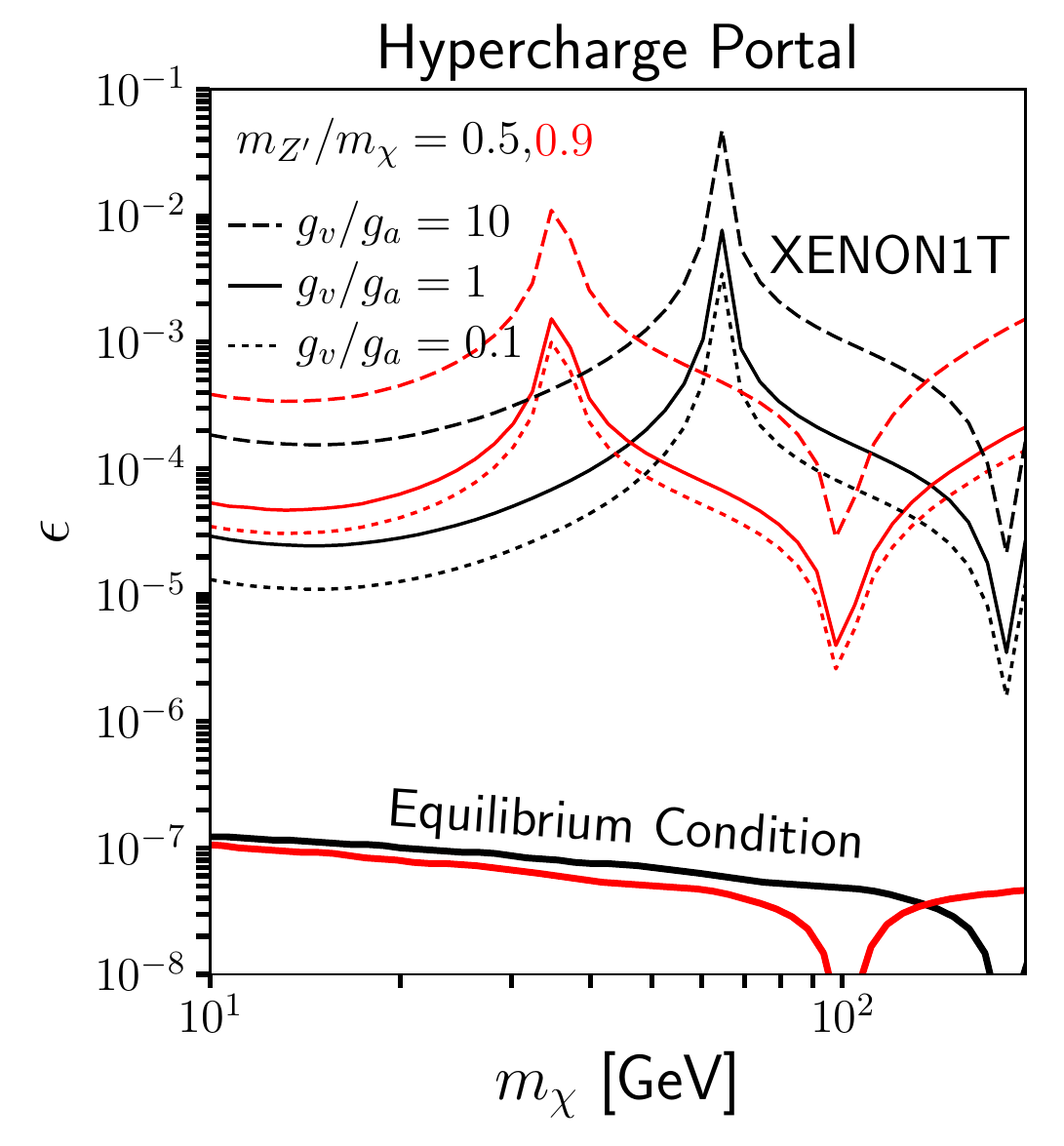}}
\subfloat{\includegraphics[width=0.48\columnwidth]{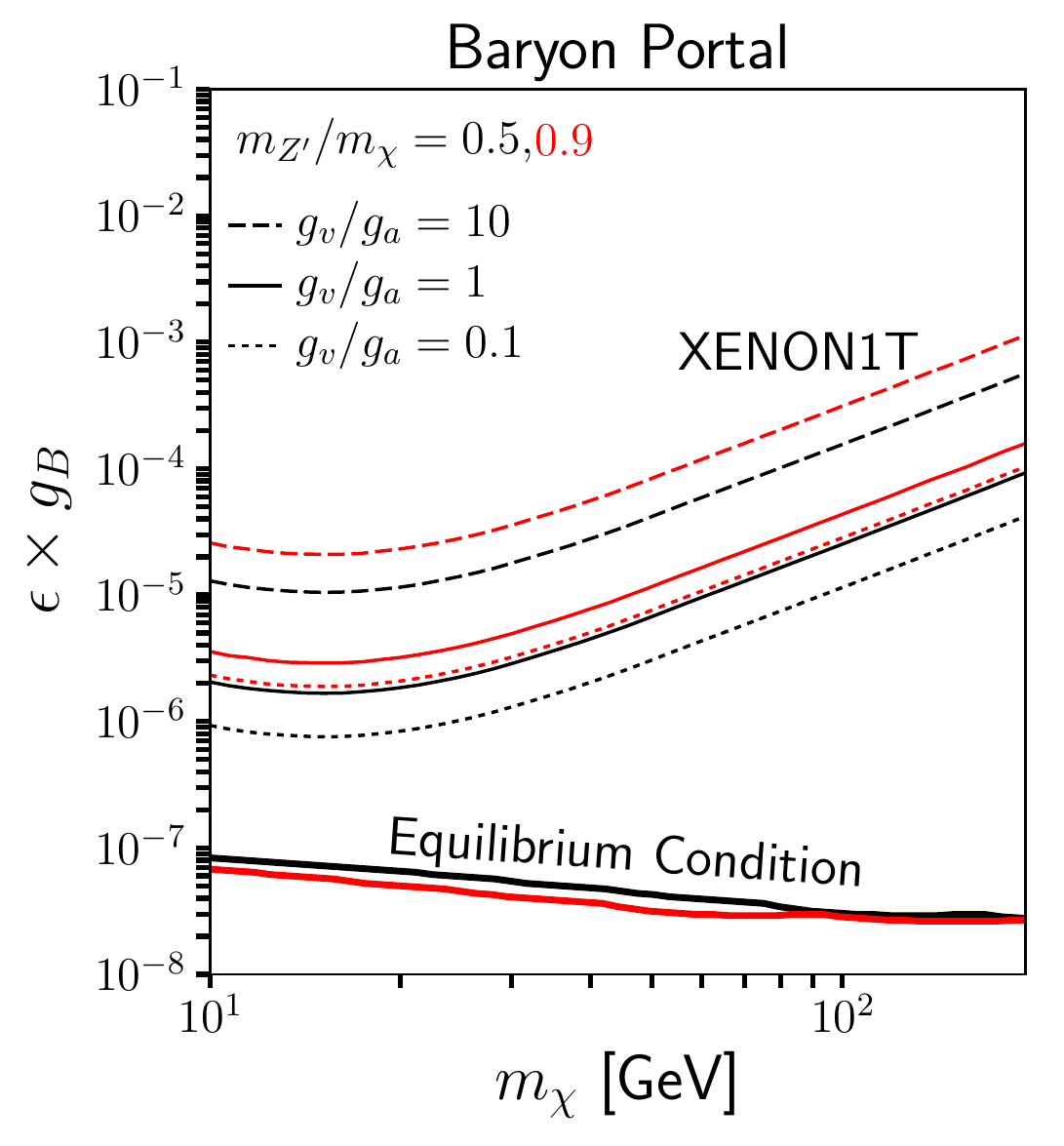}}\\
%\vspace{8mm}
\subfloat{\includegraphics[width=0.48\columnwidth]{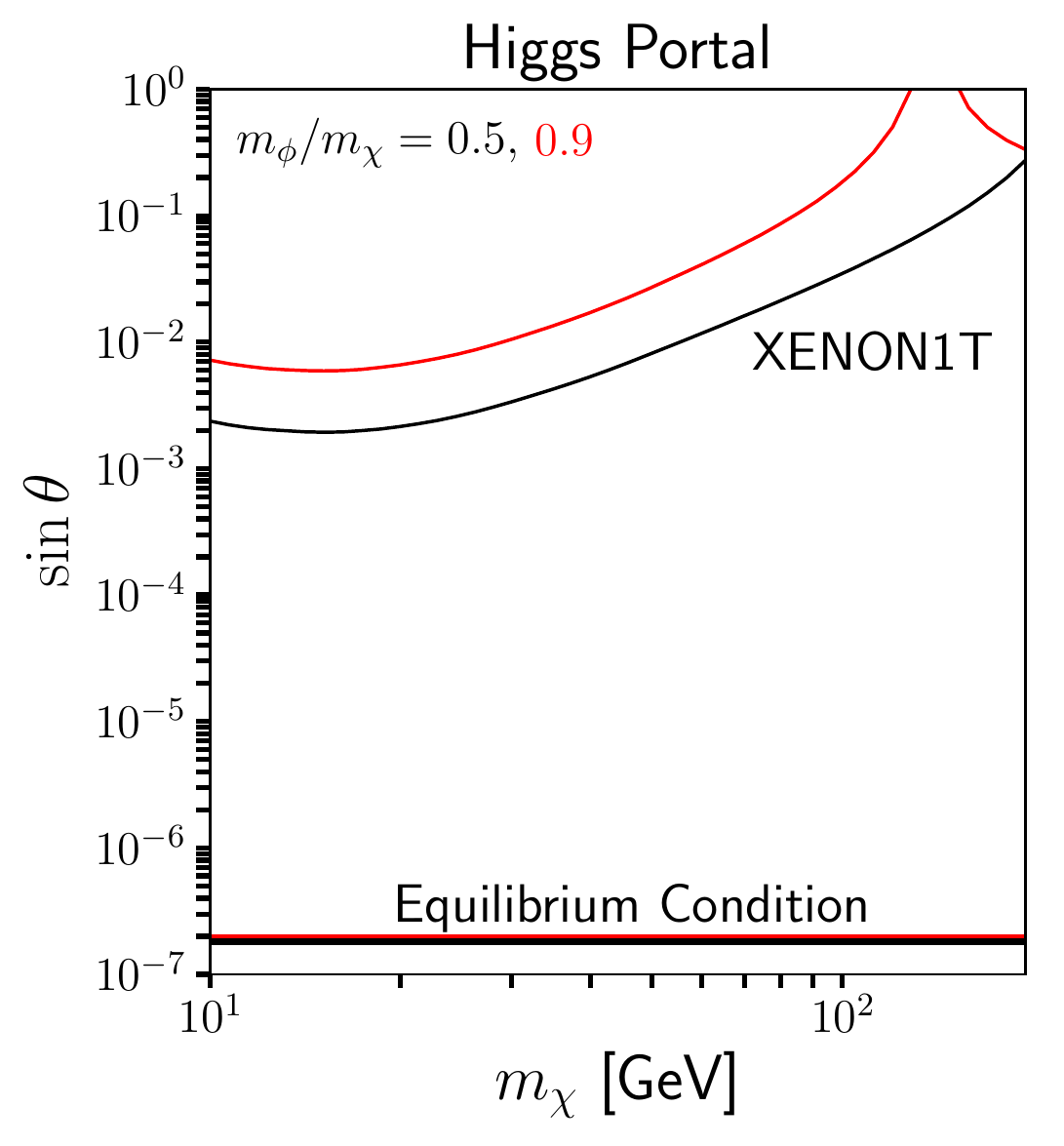}}
\caption{The thick solid lines denote the minimum values of portal coupling that are required to maintain equilibrium between the hidden and SM sectors leading up to the time of dark matter freeze-out. In the case of the Higgs portal, we plot this condition for the case of $v_{\phi} = 100$ GeV, where $v_{\phi}$ is the hidden sector vev. The thin lines represent the constraints from XENON1T, for the case of spin-independent scattering between dark matter and nuclei. Note that future experiments with sensitivity near the ``neutrino floor'' will improve on these constraints by a factor of $\sim$10-20 (in terms of the quantities shown on the $y$-axes of this figure). Accounting for $2\leftrightarrow 1$ processes and in-medium productions of dark photon from $2\leftrightarrow 2$ scatterings may alter the minimum allowed kinetic mixing by a factor of a few \cite{Hardy:2016kme,Evans:2017kti}.
\label{cosmodirect}}
\end{figure*}

To evaluate the constraint on the portal coupling that is imposed by this requirement, we calculate the scattering rates associated with the processes $Z' f \leftrightarrow \gamma f, Zf$~\cite{Berlin:2016vnh,Berlin:2016gtr} and $hh \leftrightarrow \phi \phi$~\cite{Escudero:2017yia}, for the case of portals involving spin-1 or spin-0 particles, respectively. In Fig.~\ref{cosmodirect}, we show the minimum value of the portal coupling that satisfies this equilibrium condition for the cases of the hypercharge, baryon, and Higgs portal models. For the hypercharge portal, this corresponds to a constraint on the degree of kinetic mixing, $\epsilon$, between the sectors, while in the case of the baryon portal, we derive a constraint on the degree of kinetic mixing multiplied by the $U(1)_B$ gauge coupling, $\epsilon \times g_B$. Note that accounting for $2\leftrightarrow 1$ processes and in-medium productions of kinetically mixed dark photon from $2\leftrightarrow 2$ scatterings may alter the minimum allowed kinetic mixing by a factor of a few \cite{Hardy:2016kme,Evans:2017kti}.
In the case of the Higgs portal, this condition is satisfied for $\sin \theta \gsim 2\times 10^{-7} \times (v_{\phi}/100 \, {\rm GeV})$, where $v_{\phi}$ is the value of the hidden sector vev (we plot the result for $v_{\phi}=100 \,{\rm GeV}$). In each case, we consider equilibrium to be established if the scattering rate between the two sectors exceeds the rate of Hubble expansion for any temperature between 1 GeV and 1 TeV.

If it were not for the requirement of equilibrium, the portal interaction connecting the dark matter to the SM could be extremely feeble, leaving little reason for one to be optimistic about future direct detection (or collider\footnote{We note that collider bounds can be at best roughly comparable than the optimistic direct detection bounds -- we leave a detailed collider study of these models to future work.}) efforts. But the condition of equilibrium allows us to place a lower bound on this coupling, and on the elastic scattering cross section of the dark matter with nuclei. Thus we believe it is of value to address the extent to which current and future direct detection experiments can probe these portal couplings (it is perhaps important to note, however, that direct detection and collider experiments are unlikely to ever fully exclude such models).

In Fig.~\ref{cosmodirect}, we plot the constraints on this parameter space as derived from the latest results of the XENON1T Experiment~\cite{Aprile:2017iyp} (see also Refs.~\cite{Akerib:2016vxi,Cui:2017nnn}). Whereas the curves denoting the equilibrium condition depend only on the ratio of the hidden sector particles' masses, the dark matter's elastic scattering cross section with nuclei depends on a number of other features of the model, including the spin of the dark matter candidate and its dominant annihilation diagram. The XENON1T constraints shown in this figure correspond to the most optimistic of these cases, in which this scattering occurs through a spin-independent process, without any velocity or momentum suppression. In particular, we show results for dark matter in the form of a Dirac fermion in the cases of the hypercharge and baryon portals, and for a vector dark matter candidate in the case of the Higgs portal. In the hypercharge and baryon portal examples, we show results for three different ratios of the vector and axial couplings. In the baryon portal case, we consider the limit of $m_{Z_B} \gg m_{Z'}$. Expressions for the elastic scattering cross sections in each of these models can be found in Refs.~\cite{Cline:2014dwa,Escudero:2017yia}. For each of the direct-detection curves (XENON1T), the parameters are fixed to produce the right relic abundance -- note that relic abundance does not depend on the mixing, and thus this can be independently constrained by direct detection experiments. 

From Fig.~\ref{cosmodirect}, it is clear that direct detection experiments already meaningfully constrain some of the hidden sector dark matter models we have considered in this study. 
If the direct-detection (XENON1T) curves shown in Fig.~\ref{cosmodirect} dip below the curves labeled `Equilibrium Condition', the process through which the dark matter abundance is produced can be markedly different; in this case one would not expect the annihilation cross section to be similar to either the value  produced via conventional dark matter freeze-out or the value required in order to explain the existence of the gamma-ray and anti-proton excesses. Consequently, such  models should be interpreted as interesting with regard to the astrophysical anomalies.

Furthermore, as experiments become more sensitive and approach the so-called ``neutrino floor''~\cite{Aalbers:2016jon}, we expect these constraints to improve by a factor of $\sim$10-20 (in terms of the quantities shown on the $y$-axes of this figure). This will cover a significant fraction of the parameter space that lies between the current constraints and the condition of equilibrium. One should keep in mind, however, that the cases shown are among the most optimistic, and scenarios that do not lead to unsuppressed spin-independent scattering will be much more difficult to test with current or future direct detection experiments.

\section{Summary and Conclusions}
\label{summary}

As constraints from direct detection and accelerator experiments have become more stringent, the motivation to consider dark matter candidates that do not directly couple to the particle content of the Standard Model has increased. In scenarios in which the dark matter is part of a hidden sector that does not carry any Standard Model gauge charges, direct detection and accelerator signals can be highly suppressed. In many such models, however, dark matter particles can annihilate efficiently, leading to potentially detectable gamma-ray or cosmic-ray signals. In particular, we show in this study that a wide range of hidden sector dark matter models can account for the observed features of the Galactic Center gamma-ray excess, as well as the more recently identified cosmic-ray antiproton excess.

In Sec.~\ref{models} of this paper, we described our comprehensive study of annihilating dark matter, identifying the combinations of dark matter spins, mediator spins, interaction types, and annihilation final states that allow for the efficient ($s$-wave) annihilation of dark matter particles at low-velocity (as is relevant for the case of annihilations in the Galactic Halo). These results are summarized in Tables~\ref{tab:lorentzfermion} and~\ref{tab:lorentzboson} and can be applied to a wide range of dark matter scenarios, both within the context of hidden sector models and otherwise. 

In hidden sector dark matter models, the annihilation products decay into Standard Model particles through one or more portal interactions. In this paper, we have considered decays through a variety of such portals, including the hypercharge portal and the Higgs portal, as well as through portals that connect the hidden sector to extensions of the Standard Model in which baryon number or baryon-minus-lepton number is gauged, or within the context of models with an extended Higgs sector. In each case, we calculated the branching fractions of the hidden sector annihilation products and the resulting spectrum of gamma rays and antiprotons.

We find that the observed features of the gamma-ray and antiproton excesses can simultaneously be accommodated within a wide range of hidden sector dark matter models. More specifically, models in which the dark matter's annihilation products decay through the Higgs portal, the hypercharge portal, the baryon portal or the baryon-minus-lepton portal can each produce acceptable spectra of both gamma ray and antiprotons. Although we consider constraints derived from gamma-ray observations of dwarf galaxies, we find that these observations do not significantly restrict the range of viable parameter space within this class of models.

Lastly, we have considered the prospects for direct detection in this class of hidden sector dark matter scenarios. Although the portal couplings in such models could, in principle, be extremely small, the intensity of the gamma-ray and antiproton excesses suggests that the hidden sector was in equilibrium with the Standard Model at the time of dark matter freeze-out, providing us with a way of placing a lower limit on the portal coupling. In light of this information, and within the subset of models that feature spin-independent scattering between dark matter and nuclei, the prospects for direct detection appear promising, despite the small couplings that connect the hidden Standard Model to the particle content of the Standard Model.

\section*{Acknowledgments}

We would like to thank Miguel Escudero, Tracy Slatyer, Jessie Shelton, Gordan Krnjaic, Sam McDermott, and Hitoshi Murayama for helpful discussions. This manuscript has been authored by Fermi Research Alliance, LLC under Contract No. DE-AC02-07CH11359 with the U.S. Department of Energy, Office of High Energy Physics. RKL is supported by the Office of High Energy Physics of the U.S. Department of Energy under Grant No. DE-SC00012567 and DE-SC0013999, as well as the Fermi Guest Investigator Program under Grant No. 80NSSC19K1515. SJW acknowledges support under Spanish grants FPA2014-57816-P and FPA2017-85985-P of the MINECO and PROMETEO II/2014/050 of the Generalitat Valenciana, and from the European Union’s Horizon 2020 research and innovation program under the Marie Sklodowska Curie grant agreements No. 690575 and 674896.

\bibliography{biblio}
\end{document}